\documentclass[useAMS,usenatbib]{mn2e}
\usepackage[fleqn]{amsmath}
\usepackage{graphicx}
\usepackage{color}

\title[Formation of disc galaxies around z$\sim$2]{Formation of disc galaxies around z$\sim$2}
\author[Sachdeva et al.]{Sonali Sachdeva$^{1}$, Rupjyoti Gogoi$^{2}$, Kanak Saha$^{3}$, Ajit Kembhavi$^{3}$, Somak Raychaudhury$^{3}$\\
$^{1}$Kavli Institute for Astronomy and Astrophysics, Peking University (KIAA PKU), Beijing 100871, P.R. China\\
$^{2}$Department of Physics, Tezpur University, Napaam 784028, India\\
$^{3}$Inter-University Centre for Astronomy and Astrophysics (IUCAA), Pune 411007, India}
\begin{document}

\date{in original form 2019 May 21}

\pagerange{\pageref{firstpage}--\pageref{lastpage}} \pubyear{2019}

\maketitle

\label{firstpage}

\begin{abstract}

We present combined evolution of morphological and stellar properties of galaxies on the two sides of $z$$=$2 (2.0$<$$z$$<$4.0 and 1.5$<$$z$$<$2.0) in CDFS, with ground-based spectroscopic redshifts. We perform bulge-disc decomposition on their images in {\it J} and {\it H} filters, from the 3DHST Legacy Survey obtained using HST/WFC3. Combining morphological information with stellar properties, we provide a detailed account of the formation/growth of discs and spheroids around $z$$\sim$2. The fraction of 2-component (bulge$+$disc) systems increases from $\sim$46\% for $z$$>$2 to $\sim$70\% for $z$$<$2, compensating for the fall in population of pure discs and pure spheroids. All quiescent outliers of our full sample on the main-sequence, are 2-component systems, belonging to the lower redshift range ($z$$<$2). The doubling of stellar mass of 2-component systems and decrease in their SFR by the same factor, suggests that mechanisms involved in morphological transformations are also responsible for the quenching of their star formation activity. Interestingly, while there is substantial increase in the size ($\sim$2.5 times) and mass ($\sim$5 times) of pure discs, from $z$$>$2 to $z$$<$2, pure spheroids maintain roughly the same values. Additionally, while bulge hosting discs witness an expansion in scale length ($\sim$1.3 times), their bulge sizes as well as bulge to total light ratio see no evolution, suggesting that $z$$\sim$2 is pre-dominantly a disc formation period.

\end{abstract}



\section{Introduction}
\label{sec:intro}

Deciphering the formation of the Hubble sequence of galaxies is at the heart of Astrophysics. While there are plausible theories and simulations explaining the formation of discs and ellipticals/spheroids, observational evidence to support (or discard) the theories is at present inadequate. Until recently, major mergers were proposed and widely accepted to be the primary mechanism for their formation, where mass of the merging primordials and fraction of the gas involved determines the likelihood of the presence of the two components in the merging outcome \citep{BarnesandHernquist1996,SpringelandHernquist2005,Bournaudetal2005,Hopkinsetal2009}. Lately, however, there is growing evidence that major and minor merger rates \citep{Conselice2006,Blucketal2012}, even during the crucial formative years ($z$$>$1.5), are not sufficient to explain the stellar mass assembly. Therefore, other mechanisms, mainly gas accretion, should have played a larger role \citep{Baueretal2011,Conseliceetal2013,Santinietal2014,Scovilleetal2016}. It is thus presently unresolved whether discs predominantly formed during mergers or were later accreted around pre-existing spheroids.

Such ambiguities regarding the dominant formation mechanisms can be resolved through simultaneous tracking of morphological and stellar parameters of galaxies during their formative years. Determination of both these properties at those redshifts has witnessed major advancement, since 2009, with the commissioning of {\it Wide Field Camera-3 (WFC3)} on the {\it Hubble Space Telescope}. Deep field extragalactic surveys, mainly {\it CANDELS} \citep{Groginetal2011,Koekemoeretal2011} and {\it 3DHST} \citep{Brammeretal2012,Skeltonetal2014,Momchevaetal2016}, utilizing the camera's capabilities, have produced multi-wavelength imaging and spectroscopic data for thousands of high redshift ($z$$>$1.5) galaxies at unprecedented resolution.

Following the observations, there have been quite a few attempts at discerning morphologies (and corresponding stellar parameters) of high redshift galaxies. While a couple of these were based on visual classification \citep{Mortlocketal2013,Huertas-Companyetal2015,Huertas-Companyetal2016}, broad structural decompositions, i.e., disc $+$ bulge/spheroid, have also been performed \citep{Bruceetal2012,Bruceetal2014,Langetal2014}. In these decompositions, morphology of the bulge (or spheroid) was assumed to be de-Vaucoulers (defined by S\'ersic-index equal to 4) and held fixed. An unbiased picture of structure formation requires that morphologies are measured without assumptions or generalizations and all parameters are allowed to be fitted. However, large number of free parameters coupled with low signal to noise ratio (SNR) and limited spatial resolution make unbiased structural measurements difficult. 

\citet{Margalef-Bentaboletal2016,Margalef-Bentaboletal2018} attempted such a decomposition (i.e., disc and free-bulge) for $\sim$1500 massive high redshift (1$<$$z$$<$3) galaxies using {\it Galfit} \citep{Pengetal2002}. To obtain the best fit, they probed the outcome for 27 different combinations of initial parameters. Even after such a thorough process and well defined constraints, they discovered that for a significant fraction ($\sim$60\%) of the sample, the outcomes were unrealistic, i.e., non-physical. Also, the results were riddled with the concern that the fit might have converged to a local minimum, not a global one.

This lack of robustness in estimating high redshift morphologies has led to contradictory findings regarding their presence and growth (or destruction) at various redshift epochs. For example, some state that the disc population was negligible at $z$$>$2, yet others report that the disc population has highest relative number density at all epochs from $z$$=$3 to $z$$=$1, yet others find that over the same period, galaxies move from being disc dominated to increasingly bulge dominated \citep[e.g.,][]{Mortlocketal2013,Bruceetal2014,Margalef-Bentaboletal2016}.

We propose that in such a scenario, computation and fitting of galaxies' mean isophotal radial intensity profile will enable us to obtain more accurate parameters \citep{Jedrzejewski1987}. The benefits of fitting the 2D image rather than the 1D averaged profile are indubitable, mainly because 2D image fitting algorithms are better equipped to account for inherent shape and presence of non-axis-symmetric features in the galaxy \citep{ByunandFreeman1995,deJong1996}. However, if the main aim is restricted to obtaining the underlying bulge and disc parameters, 1D fitting offers significant advantage in terms of increase in SNR obtained by smoothing over irregularities and noise \citep{Kent1986}. Even with 1D fitting, the problem of large number of fitting parameters and degenerate outputs persists. Based on ``marking the disc" method \citep{Freeman1977,Kormendy1977}, we have developed an algorithm that breaks-up the fitting process into multiple steps, reducing the number of free parameters, leading to realistic and robust decomposition results \citep{SachdevaandSaha2016,Sachdevaetal2017}. In modification to past implementations \citep[e.g.,][]{LaubertsandValentijn1989}, we deal with each profile in an non-automated manner. The disc is fitted interactively avoiding irregularities which can lead to unrealistic parameters \citep{Giovanellietal1994}. After fixing disc parameters, full function (i.e., free-bulge + disc) is fitted to the full profile to ensure that free-bulge accounts for only that light at the centre which is in addition to the underlying disc. In \citet{SachdevaandSaha2018} we have demonstrated through simulations that our procedure is adept at computing the actual global minimum solution for the galaxy's decomposition.

\begin{figure}
\mbox{\includegraphics[width=65mm]{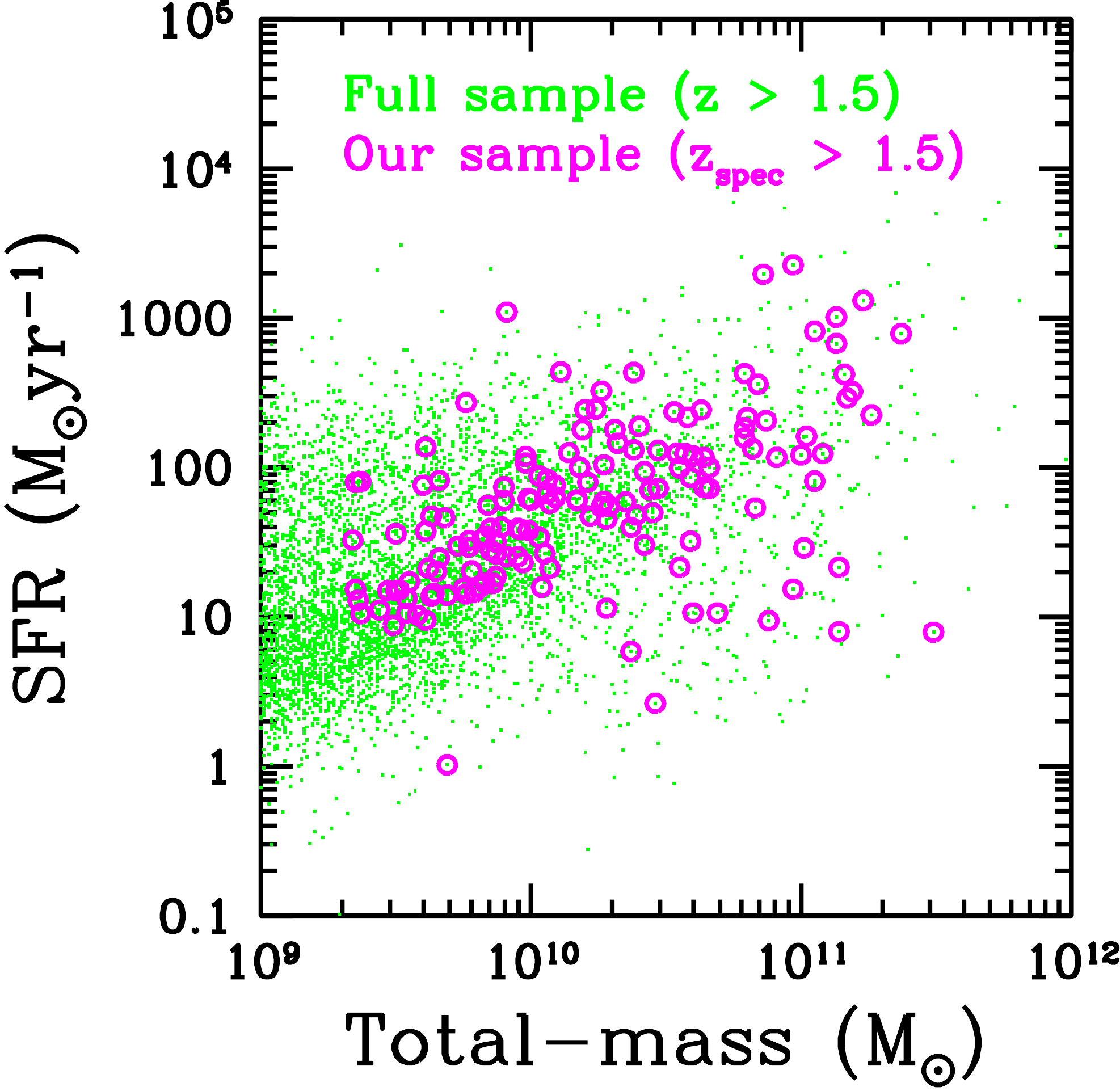}}
\caption{{\bf Sample selection:} Distribution of our sample, i.e., galaxies with ground based spectroscopic redshifts, on the main-sequence, is shown with respect to the distribution of the full sample, i.e., all galaxies in CDFS with stellar mass and SFR measurements and $z$$>$1.5, from CANDELS and 3DHST Treasury Program \citep{Groginetal2011,Skeltonetal2014}.}
\label{fullcomparison}
\end{figure}

In this paper, we have applied this morphology decomposition procedure to all high redshift ($z$$>$1.5) galaxies in {\it Chandra Deep Field South (CDFS)} with ground based spectroscopic redshifts \citep{Wuytsetal2008}. In Fig.~\ref{fullcomparison} distribution of this selection on the main sequence with respect to the full sample (i.e., all galaxies in CDFS with stellar mass and star formation rate measurements with $z$$>$1.5) from CANDELS and 3DHST Treasury Program \citep{Groginetal2011,Skeltonetal2014}, is presented. This selection, for $M_{\odot}$$>$$10^{9.2}$, appears to be an unbiased representation of the entire sample. Ground based spectroscopic redshifts ensure that dense and wide SEDs of these galaxies, obtained with the homogeneous combination and interpretation of 127 distinct data sets by 3DHST Legacy Survey \citep{Momchevaetal2016}, will result in the most accurate stellar parameters \citep{Whitakeretal2014}.

There is growing observational evidence that galaxies witnessed intense activity around z$\sim$2, not only in terms of their internal processes (i.e., structural transformations and star formation efficiency), but also in terms of their external interactions (i.e., accretion rate and merger frequencies). Since the cosmic star formation rate density peaks at around this epoch ($z$$\sim$2) and declines exponentially thereafter (e-folding time-scale 3.9 Gyrs), galaxies have gained more than half of their present stellar mass before $z$$\sim$1.5 \citep{Karimetal2011,MadauandDickinson2014}. Galaxies at $z$$\sim$2 have 5 times higher star formation efficiency (than local ones) leading to shorter gas depletion time-scales, thus, gas accretion rates should have been significantly higher during this epoch to save galaxies from becoming red and dead \citep{Santinietal2014,Scovilleetal2016,Putman2017}. In addition to that, although merger rates reported in literature differ significantly, there is consensus that merger frequency peaks at around cosmic noon, i.e., from $z$$\sim$1.5 to $z$$\sim$2.5, for low mass galaxies ($M_*$$<$$10^{10}$$M_{\odot}$) and increases up to $z$$\sim$3 for high mass galaxies \citep{Jogeeetal2009,Blucketal2012,Conselice2014}. Structurally also, galaxies have been observed to depict significant transformation as we move from $z$$\sim$3 to $z$$\sim$1.5, where $z$$\sim$2 is marked as an epoch of transition \citep{Bruceetal2012,Mortlocketal2013,Huertas-Companyetal2015}.

To examine which internal/external physical processes, whether same or distinct, are responsible for transforming the morphology and regulating stellar activity, it is essential to study the two properties of the same sample in the most accurate manner. We will investigate the change in morphological parameters and stellar properties of galaxies in our sample (i.e., all galaxies in CDFS with ground based spectroscopic redshifts) on the two sides of $z$$=$2 (i.e., 1.5$<$$z$$<$2.0 and 2.0$<$$z$$<$4.0) in rest-frame {\it B}-band. The first phase (1.5-2.0) covering the age of the Universe from 3.3 to 4.3 Gyrs, will help us analyse the after-effects of intense ``in-situ" and ``ex-situ" processes witnessed by the galaxies at $z$$\sim$2. The second phase (2.0-4.0) covering a longer period from 1.6 to 3.3 Gyrs, will enable us to create a solid comparison data set of properties of galaxies before $z$$\sim$2. In terms of morphology, we compute both parametric (e.g., luminosity, size, S\'ersic-index, etc.) and non-parametric (Petrosian-radius, Concentration and Asymmetry index) measures for the full galaxy through 2D image fitting. In addition to that, we apply our decomposition procedure on 1D profiles to compute bulge and disc defining parameters. Selection of the sample and computation of morphological and stellar parameters is detailed in Section~2. Results are presented in Section~3 and major findings are discussed in Section~4. We consider a flat $\Lambda$-dominated Universe with $\Omega_{\Lambda}=0.714$, $\Omega_m=0.286$ and $H_o=69.6$ km sec$^{-1}$ Mpc$^{-1}$. All magnitudes are in the AB system.


\section{Data}
\label{sec:data}

\subsection{Sample selection}

Our sample consists of 180 galaxies in the GOODS-CDFS field which have ground-based measurements of spectroscopic redshifts, and have $z_{spec}$$>$1.5. The redshifts are from the FIREWORKS catalogue, which is a compilation of all ground based spectroscopic redshift measurements in CDFS \citep{Wuytsetal2008}. To examine their properties in rest frame {\it B}-band, we obtain {\it HST/WFC3} {\it J(F125W)} and {\it H(F160W)} filter images for these galaxies from CANDELS and 3DHST Treasury Program\footnote{Based on observations taken by the 3D-HST Treasury Program (GO 12177 and 12328) with the NASA/ESA HST} \citep{Groginetal2011,Koekemoeretal2011,Skeltonetal2014}. The images have been pre-processed for dark-sky subtraction, flat-fielding, cosmic-ray removal, etc.

Using RA, Dec information from 3DHST photometric catalogue \citep{Skeltonetal2014}, which itself has been matched with the FIREWORKS catalogue, we extract a 10 arcseconds cutout for each galaxy in F125W (for 1.5$<$$z$$<$2.0) and F160W (for $z$$>$2.0) filters. All galaxies are well within 2 arcseconds radius, thus, 10 arcseconds cutout leaves more than $\sim$60\% of the space for sky estimation. Running {\it Source Extractor} \citep{BertinandArnouts1996} on each cutout, we first select a very low threshold and broad filter to mask all sources along with their extended/diffused light \citep[as detailed in][]{AkhlaghiandIchikawa2015}. Using the remaining pixels, we compute the sky value for each cutout. After that, we run {\it Source Extractor} with normal threshold and filter size but with a low de-blending value to create masks for neighbouring sources. Analysing each segmentation map individually, we carefully de-mask our source of interest. These mask images are used in 2D image fitting (using {\it Galfit}) and in isophotal intensity profile generation.  

Note that for accurate computation of non-parametric measures (Petrosian-radius, Concentration, Asymmetry index) it is essential that masked out pixels are replaced with correct sky values and sigma. This is because the number of pixels involved in each radial step should not vary for different sources \citep{Conselice2003}. Best way to achieve that is to use {\it imedit} task of {\it IRAF} in an interactive manner. This task creates an annular disc around a selected object (where the size and distance of the disc is determined by the user) and replaces the pixels of that object with an average of pixel values on the annular disc. We have carried out this procedure on each cutout before the computation of non-parametric measures. 

\subsection{Single S\'ersic fitting}

After obtaining the cutouts and cleaning the images of neighbouring light, we fit each galaxy image in our sample with a single S\'ersic component using {\it Galfit} \citep{Pengetal2002} to obtain their total luminosity, size and morphology. The intensity distribution is given by,

\begin{equation} 
I_{Sersic}(r)=I_e\exp[-b_n((\frac{r}{r_e})^{1/n}-1)],
\end{equation}

\noindent where, $r_e$ is the effective radius of the galaxy, $I_e$ is the intensity at that radius and $n$ is the S\'ersic index. The initial input values are estimated from {\it phot} and {\it imexam} tasks of {\it IRAF}, utilizing their functionality of computing the number of counts and pixels inside selected radii. Note that while {\it Galfit} might not be very efficient to fit two components at higher redshifts, it gives robust results in one-component fitting due to lesser number of free parameters \citep{Sachdeva2013}. Additionally, we fit each image individually, continually examining the residual image for the percentage of residual flux along with the reduced $\chi^2$ value, to converge to best results. This provides global (i.e., for the full galaxy) parametric values of magnitude (or luminosity), size and S\'ersic index. 

Using redshift information, cosmological distance computations and corresponding K-corrections, we compute absolute, i.e., intrinsic parameters for the galaxy in rest-frame {\it B}-band. The equations involved are the same as those given in \citet{GrahamandDriver2005,Sachdeva2013,Sachdevaetal2015}. We thus obtain absolute magnitude $M_g$, half-light-radius $R_{e,g}$, surface brightness at effective radius $SB_{e,g}$ and average surface brightness inside effective radius $<$$SB_{e,g}$$>$, in addition to the S\'ersic index $n_g$, in rest-frame {\it B}-band. Note that subscript ``g" denotes that these parameters are global, i.e., for the full galaxy.

\subsection{Profile decomposition}       

We next perform decomposition of each galaxy's intensity profile to obtain separate bulge and disc defining parameters. The intensity profile is obtained using the {\it ellipse} task of {\it IRAF}. It iteratively finds the best fitting isophote at successively increasing semi-major-axis distance, varying parameters such as intensity, position-angle, ellipticity and the central coordinates \citep{Jedrzejewski1987}.

Each intensity profile has been deconvolved with the PSF profile. We found that deconvolution affects only the inner values (up to 3 pixels, $<$0.1 arcseconds) of the profile. The difference in the convolved and deconvolved values is found to be susceptible to the ellipticity and position angle of the galaxy \citep{Trujilloetal2001,Ciambur2016}, leading to significant difficulties in accounting for it in a robust manner. Fitting each profile individually we find that inclusion (or exclusion) of the inner 0.1 arcsecond values does not affect the computation of our bulge parameters.    

Each log-intensity radial profile is fitted following the steps described. First, through visual inspection we recognize that part (or range) of the profile which represents only the underlying disc, i.e., shows exponential behaviour and is thus not contaminated by the light of other components or features. Fitting the disc function $I_{disc}$,

\begin{equation} 
I_{disc}(r)=I_o\exp(-\frac{r}{r_d}),
\end{equation}

\noindent over that range (using Levenberg-Marquardt algorithm), we obtain precise (i.e., non-degenerate) values of the central intensity $I_o$ and scale length $r_d$ of the disc component of the galaxy. Keeping these parameters fixed, we fit the full profile with the full intensity function $I_{full}$,

\begin{equation} 
I_{full}(r)=I_{disc}(r)+I_{Sersic}(r),
\end{equation}

\noindent where the equation for $I_{Sersic}$ is as given in the previous section. Since $I_{disc}$ parameters are already in place, in an attempt to fit $I_{full}$ to the full profile, the algorithm (Levenberg-Marquardt) automatically uses $I_{Sersic}$ to fit only that intensity at the centre which is additional to the underlying disc, i.e., the bulge. We thus obtain all bulge parameters, i.e., the S\'ersic index ($n_b$), effective radius ($r_{eb}$) and intensity at that radius ($I_{eb}$), in addition to the disc parameters obtained earlier. Again employing redshift information, cosmological distances and K-corrections, we derive intrinsic (or absolute) parameters defining the bulge and disc components in rest-frame {\it B}-band.

Disc galaxies with higher inclination, i.e., more edge-on than face-on, might appear brighter if they are optically thin because emission from all stars adds up in a smaller projection area. This effect is notable when galaxies are observed in near and far IR rest-frame wavelengths because higher wavelength light passes largely undisturbed through dust clouds \citep{Graham2001}. The distribution of our rest-frame optical sample on the central disc intensity ($I_o$) and disc scale length ($r_d$) plane did not depict any correlation with the ellipticity of the disc, suggesting that galaxies in our sample are optically thick. We have, therefore, not applied any inclination correction on the central disc intensities in our sample. 

Division of the fitting process into multiple steps reduces the number of free parameters, thereby, increasing the accuracy. In our earlier work \citep{SachdevaandSaha2018}, we demonstrated the exactness of this process through simulations. We created model galaxies with a range of bulge-total ratios, magnitudes, sizes, ellipticities and S\'ersic indices, using {\it Galfit}. Then we performed our full fitting procedure on these model galaxies, first obtaining their intensity profile from {\it ellipse} task of {\it IRAF} and then fitting functions on that profile. We found that when both components, i.e., bulge and the disc, are present in good measure (i.e., 0.05$<$B/T$<$0.95), our process recovers the parameters with two decimal accuracy. For large axis-ratios, it falls to one decimal accuracy, still well within the error range associated with these parameters.

\begin{figure*}
\mbox{\includegraphics[width=180mm]{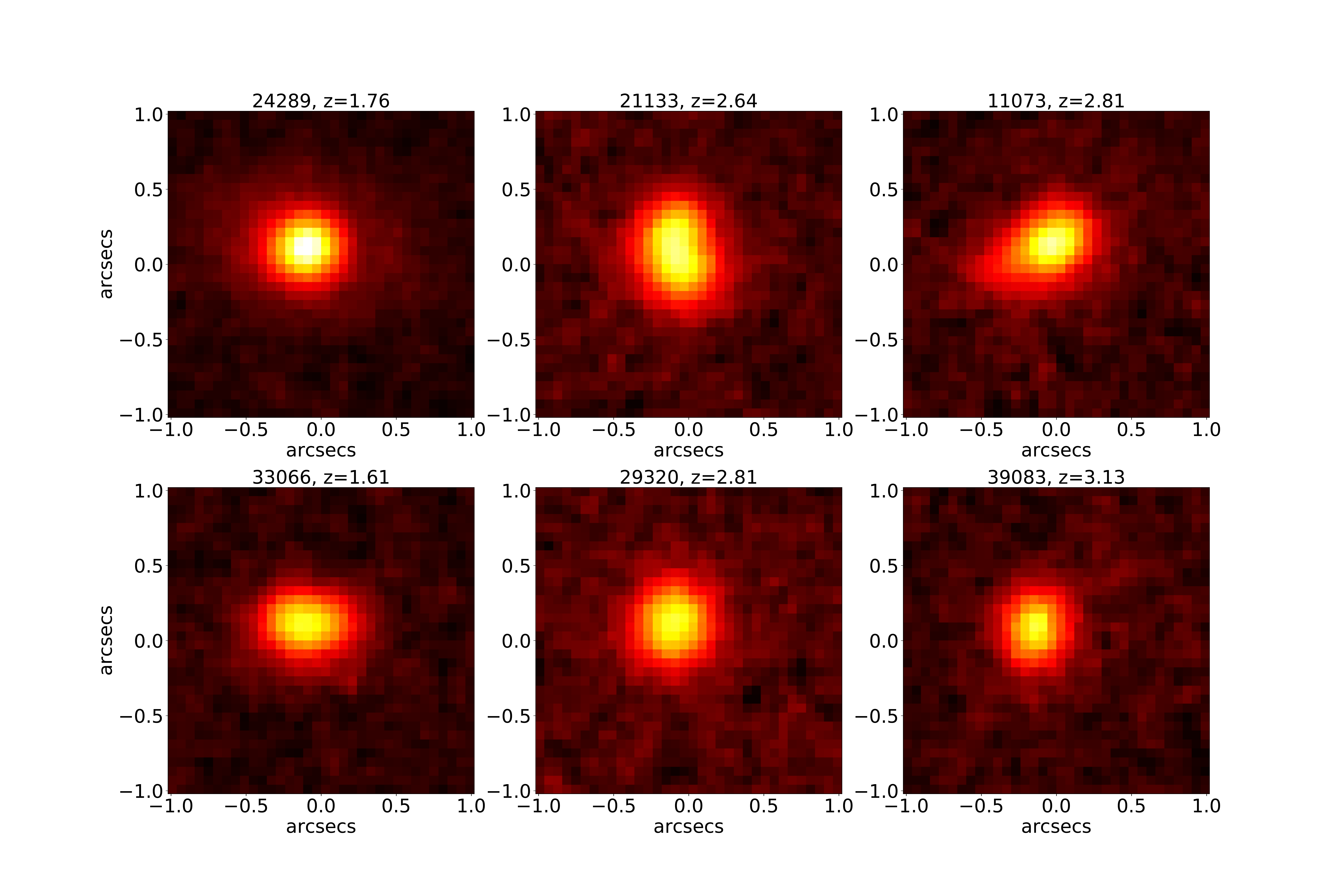}}\\
\mbox{\includegraphics[width=45mm]{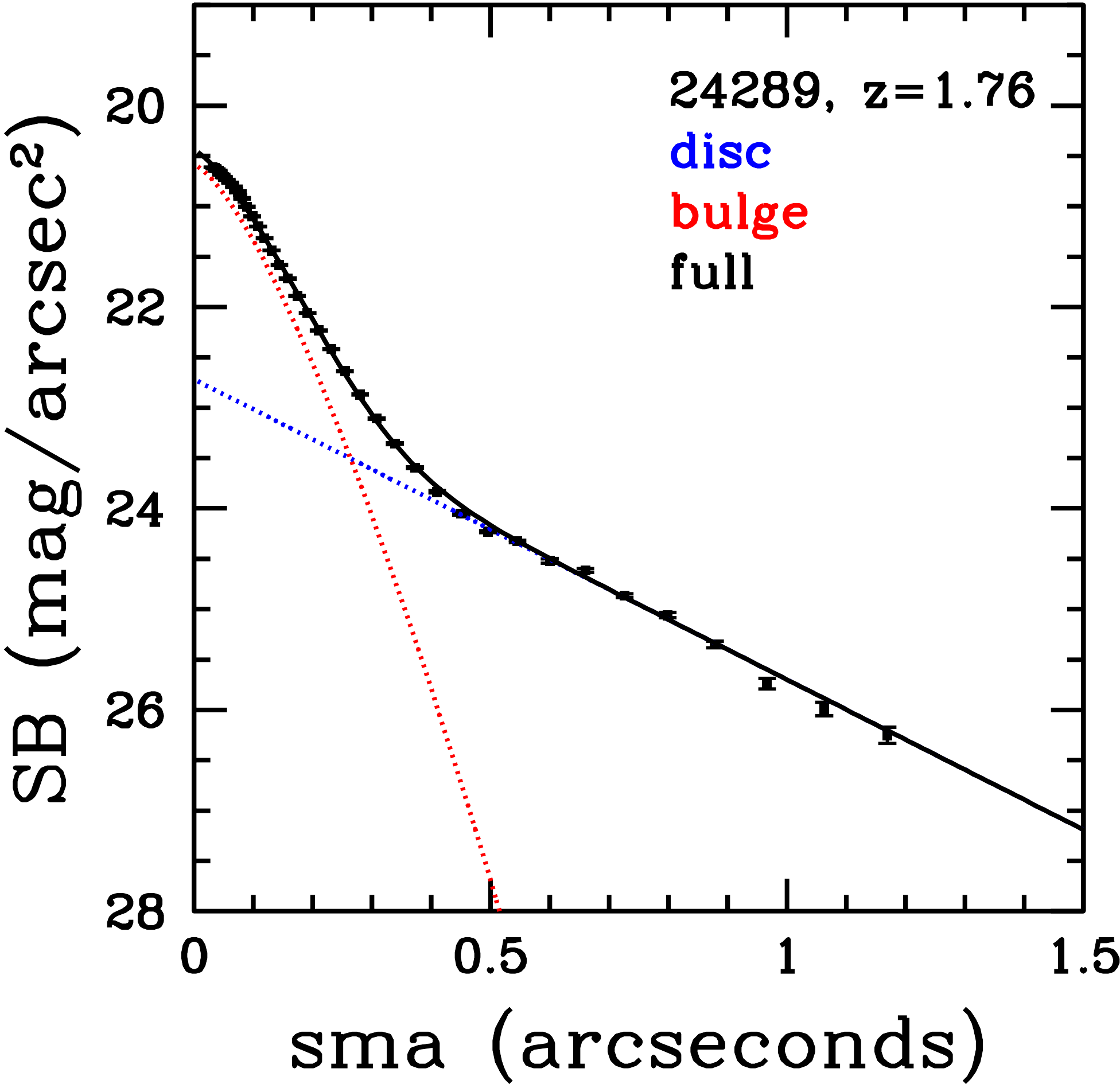}}
\mbox{\includegraphics[width=45mm]{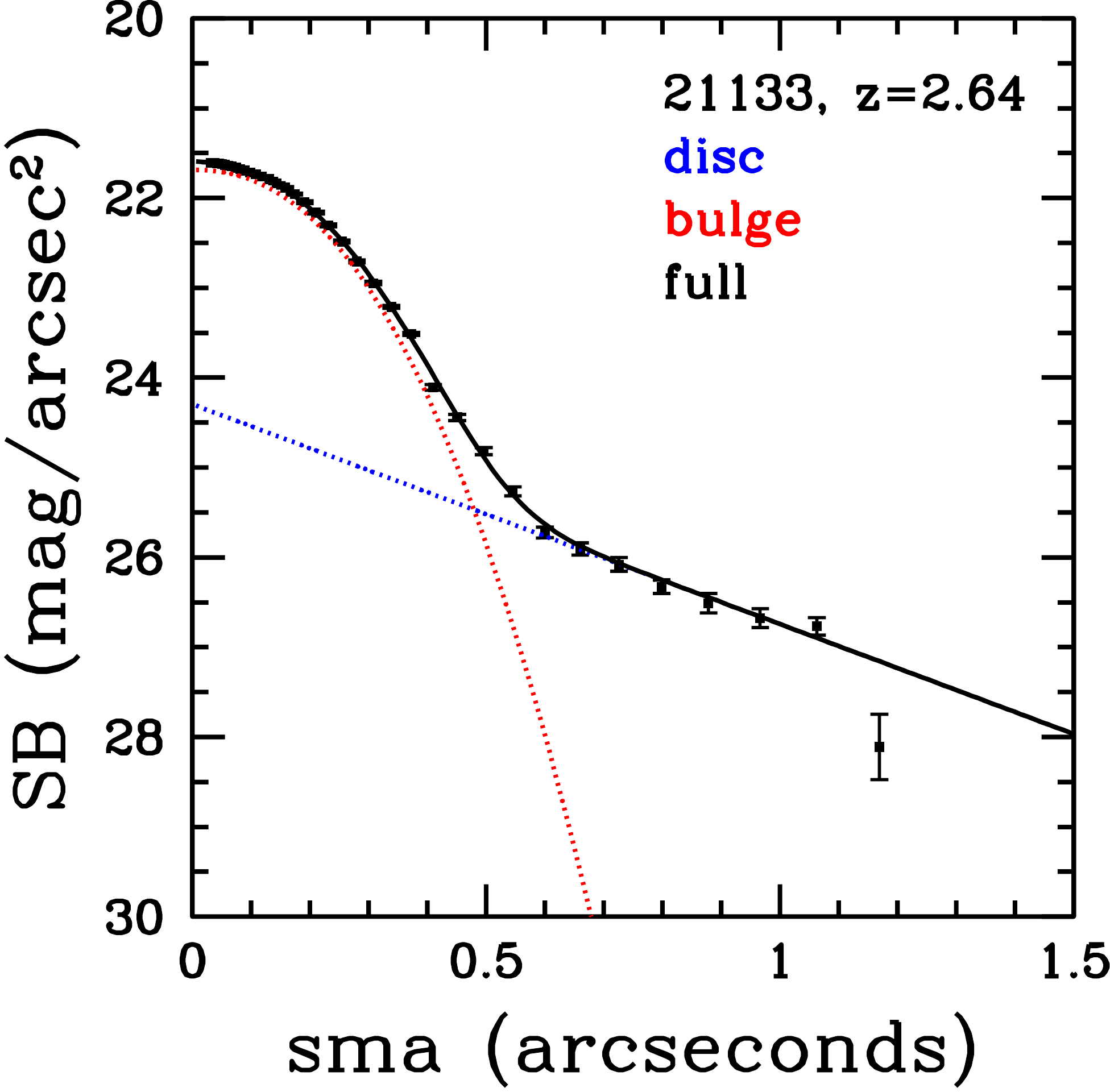}}
\mbox{\includegraphics[width=45mm]{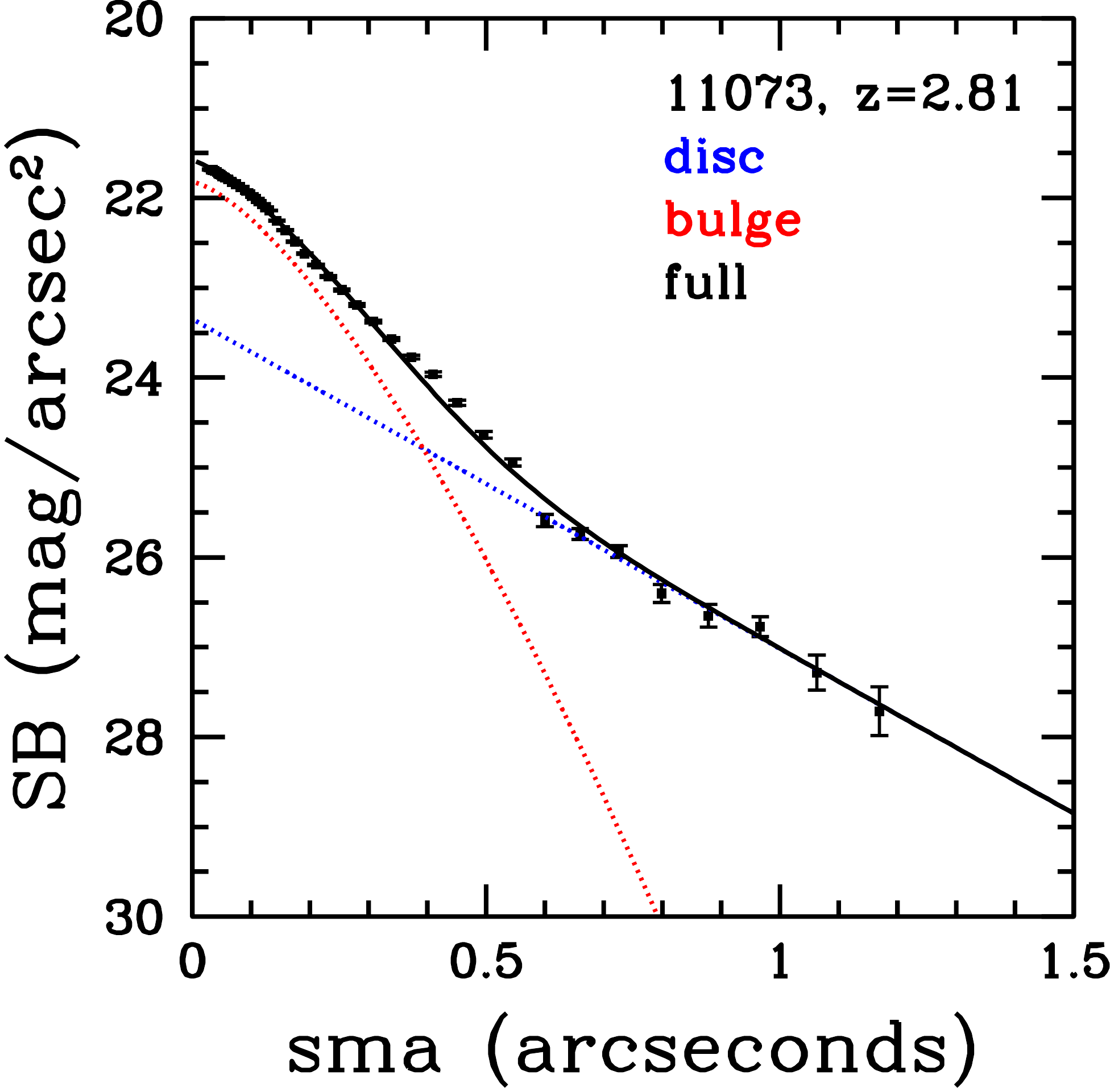}}\\
\mbox{\includegraphics[width=45mm]{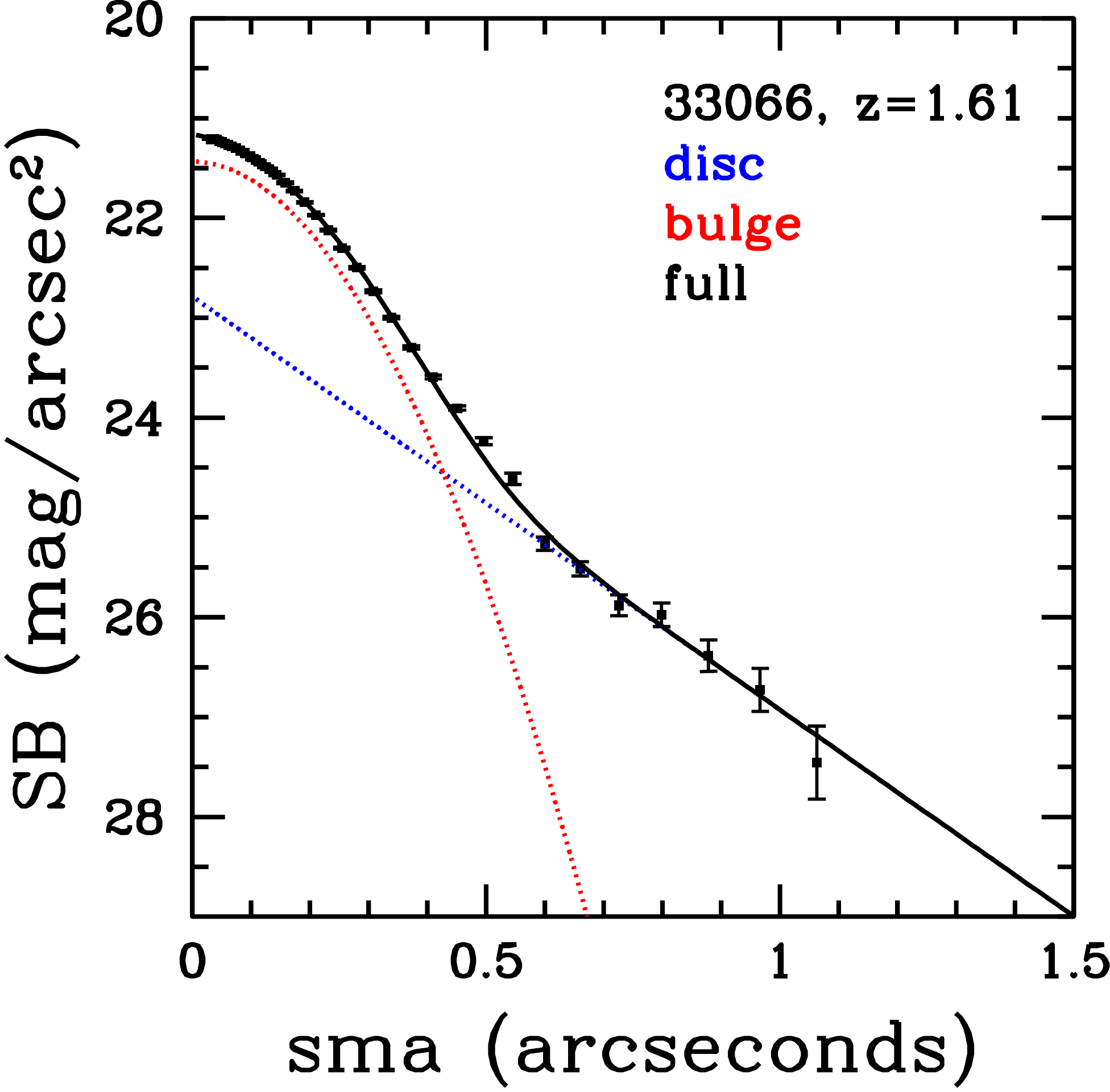}}
\mbox{\includegraphics[width=45mm]{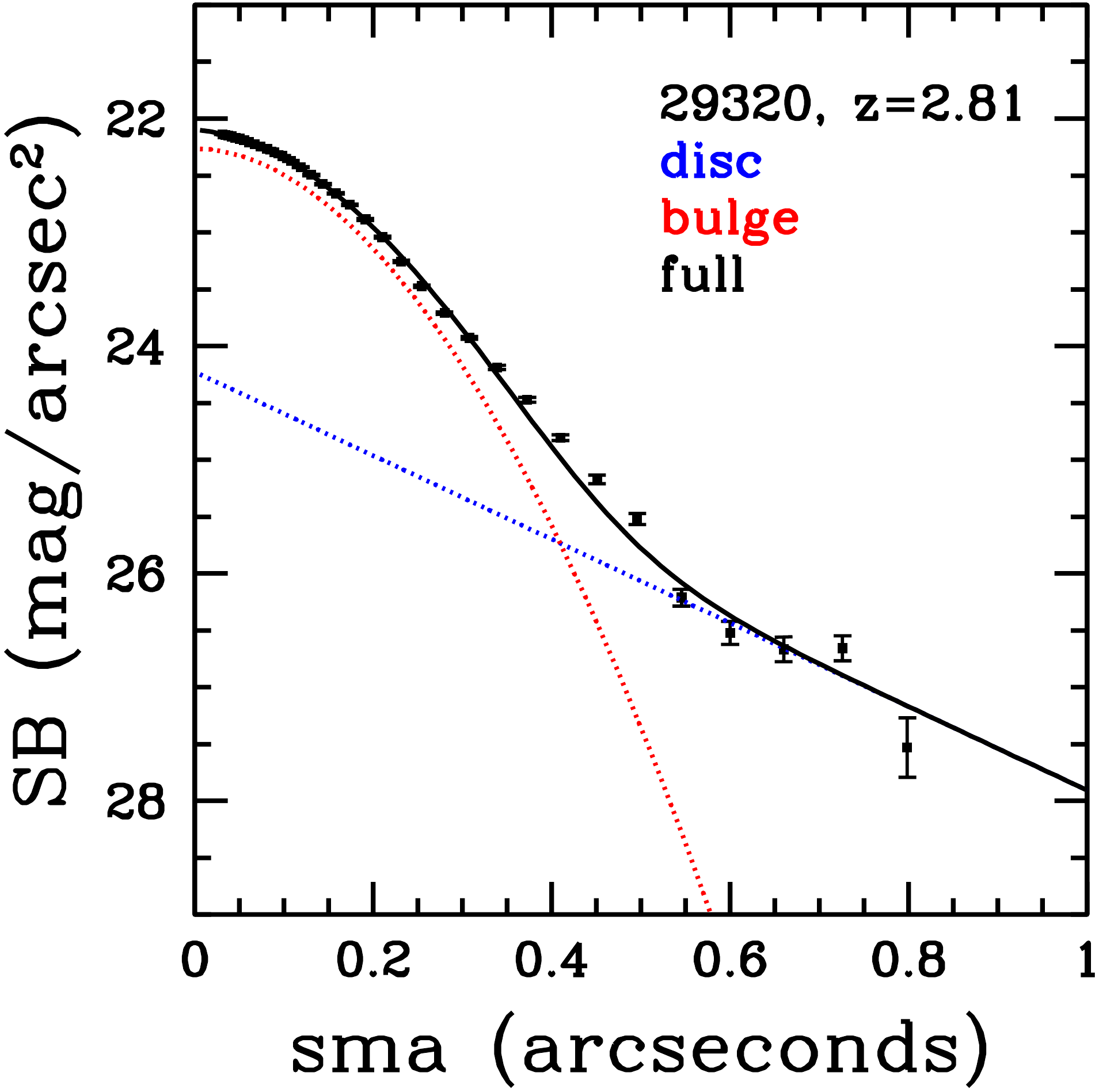}}
\mbox{\includegraphics[width=45mm]{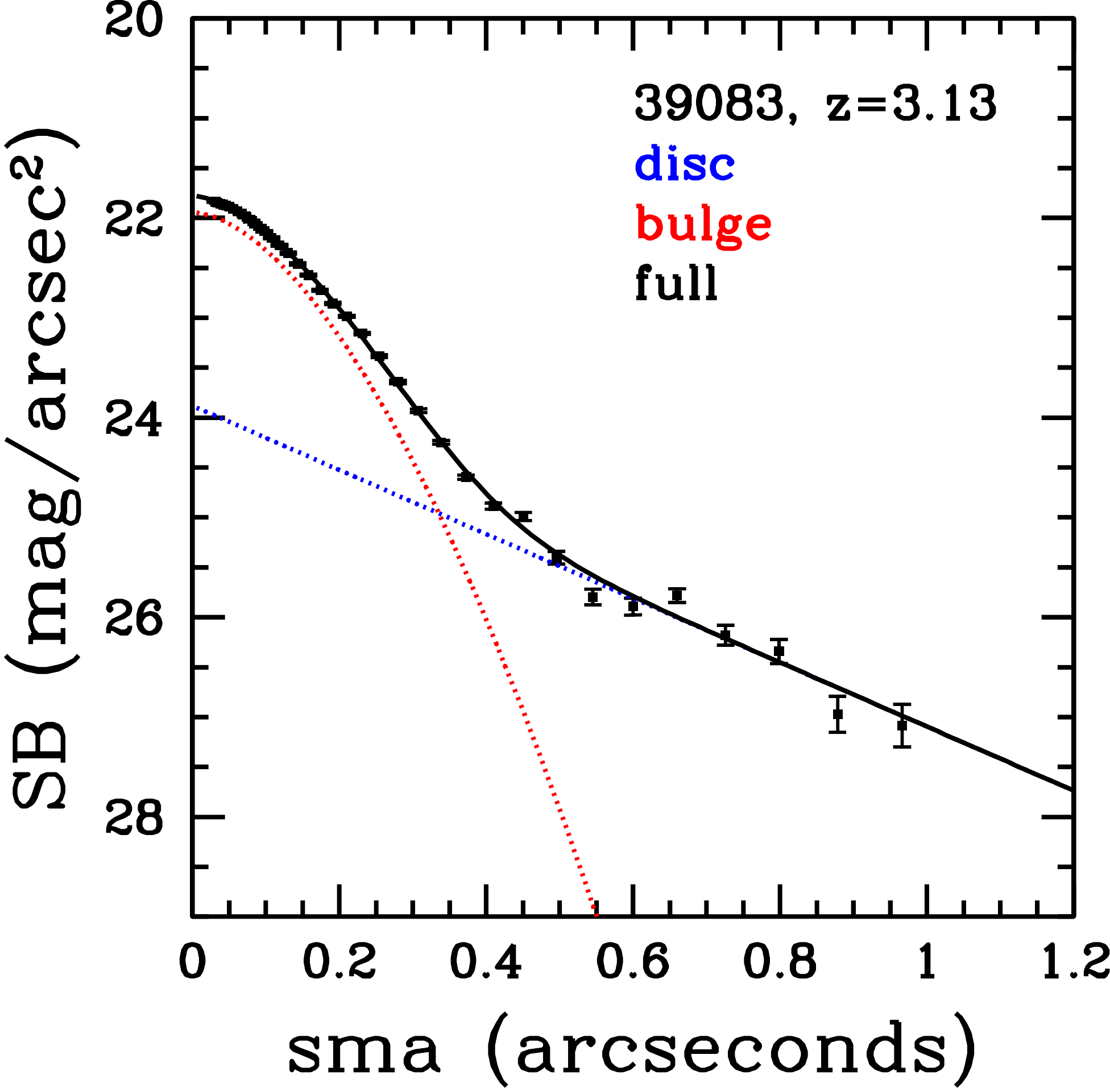}}
\caption{{\bf Galaxies with 2-components:} The images, 2.04" X 2.04" in size, are focussed on the inner 1 arcsecond radius view of the galaxies. Since each pixel is 0.06" in size, these are 34 X 34 pixel cutouts. These galaxies are examples of the sample which was well fitted by 2 components, i.e., bulge and host disc. Their corresponding surface brightness profiles fitted with the combination of two functions is also presented. The ID is according to the unique identifier from \citet{Skeltonetal2014}. Their spectroscopic redshifts have also been marked.}
\label{profilefitting1}
\end{figure*}

\begin{figure*}
\mbox{\includegraphics[width=180mm]{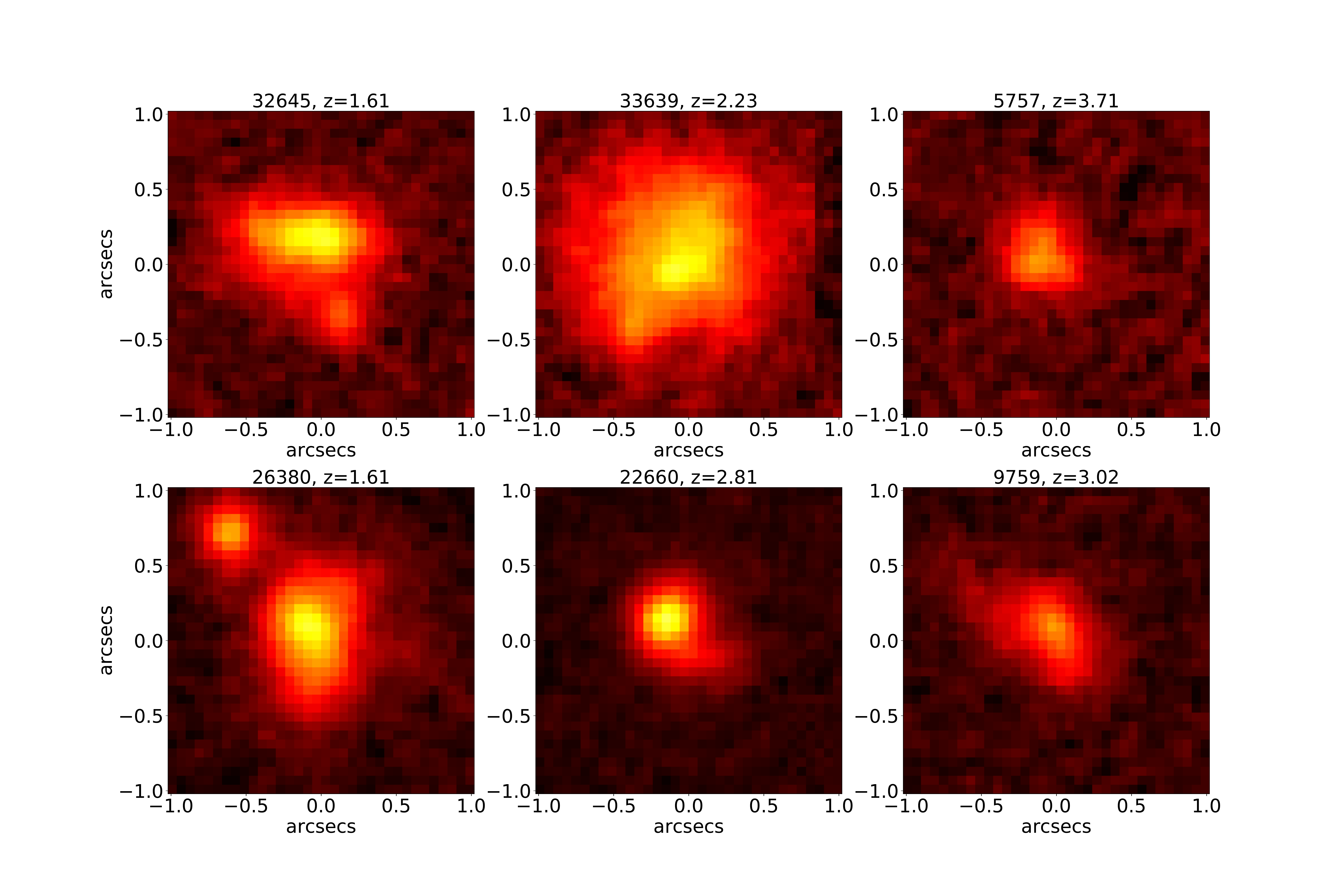}}\\
\mbox{\includegraphics[width=45mm]{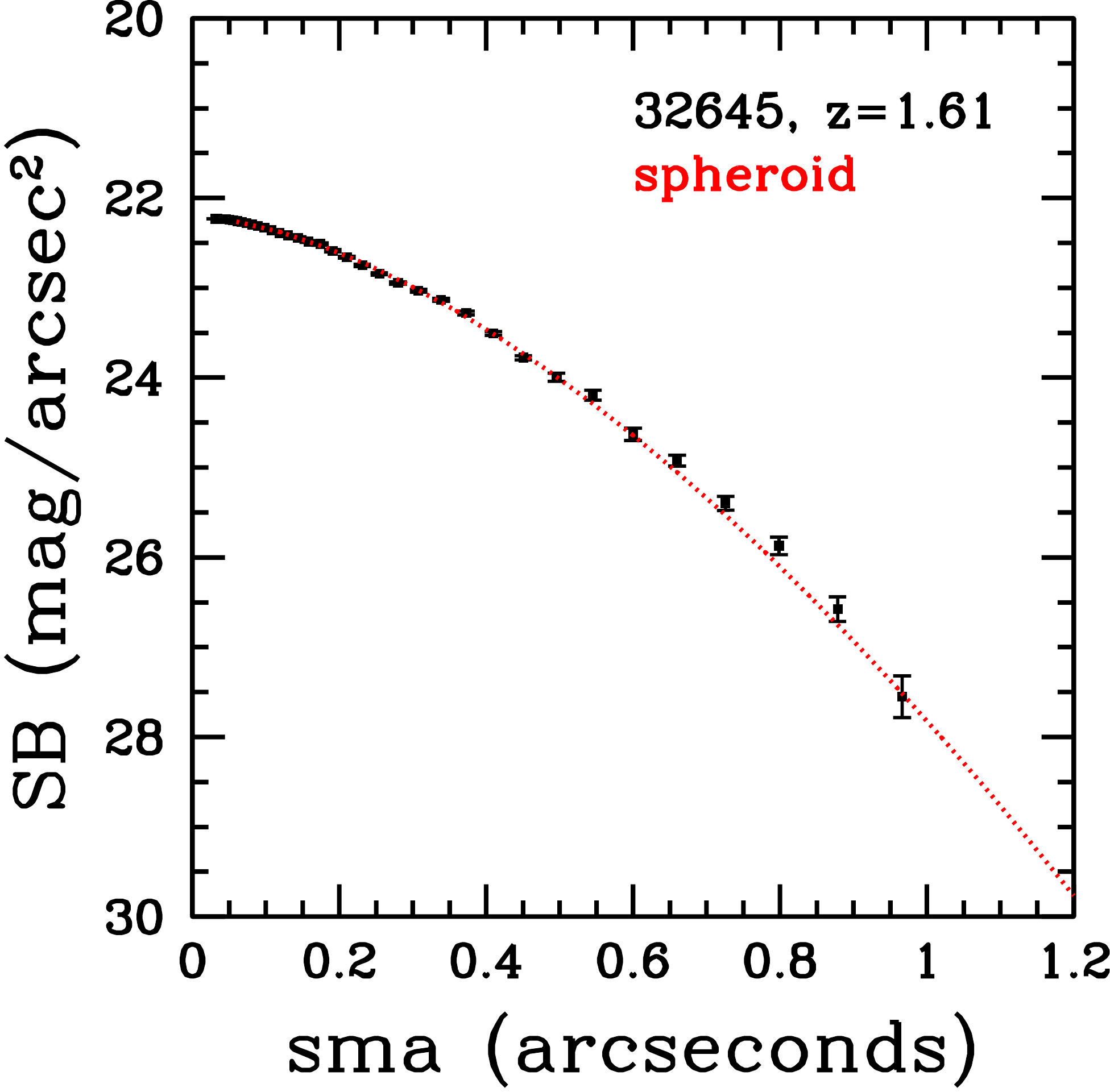}}
\mbox{\includegraphics[width=45mm]{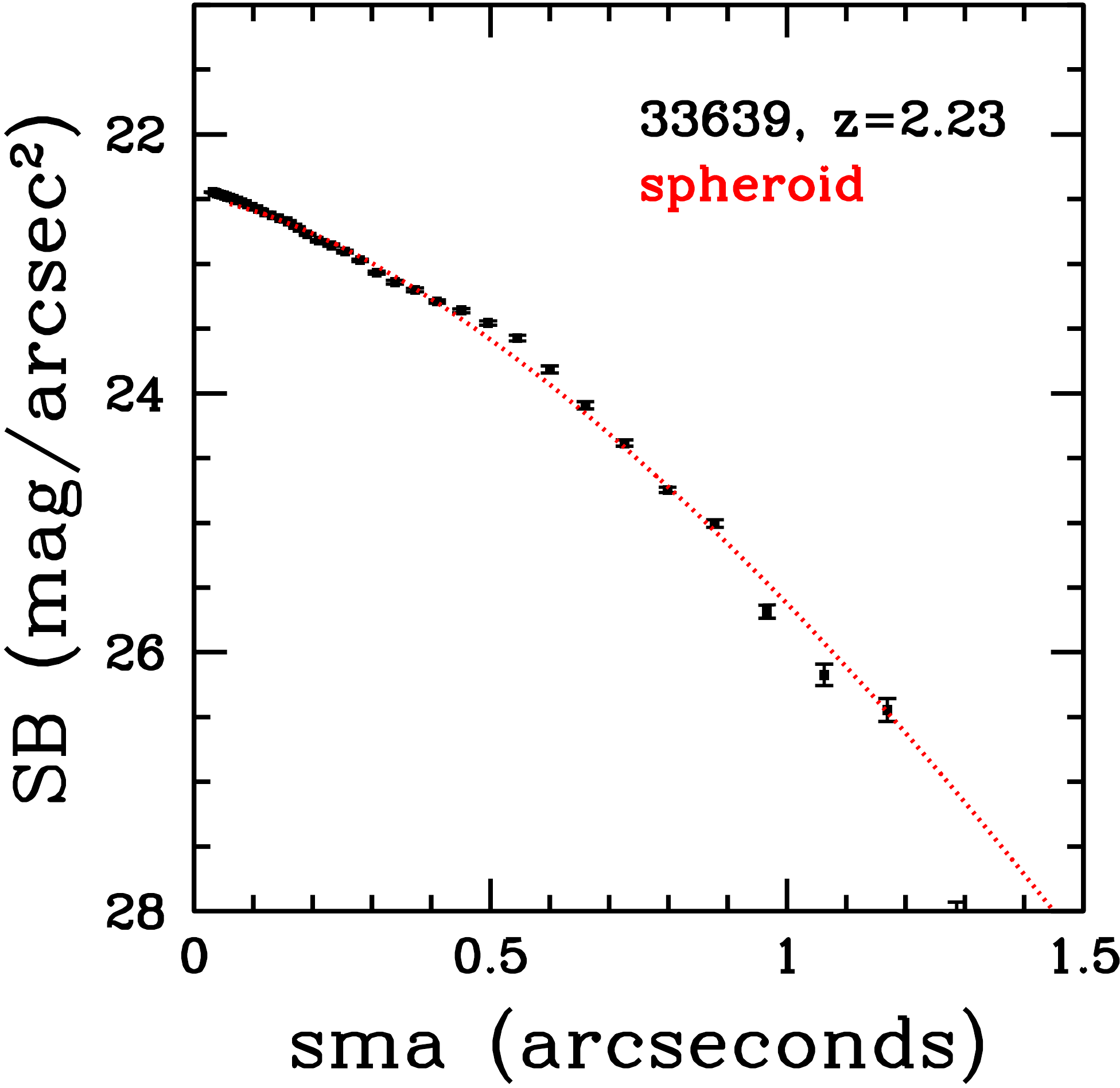}}
\mbox{\includegraphics[width=45mm]{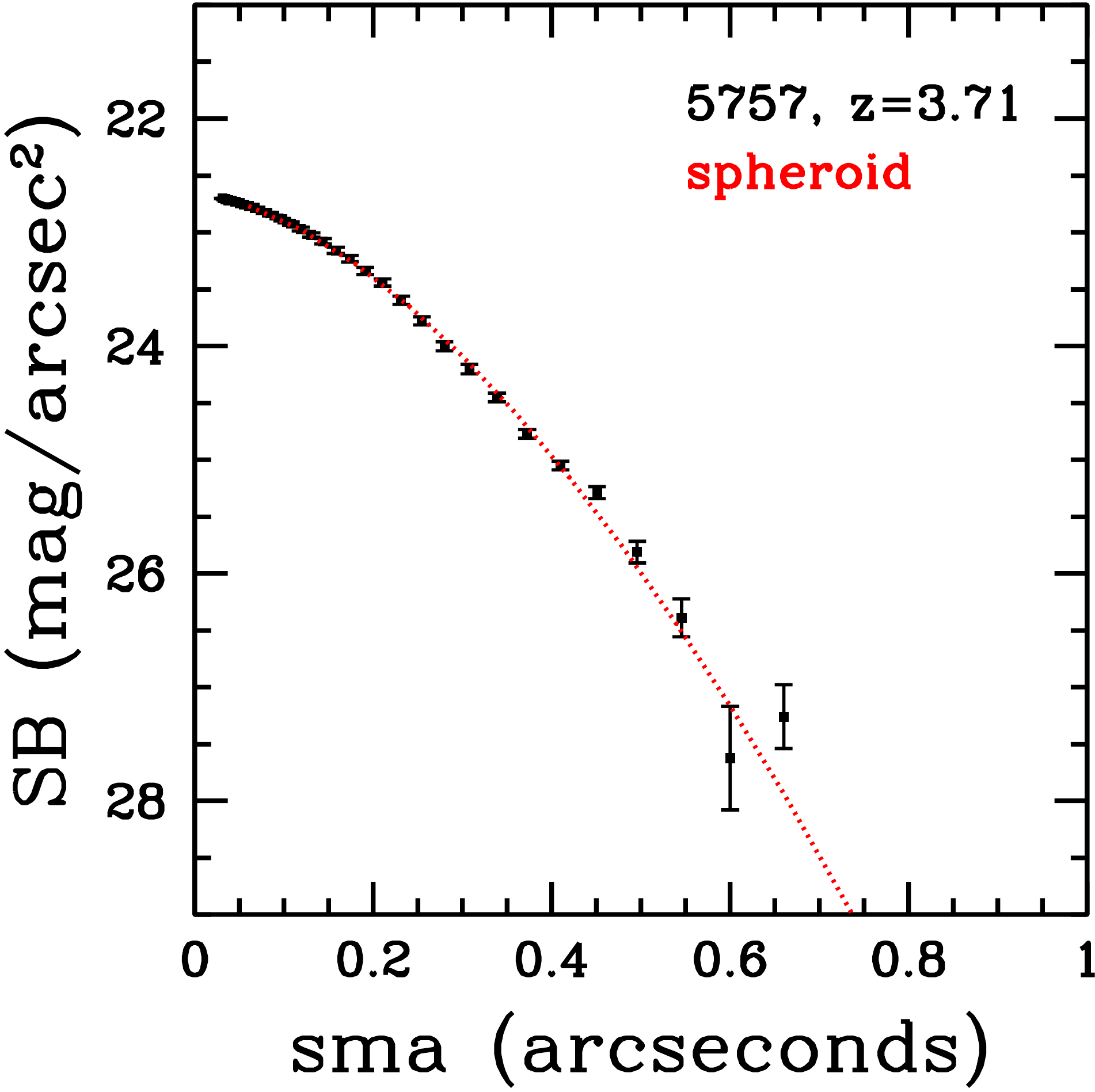}}\\
\mbox{\includegraphics[width=45mm]{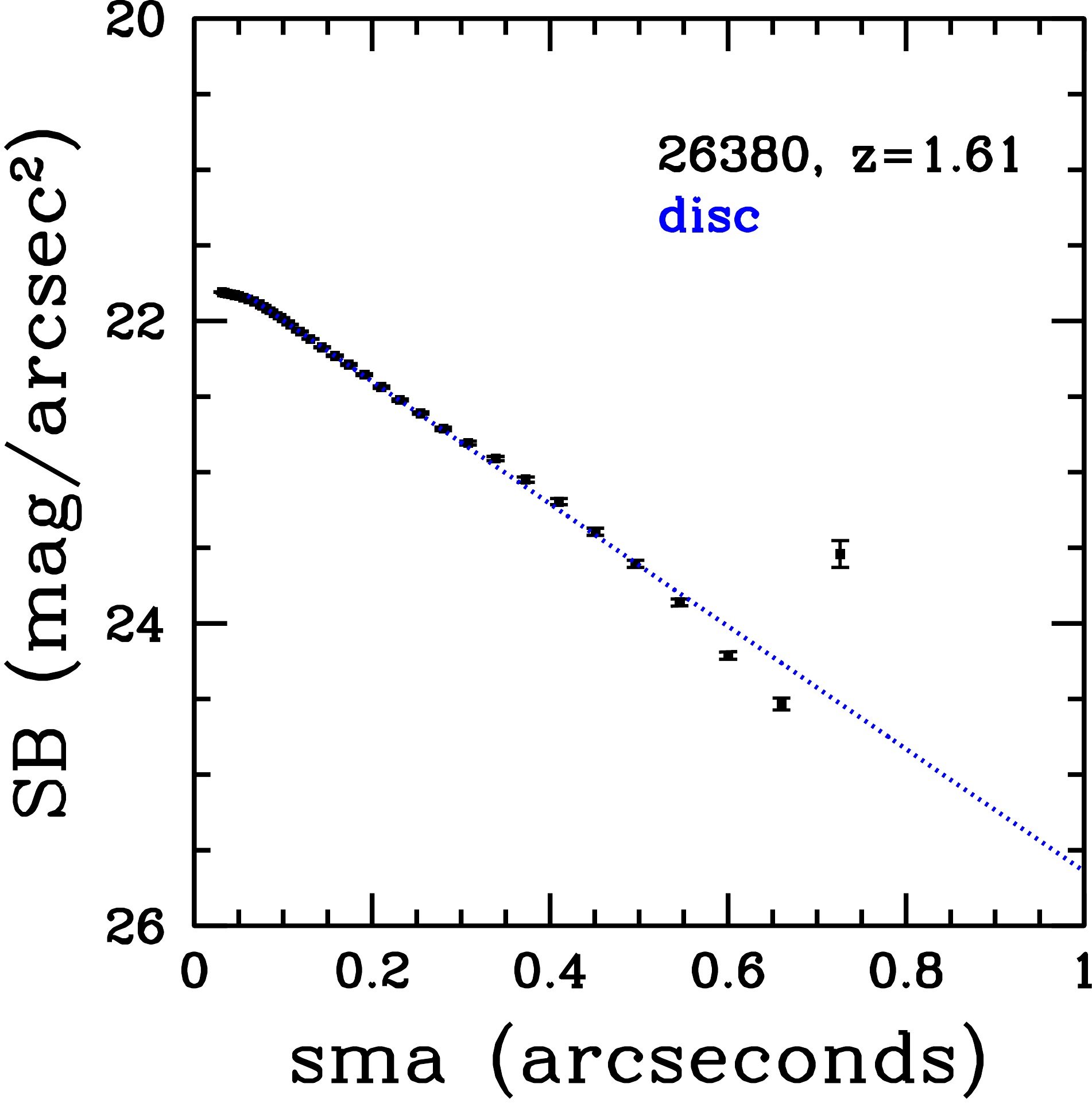}}
\mbox{\includegraphics[width=45mm]{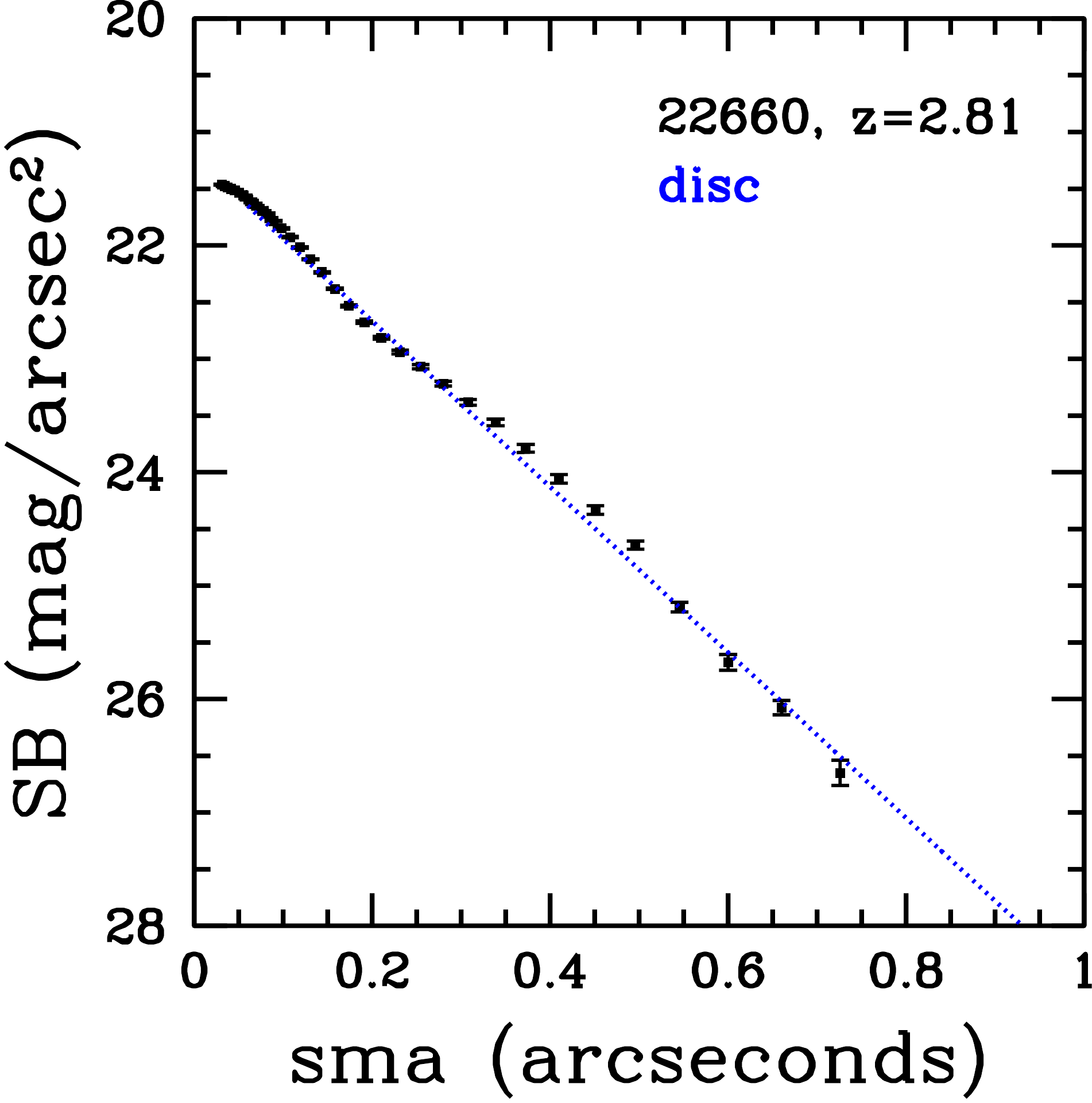}}
\mbox{\includegraphics[width=45mm]{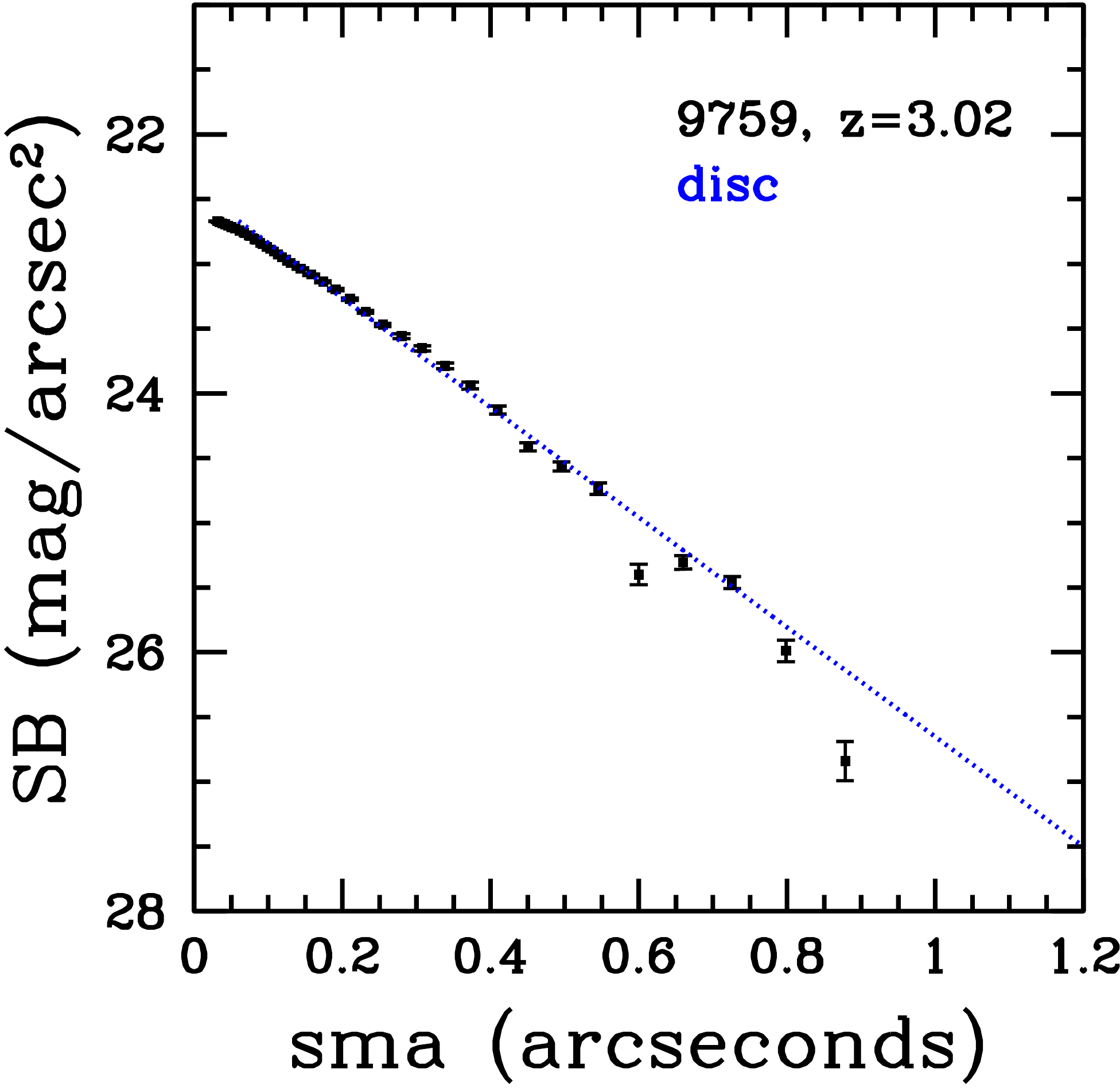}}
\caption{{\bf Galaxies with single component:} The images, 2.04" X 2.04" in size, are focussed on the inner 1 arcsecond radius view of the galaxies. Since each pixel is 0.06" in size, these are 34 X 34 pixel cutouts. These galaxies are examples of the sample which was well fitted by a single component, i.e., either a free S\'ersic function ({\it upper panel}) or an exponential function ({\it lower panel}). Their corresponding surface brightness profiles fitted with these functions is also presented. The ID is according to the unique identifier from \citet{Skeltonetal2014}. Their spectroscopic redshifts have also been marked.}
\label{profilefitting2}
\end{figure*}

In Fig.~\ref{profilefitting1} and Fig.~\ref{profilefitting2} we present examples of some of the galaxy images (mosaics focussed on the inner 1" radius view) and their corresponding surface brightness profiles, fitted through our procedure. Note that only those isophotal values are fitted (and are shown) for which stop-code is zero, i.e., proper (physical) solution is found without exceeding the maximum number of iterations. The error bars on each isophote are as computed by the {\it ellipse} task accounting for background uncertainties and standard deviation of fitted parameters (i.e., intensity, ellipticity, position angle and centroids) at that isophote. 

We find that for most of the galaxies ($\sim$55\%), the profiles are well fitted with 2 functions, i.e., a free S\'ersic and an exponential (Fig.~\ref{profilefitting1}). However, for a substantial number of galaxies ($\sim$45\%), a single component, i.e., either an exponential disc ($\sim$27\%) or a free S\'ersic function ($\sim$18\%) fitting, provides a better representation to the light profile (Fig.~\ref{profilefitting2}). Each profile has been fitted individually with multiple checks made to select the combination of functions which represent the profile in the most accurate manner.

\subsection{Non-parametric measures}

High redshift morphologies may differ significantly from their local stable counterparts \citep{Bruceetal2012,Mortlocketal2013}. Since all parametric measures are based on an empirical understanding of the local morphologies, application of the same measures to higher redshifts might bias our computations. It is thus imperative to explore morphologies using non-parametric measures, in addition to their parametric analysis. Non-parametric values can then be compared with the corresponding parametric values (e.g., of size, concentration), to ascertain the accuracy and validity of both the measurements.

Based on the algorithm defined by \citet{Conselice2003}, we compute the Petrosian radius ($R_p$), Concentration ($C$) and Asymmetry index ($A$) for all galaxies in our sample. Steps involved in the computation of each of these parameters for a given galaxy are detailed out in \citet{Sachdevaetal2015}. The Petrosian radius provides an estimate of the full size of the galaxy, unbiased by its inherent brightness and our observational capabilities. Its computation, based on extrapolation, involves measuring the ratio of intensity at a given isophote and inside that isophote, for successively increasing radii. The process is halted when the algorithm reaches a desired, empirically determined ratio, 

\begin{equation}
\eta(r_p)=\frac{I(r_p)}{<I(<r_p)>}=0.2,
\end{equation}

\noindent at some radius $r_p$. The intensity at $r_p$ is 20\% of the intensity inside $r_p$. This radius is then multiplied by an empirically set multiple,  

\begin{equation}
R_P=1.5 \times r_p,
\end{equation}

to obtain an estimate for the total radius $R_p$. Computation of Concentration $C$ follows from this value. Considering counts (or flux) inside this radius $R_p$ to be the total flux, we find those two isophotal radii which contain 80\% and 20\% of the total flux. The logarithm of their ratio gives the concentration index,

\begin{equation}
C=5 \times \log_{10}(\frac{r_{80percent}}{r_{20percent}}),
\end{equation}

\noindent where, 5 is again an empirically set multiple. The greater the fraction of light contained inside, the smaller will be $r_{20percent}$ compared to $r_{80percent}$ and larger will be the concentration value. 

Asymmetry index ($A$) is a measure of flux (absolute value) in the residual image (normalized w.r.t. the original image), where the residual image is obtained by subtracting the rotated ($180^o$) image from the original image. An important requirement of this measurement is that the centre of rotation must coincide with galaxy's symmetry centre because even a small departure can result in an over-estimation of galaxy's asymmetry. To achieve that, centre for rotation is iteratively chosen (by our code) to be that pixel which results in minimum flux in the residual image.

Accuracy and sensitivity level of these parameters ($R_p$, $C$, $A$) towards changes in total size, concentration and asymmetry were explored through detailed simulations in \citet{Sachdevaetal2015}. This involved creating a set of similar galaxies with variation in only that property which is correlated with the parameter under examination. To check for Petrosian radius, half light radius was varied; for Concentration, S\'ersic index was varied; and for Asymmetry, we gradually added and varied the amplitude of Fourier modes in model images. We found errors in the range of 4-5\% for Petrosian-radius, 2-3\% for Concentration and 5-10\% for Asymmetry, ensuring that these parameters respond quite well to small internal changes.

\subsection{Stellar parameters}

For accurate estimation of stellar parameters, of a given galaxy, the most crucial input to Stellar Population Synthesis (SPS) models is its Spectral Energy Distribution (SED), along with redshift. To obtain a galaxy's full wide SED, along with all dense features, requires the combination of observations from different telescopes and filters, catering to various wavelength regions, in a homogeneous manner. The 3DHST project is such an attempt which boasts of homogeneous combination and interpretation of 127 distinct data sets, in 5 extragalactic fields \citep{Skeltonetal2014}. For galaxies in GOODS-CDFS, there are photometric fluxes from 26 broad bands and 14 medium bands, covering 0.3-8 $\mu$m, aiding creation of dense and wide SEDs.
 
For our sample, i.e., galaxies with ground based spectroscopic redshifts, there is no redshift measurement requirement. The SEDs and redshifts are directly given as an input to the FAST code \citep{Krieketal2009} to obtain stellar masses. Here the SPS model library is from Bruzual and Charlot (2003), Initial Mass Function (IMF) is from \citet{Chabrier2003} and solar metallicity is assumed. Star Formation History (SFH) is assumed to be exponentially declining with minimum e-folding time of 10 Myrs and minimum age of 40 Myrs with 0$<$Av$<$4 mag. The dust attenuation law is from \citet{Calzettietal2000}.

In addition to stellar masses, we also have Star Formation Rate (SFR) information for the galaxies in our sample. Computation of SFRs also benefits from the homogeneous combination of distinct data sets, leading to a wholistic picture of stellar emissions. \citet{Whitakeretal2014} utilized this to compute SFRs in the following manner,

\begin{equation}
SFR [M_{\odot}/yr] = 1.09 \times 10^{-10} \times (L_{IR} + 2.2 \times L_{UV}) [L_{\odot}],
\end{equation}

\noindent where, $L_{IR}$ is the bolometric luminosity (8-1000 $\mu$m) of completely obscured population of young stars, $L_{UV}$ is the integrated luminosity (1216-3000 \r{A}) emitted by unobscured stars, and the multiplication factor 2.2 accounts for extra ($<$1216, $>$3000 \r{A}) unobscured light. We thus have stellar masses and SFR and thereby specific SFR (sSFR) for all the galaxies in our sample.   

\subsection{Overall sample}

Out of our total sample of 180 galaxies with ground-based spectroscopic redshifts, single S\'ersic fitting (using {\it Galfit}) and profile decomposition was successfully performed on 177 galaxies. For three galaxies, there was no convergence of results, either from image fitting or profile fitting. Other than that, we found two pairs of galaxies which gave exactly the same fitting results, probably due to overlap or because they were actually a single entity. We thus chose only one from each pair which was a more likely candidate, reducing our sample to 175 galaxies. For four sources, the bulge S\'ersic index in profile fitting was below 0.3, leading to non-physical values of surface brightness.
 
Overall, in terms of parametric morphological measures, we have single S\'ersic (image fitting) values for 175 galaxies and bulge-disc decomposition (profile fitting) values for 171 galaxies. Non-parametric morphological measures (Concentration, Asymmetry and Petrosian radius) were obtained for all the galaxies in our sample. While stellar mass was available for all the galaxies, SFR was only available for $\sim$90\% of the sample. In the next section, we present a comparison of morphological and stellar properties of galaxies in the two redshift ranges, i.e., 1.5$<$$z$$<$2.0 and 2.0$<$$z$$<$4.0.

\section{Results}

\subsection{Morphological transformation}

Out of the full sample of 180 galaxies, the intensity profile was fitted successfully for 171 galaxies. Out of this sample of 171, for 31 galaxies there was no visually detected disc (or exponential) range. Their full light profile was best fitted with a single S\'ersic function. We, thus, characterize them as ``pure spheroids". Our claim of the ``non-detection" of the disc does not imply the absence of the disc. It remarks that even if the disc is present, it is not prominent enough to be detected (i.e., has $SB_o$$>$21.3 mag/arcsec$^2$) and contributes to less than $\sim$5\% of the galaxy's total detected light.

Being at higher redshifts, thus, more diffused and less concentrated, pure spheroids' S\'ersic indices are quite low at around $\sim$0.6. Note that the PSF is found to affect only the innermost values (upto 3 pixels) and the spheroid function, for each galaxy, is fitted on nearly $\sim$50-60 pixels. Thus, would not have resulted from the fitting of a non-resolved Gaussian at the centre. Pure spheroid population drops, from $\sim$22\% (for $z$$>$2.0) to $\sim$10\% (for $z$$<$2.0), with time. 

Interestingly, 47 galaxies, out of 171, were well fitted by a single exponential function (or disc) only. This is remarkable because the survival of fragile discs at high redshifts, where mergers are ever more frequent, poses challenge to the accepted theory of galaxy formation \citep{Kormendyetal2010,PeeblesandNusser2010}. While the azimuthally averaged isophotal intensity profile is exponential, there could be features which got submerged due to this averaging. We call them ``pure discs" for the ease of nomenclature and don't claim anything more. Like pure spheroids, pure discs are more numerous in the higher redshift bin ($z$$>$2.0) constituting $\sim$32\% of the sample, than in the lower bin ($z$$<$2.0) where it falls to $\sim$20\%.

The remaining 93 (i.e., 171$-$47$-$31) galaxies were well fit by two components (bulge $+$ disc). Due to the falling fraction of pure spheroids and pure discs, the fraction of 2-component systems witnesses a substantial increase, starting from $\sim$46\% (for $z$$>$2.0), it becomes $\sim$70\% of the total galaxy population in the lower redshift bin ($z$$<$2.0).

\begin{figure*}
\mbox{\includegraphics[width=65mm]{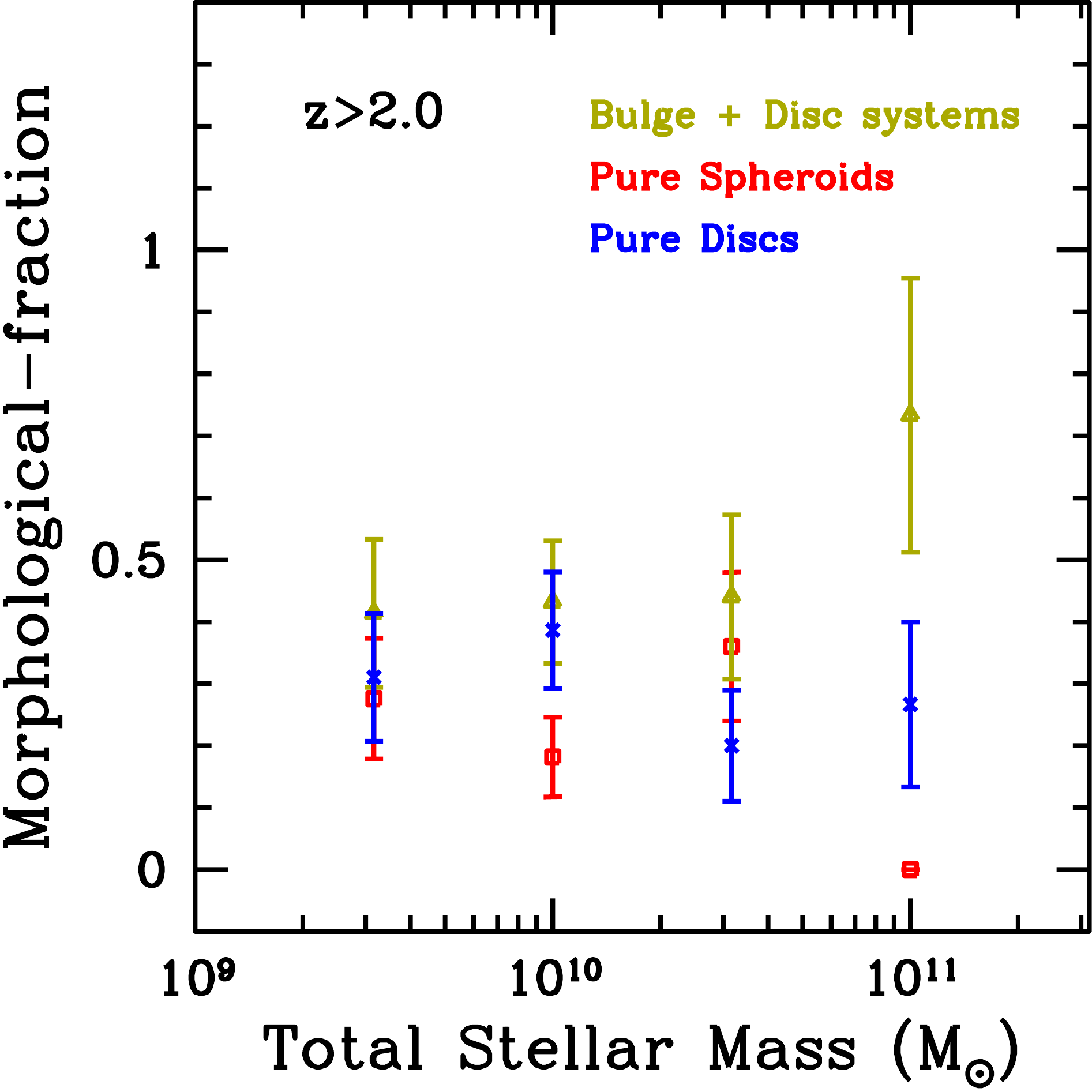}}
\mbox{\includegraphics[width=65mm]{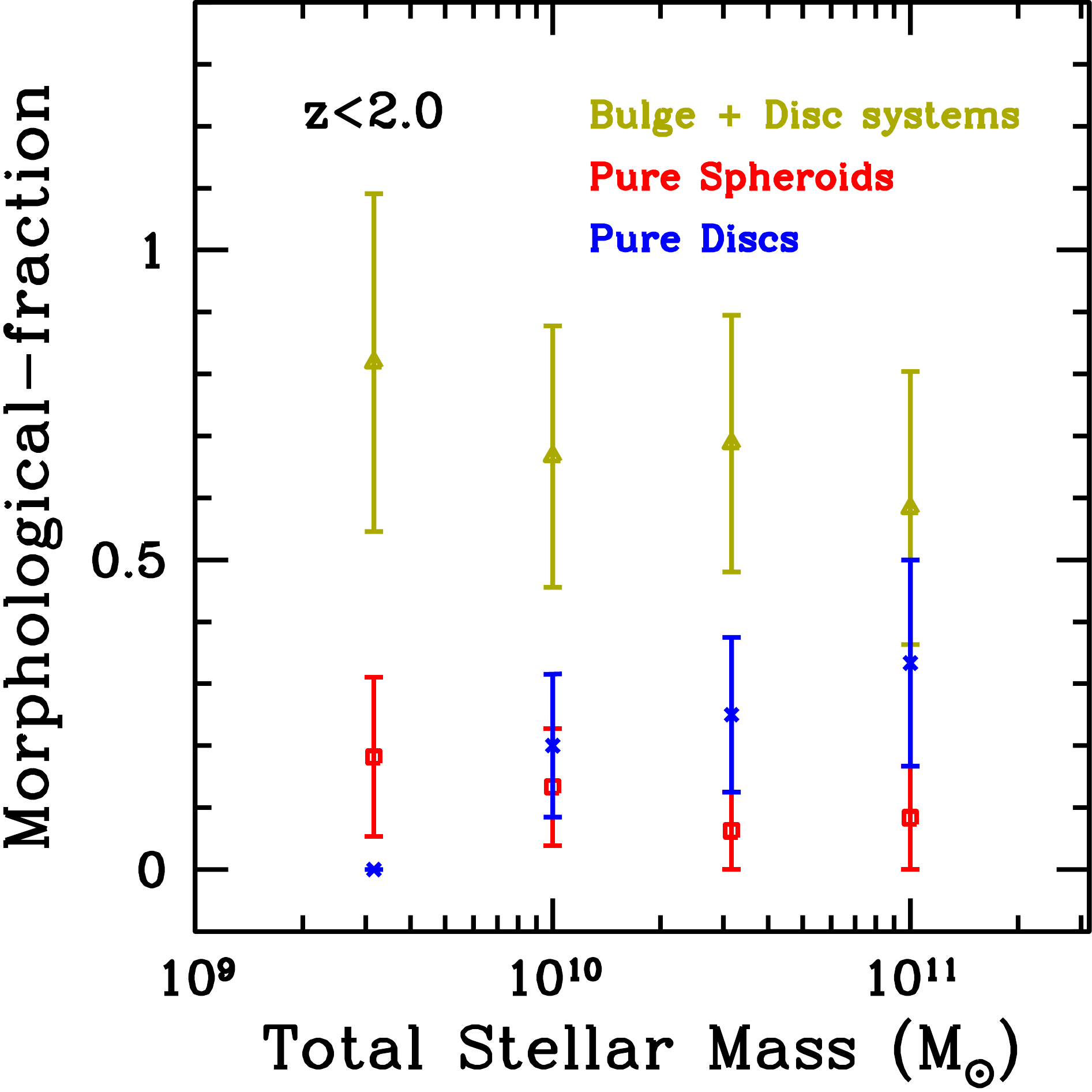}}
\caption{Morphological fraction of galaxies, as a function of their total stellar mass, is presented for the two redshift ranges.}
\label{morph-frac}
\end{figure*}

In Fig.~\ref{morph-frac}, we present the morphology fractions in each stellar mass bin for the two redshift ranges. For $z$$>$2, all morphologies have comparable fractions and 2-component systems dominate only in the most massive range (i.e., $>$ $10^{10.5}$$M_{\odot}$). The domination of 2-component systems in the most massive range is not driven by the fact that more massive galaxies supporting larger sizes would have been more susceptible for decomposition. An examination of Petrosian radii (known to be an unbiased measure of total size) of galaxies reveals that for 1-component systems, i.e., pure discs and pure spheroids, the smallest radius at $z$$>$2 is 4.6 and 4.9 kpc respectively. From the sample of 53 galaxies decomposed into 2-components at $z$$>$2, 6 galaxies have radii smaller than 4.9 kpc and 3.7 kpc is the minimum value.

For $z$$<$2, 2-component systems dominate in each stellar mass bin, especially towards the lower mass range (i.e., $<$ $10^{10}$$M_{\odot}$).
These patterns suggest that transformation of galaxies, from one component to 2-component systems, has been most significant at the lower end of the mass spectrum. We explore the effects of this transformation, i.e., the probable growth of substantial disc around pure spheroids and the probable growth of substantial bulge at the centre of pure discs, on the scaling relations and stellar properties of these morphologies.

\subsection{Scaling relations for bulges and host-discs}

\begin{figure*}
\mbox{\includegraphics[width=55mm]{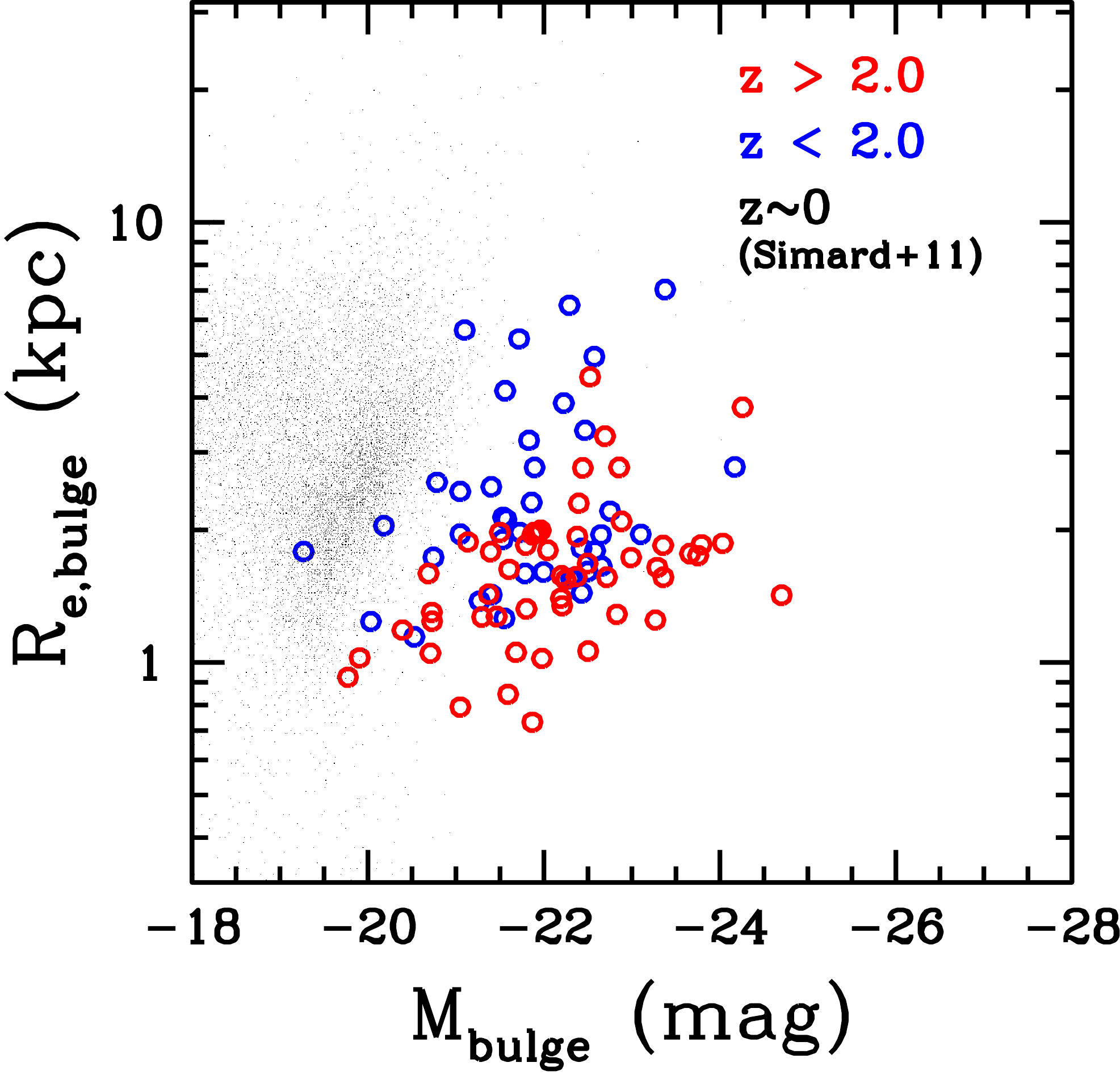}}
\mbox{\includegraphics[width=55mm]{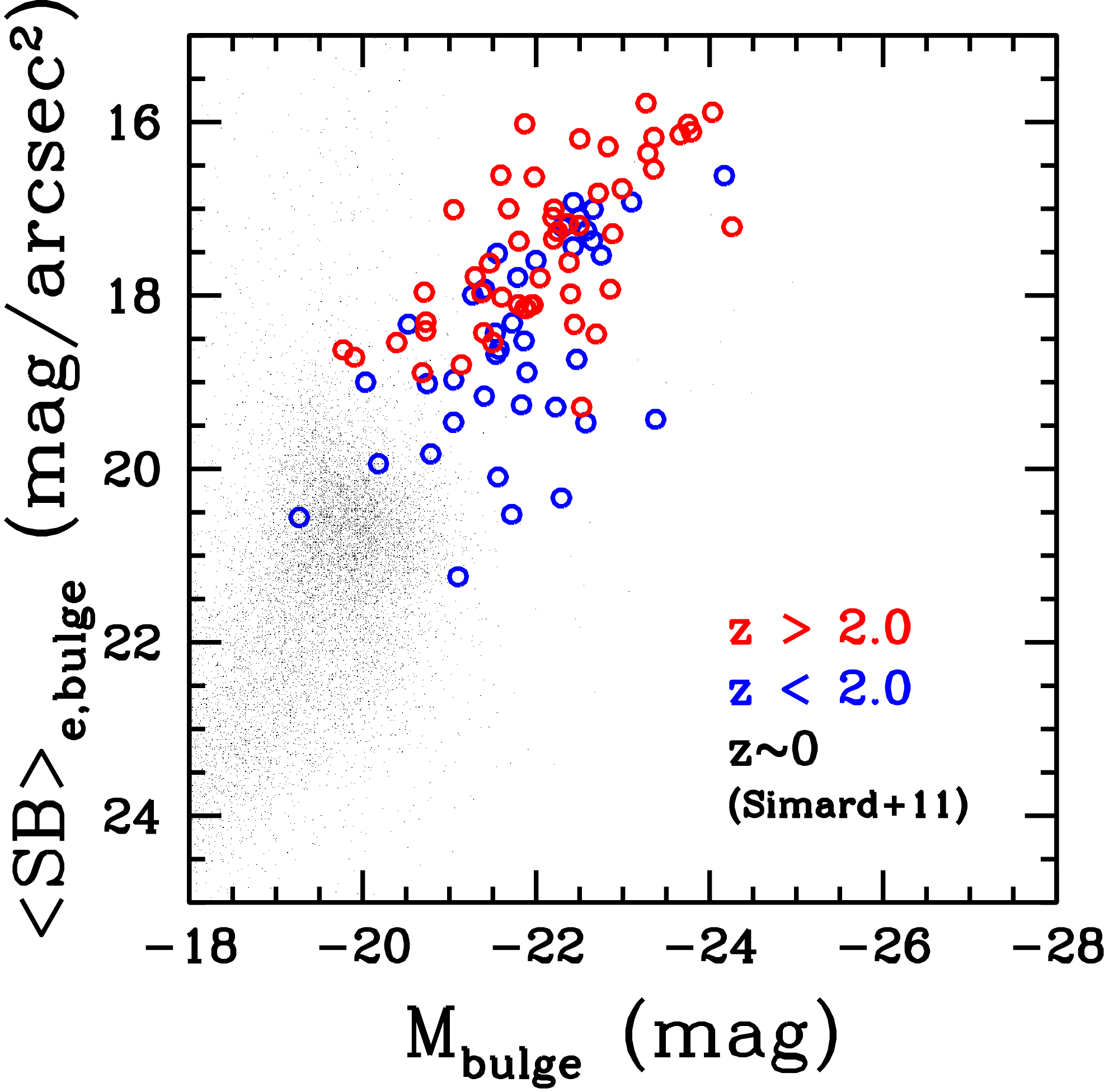}}
\mbox{\includegraphics[width=55mm]{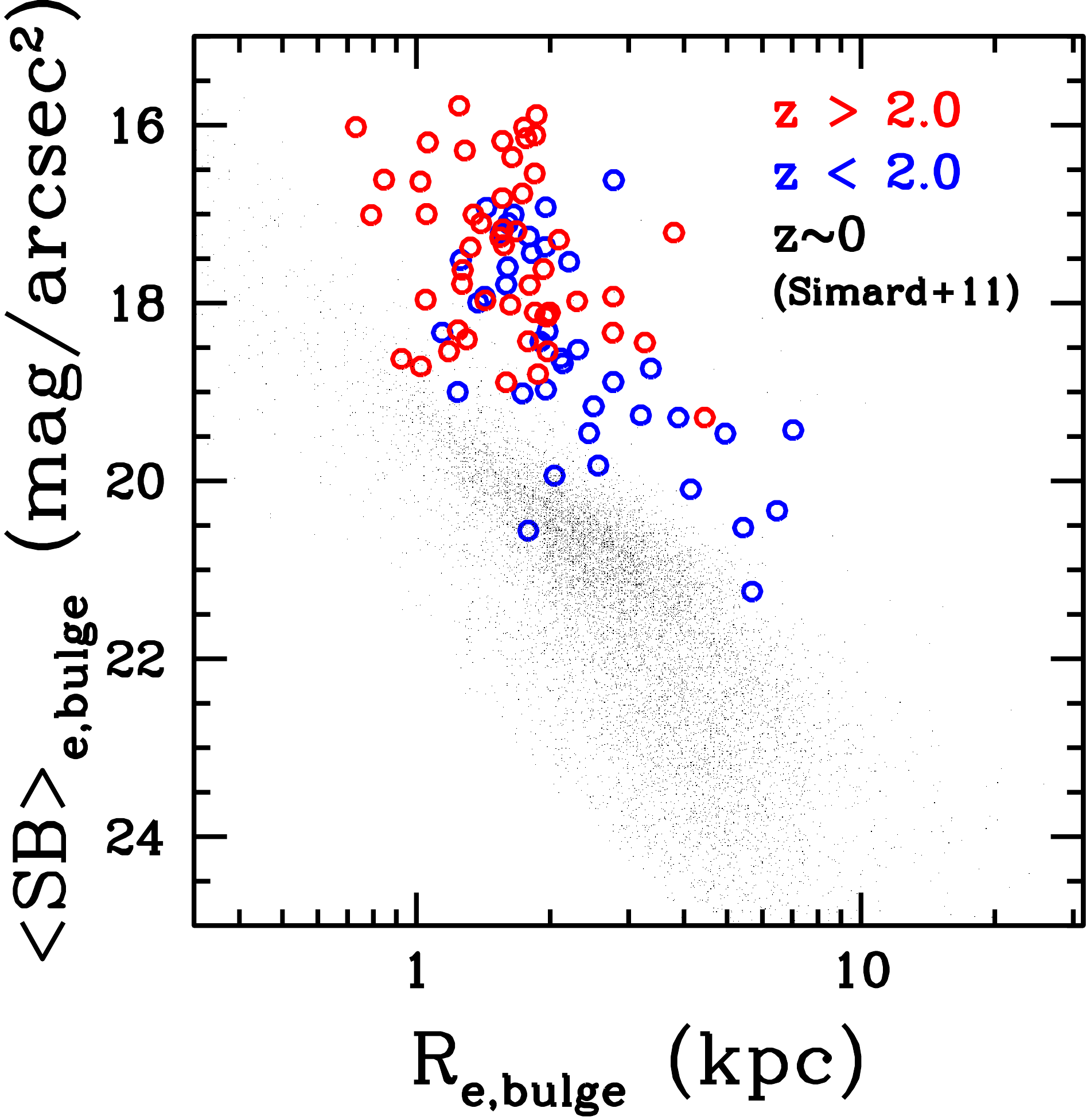}}\\
\mbox{\includegraphics[width=55mm]{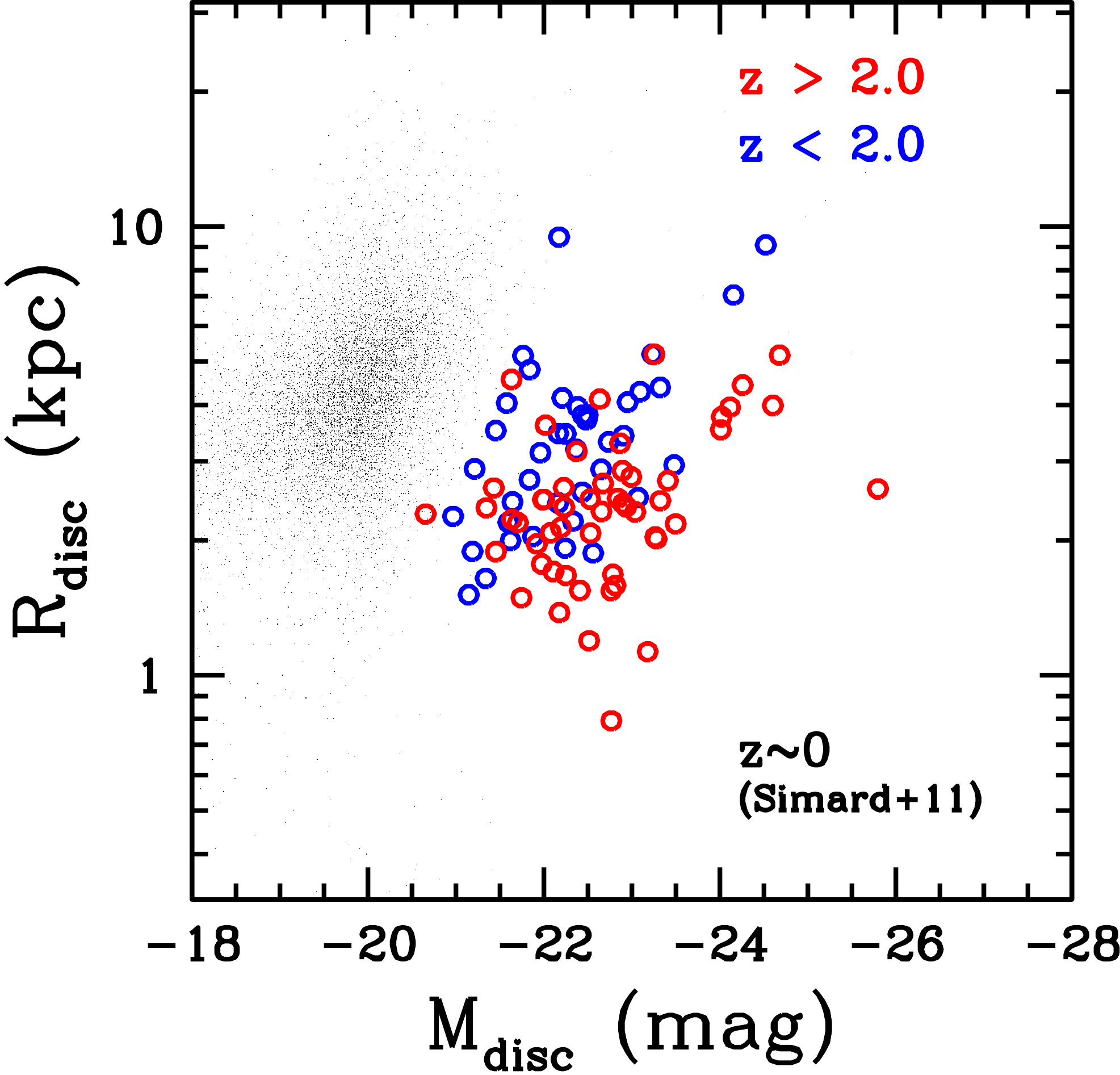}}
\mbox{\includegraphics[width=55mm]{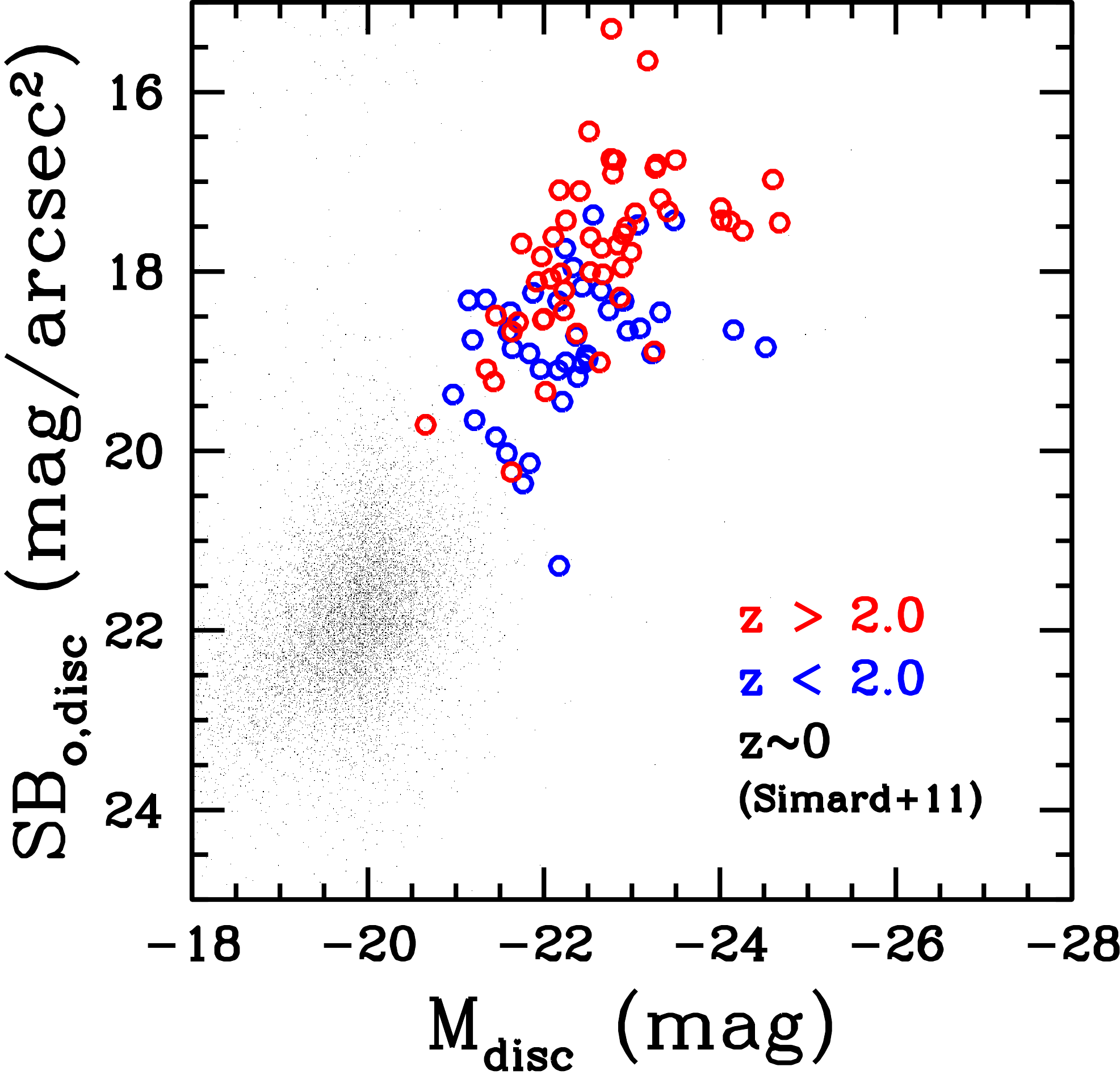}}
\mbox{\includegraphics[width=55mm]{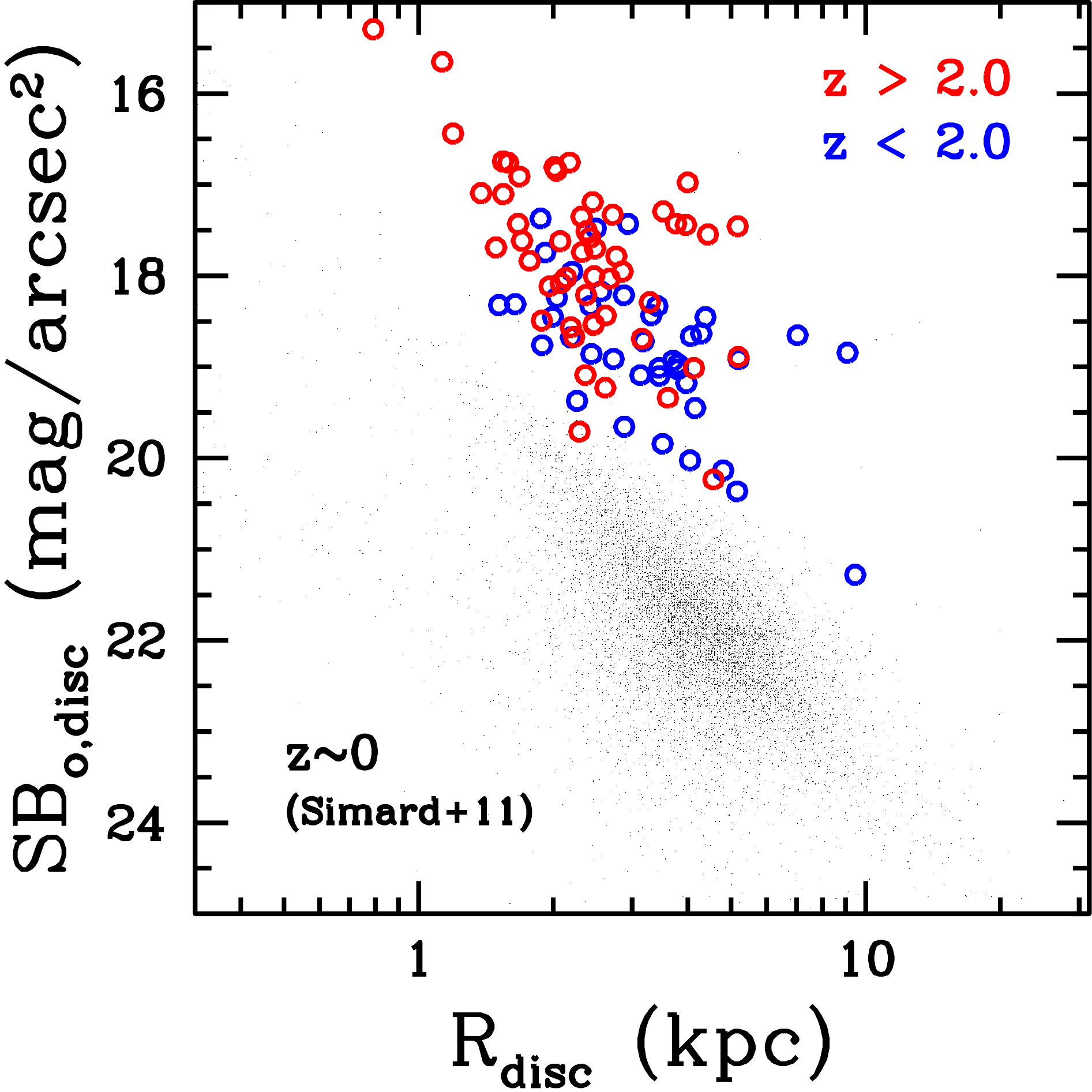}}
\caption{{\bf Scaling Relations:} {\it Upper panel:} Correlations between the absolute magnitude $M_{bulge}$, effective radius $R_{e,bulge}$ and average effective surface brightness $<$$SB$$>$$_{e,bulge}$ of the bulges in our sample compared to their distribution for the most luminous galaxies in the local Universe. {\it Lower panel:} Correlations between the absolute magnitude $M_{disc}$, scale length $R_{disc}$ and central surface brightness $SB_{o,disc}$ of host-discs in our sample compared to their distribution for the most luminous galaxies in the local Universe.}
\label{scaling-relations}
\end{figure*}

In Fig.~\ref{scaling-relations} we present the correlations of absolute magnitude, intrinsic size and intrinsic surface brightness of bulges and host-discs at the two redshift ranges probed in this work ($z$$<$2.0, $z$$>$2.0). We compare their distribution to that of bulges and host-discs in the local Universe. The local sample is derived from the extensive bulge-disc decomposition sample (82425 galaxies) of \citet{Simardetal2011}. On this local sample, we apply a total absolute magnitude cut ($M_B$$<$-20, in r.f. {\it B}-band) to select the most luminous $\sim$10\% (10971 galaxies) of the total. Since galaxies in the local are significantly fainter and dimmer than their high redshift counterparts, we are drawing a comparison only with the most luminous galaxies in the local. Even while comparing with the most luminous galaxies in the local, we find that both bulge and host-disc component have undergone significant decrease in luminosity ($\sim$3-4 mag) and brightness ($\sim$3-4 mag/arcsec$^2$), along with near doubling of their sizes, from the higher redshift ranges to the present epoch. If the entire local sample is included, it will further enhance the decrease observed in the luminosity and surface brightness of high redshift galaxies.

\begin{figure*}
\mbox{\includegraphics[width=55mm]{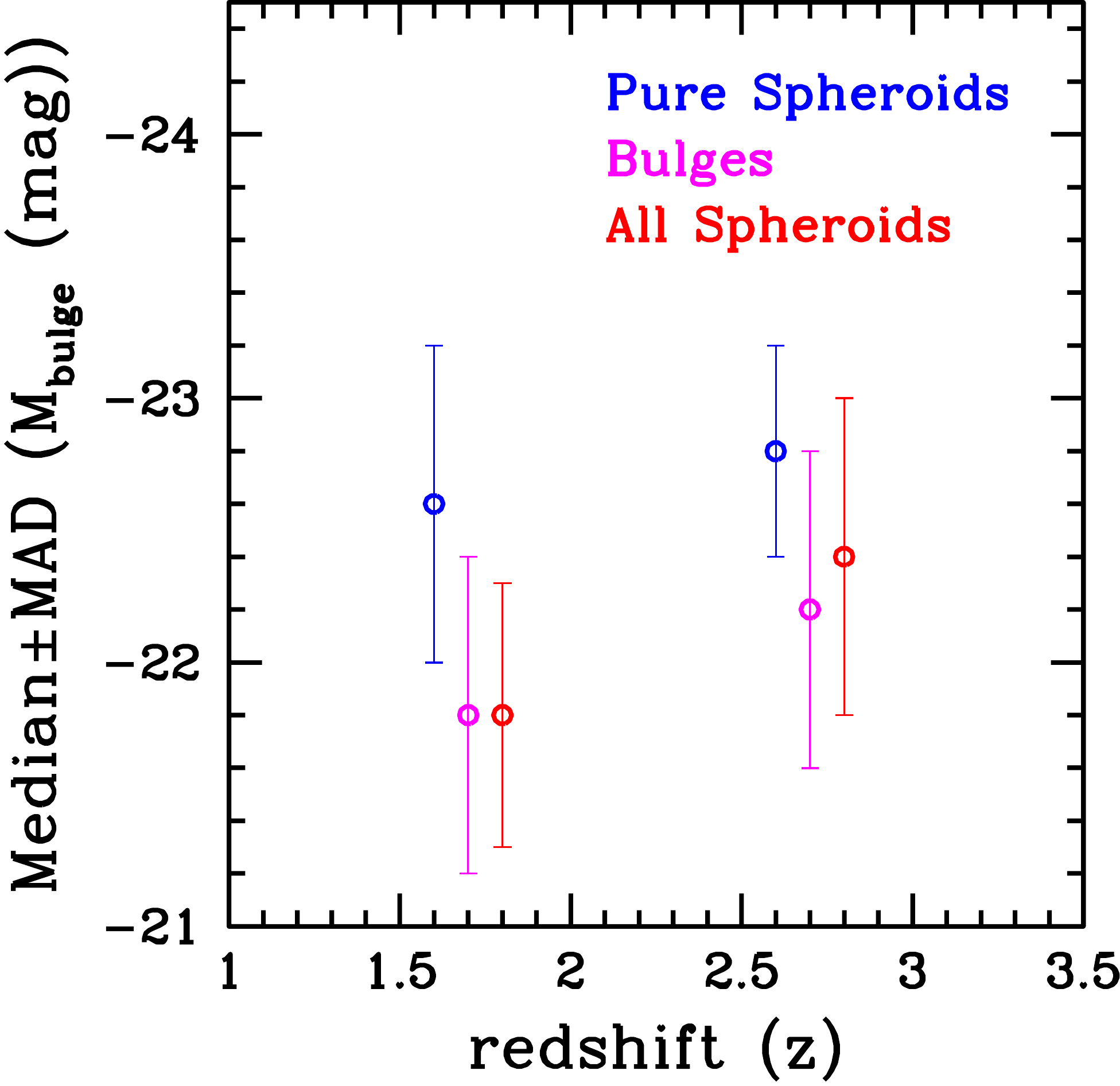}}
\mbox{\includegraphics[width=55mm]{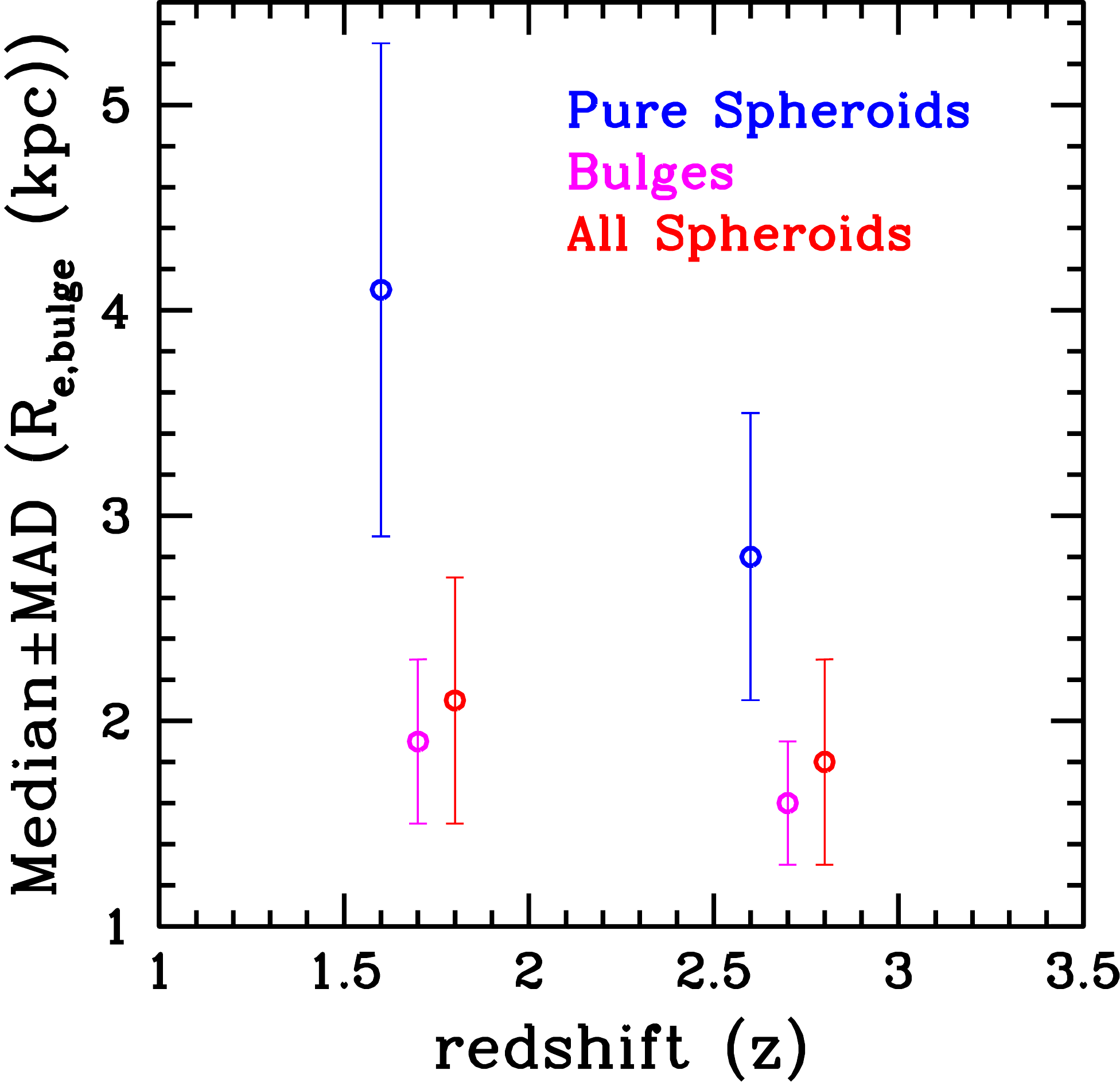}}
\mbox{\includegraphics[width=55mm]{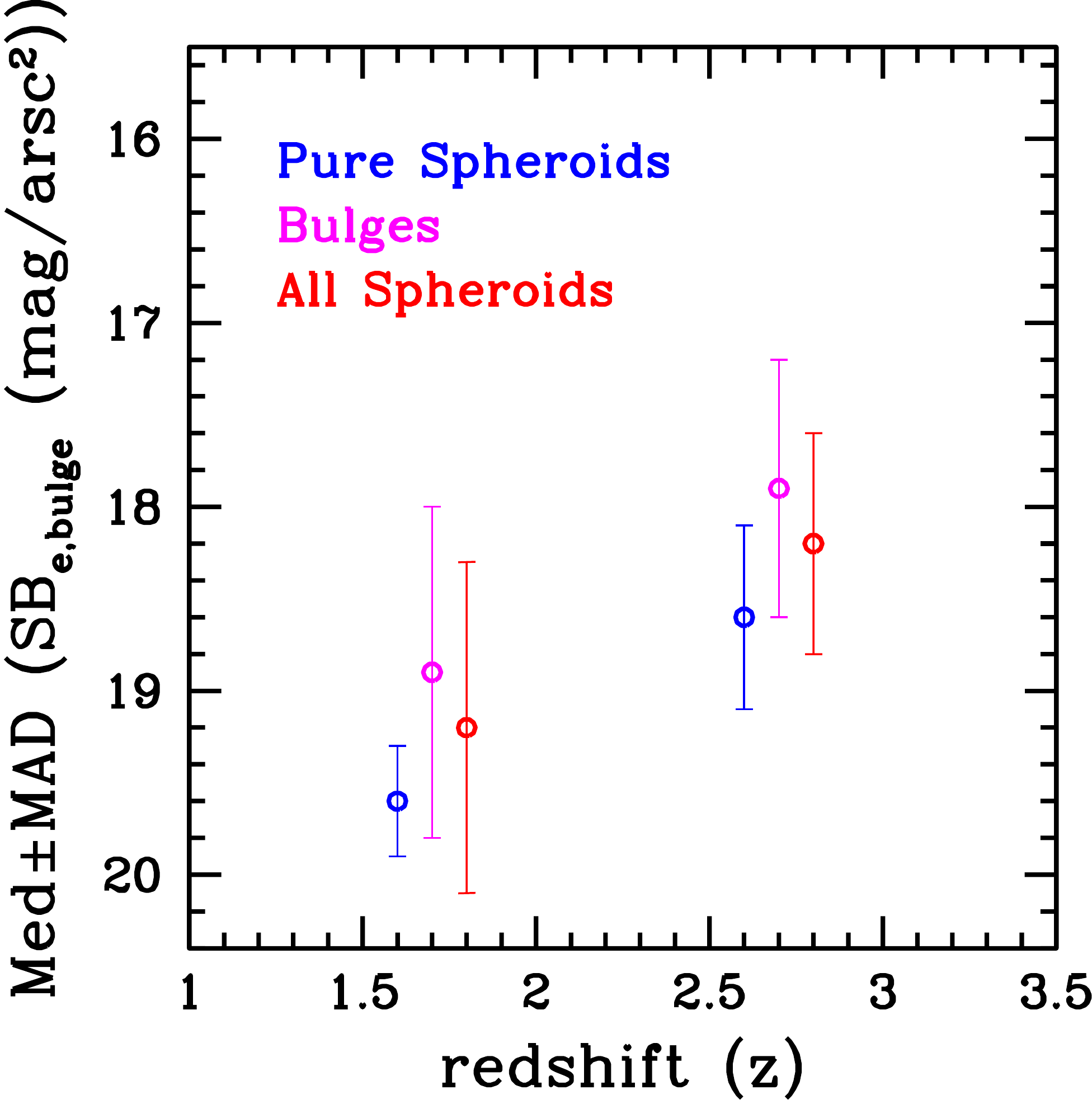}}\\
\mbox{\includegraphics[width=55mm]{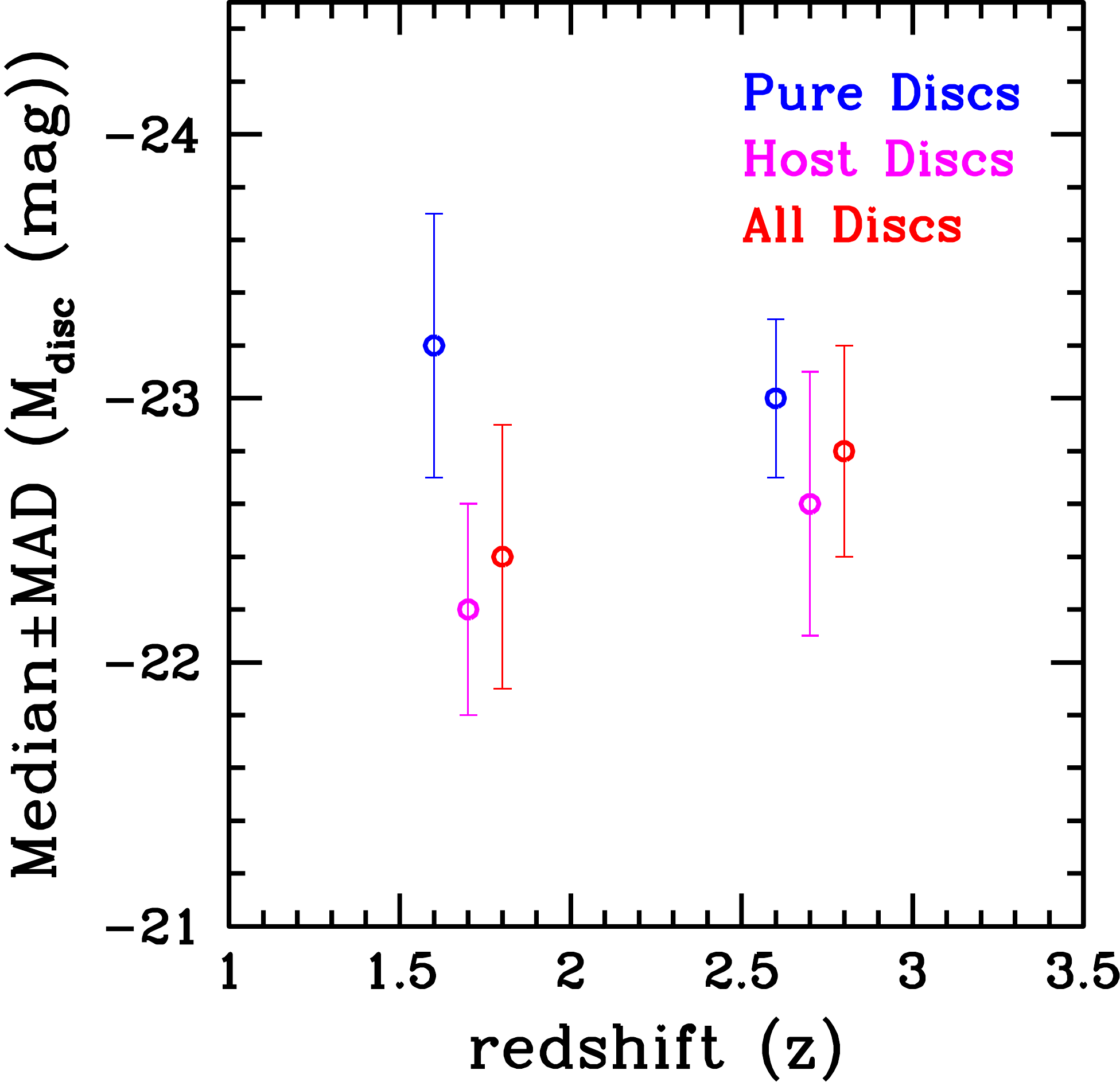}}
\mbox{\includegraphics[width=55mm]{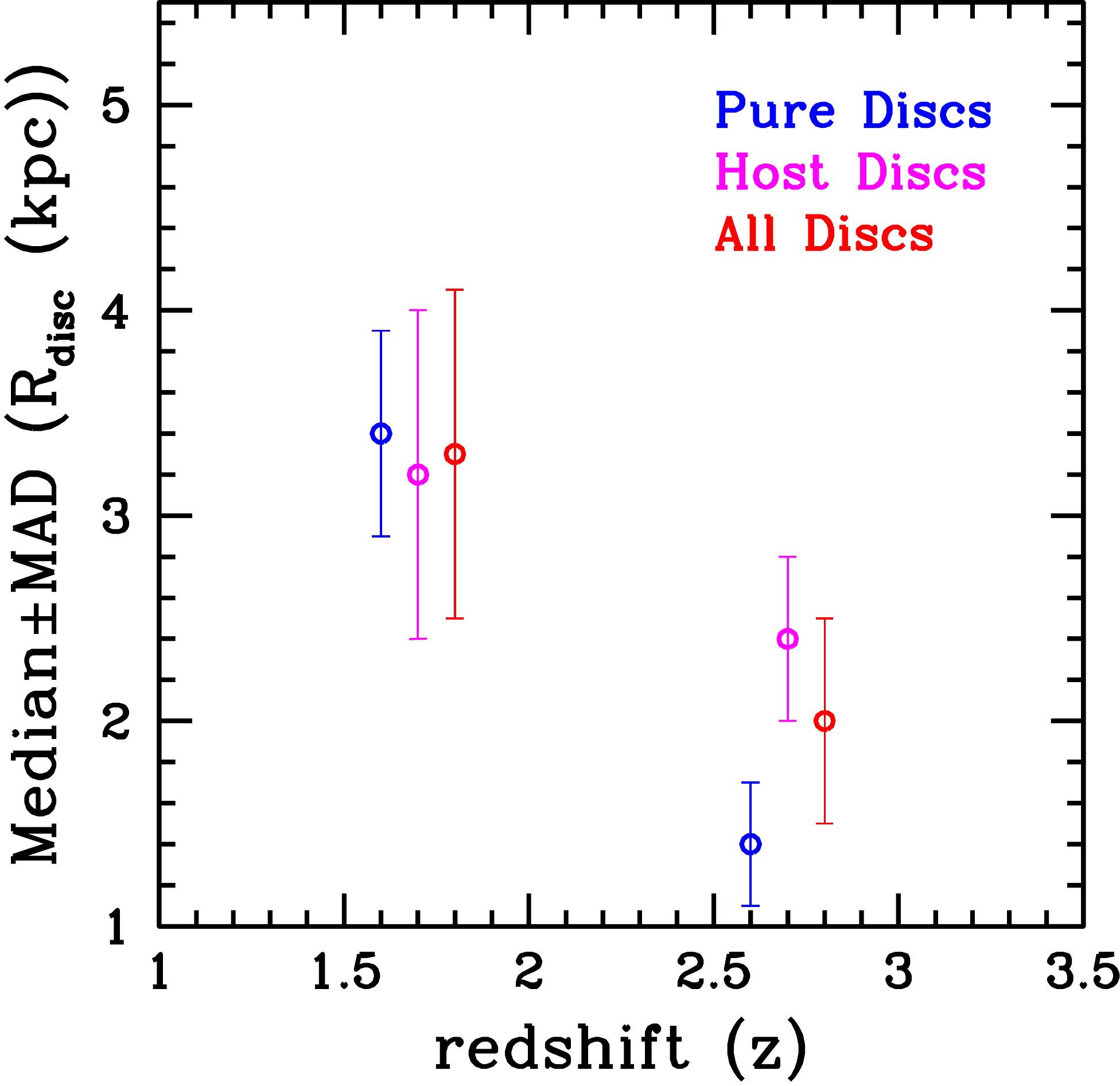}}
\mbox{\includegraphics[width=55mm]{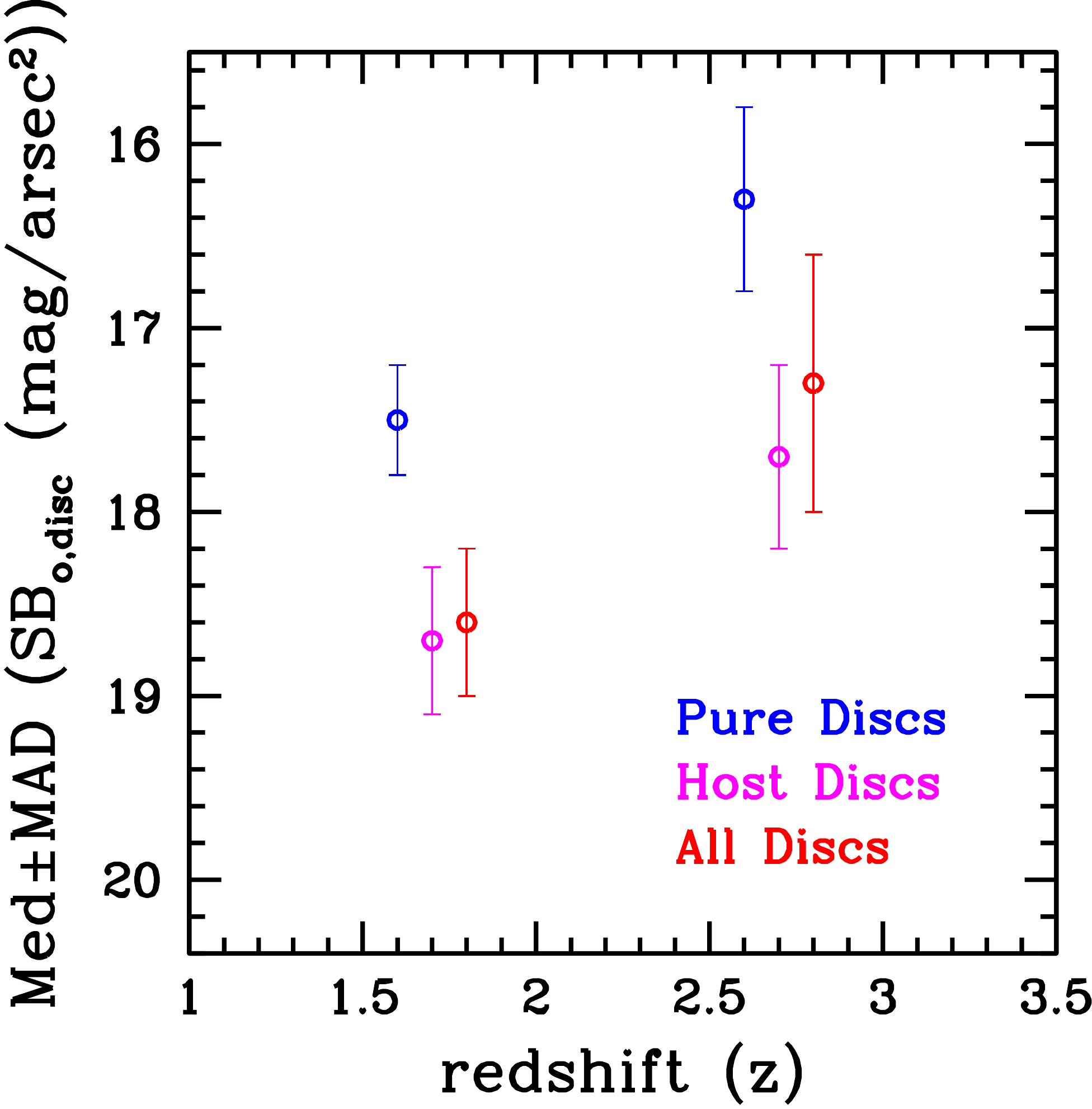}}
\caption{{\bf Median values:} {\it Upper panel:} Median values of absolute magnitude $M_{bulge}$, effective radius $R_{e,bulge}$ and average effective surface brightness $<$$SB$$>$$_{e,bulge}$ of the bulges, spheroids and the two combined, are marked for the two redshift ranges. {\it Lower panel:} Median values of the absolute magnitude $M_{disc}$, scale length $R_{disc}$ and central surface brightness $SB_{o,disc}$ of host-discs, pure discs and the two combined, are marked for the two redshift ranges. The error bars in each range refer to the median absolute deviation (MAD) values of that sample.}
\label{median-param}
\end{figure*}

While high redshift distributions are markedly distinct from the local, in Fig.~\ref{scaling-relations} transition signs are apparent in the distribution of the two high redshift range samples ($z$$<$2.0 and $z$$>$2.0) as well. These transition signs become more evident from Median and Median Absolute Deviation (MAD) values, presented in Fig.~\ref{median-param}, for the two redshift ranges. It shows that both bulge as well as the host-disc component get fainter ($\sim$0.5 mag) and dimmer ($\sim$1 mag/arcsec$^2$), as we move to $z$$<$2. Interestingly, while host-discs witness an increase in their scale length by a factor of $\sim$1.3, their bulges maintain roughly the same size, as we reach $z$$\sim$1.5. Thus, while for host-discs, size expansion appears to be a crucial factor in causing the dimming, the dimming of bulges appears to be driven only by the fall in their star formation activity.

\begin{figure*}
\mbox{\includegraphics[width=55mm]{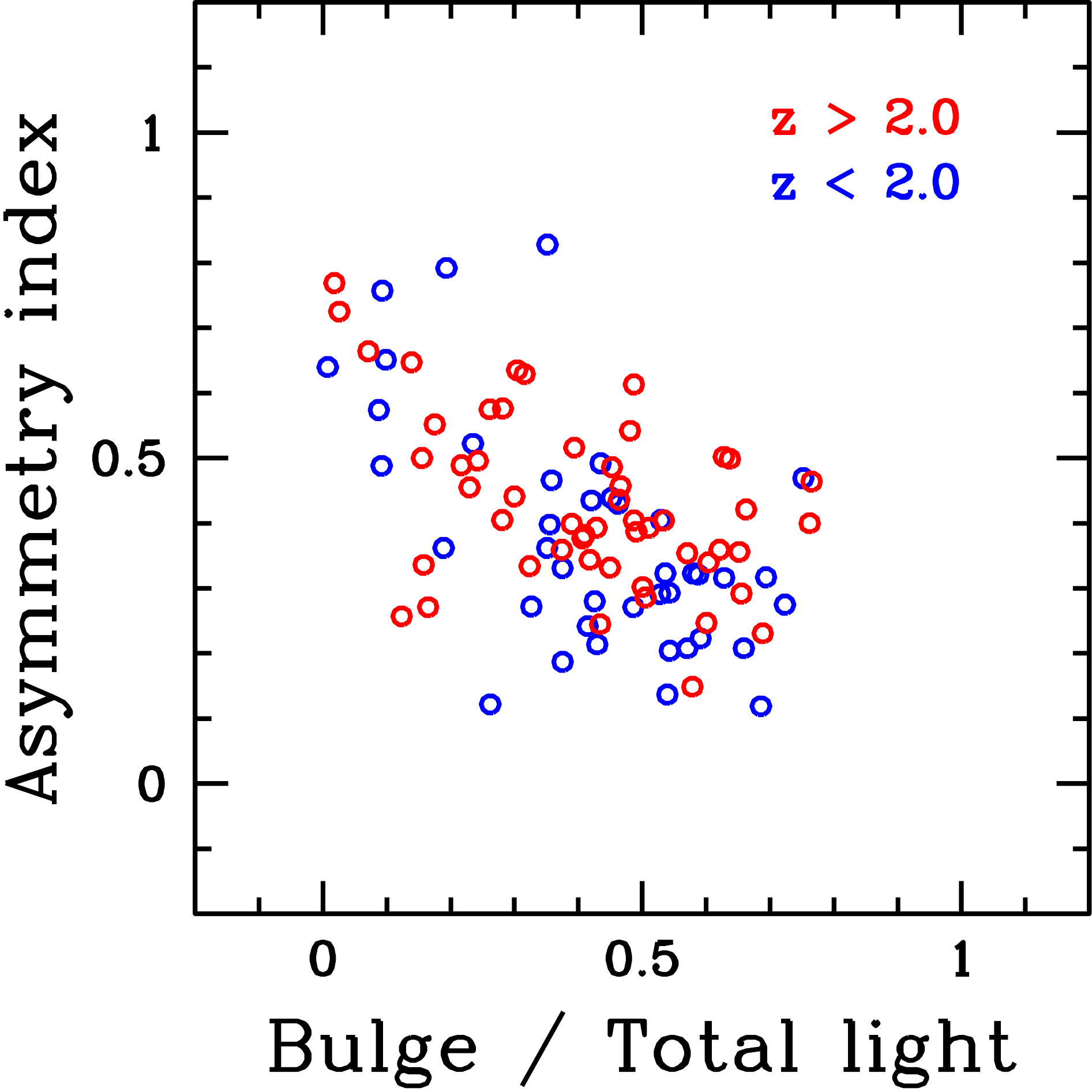}}
\mbox{\includegraphics[width=55mm]{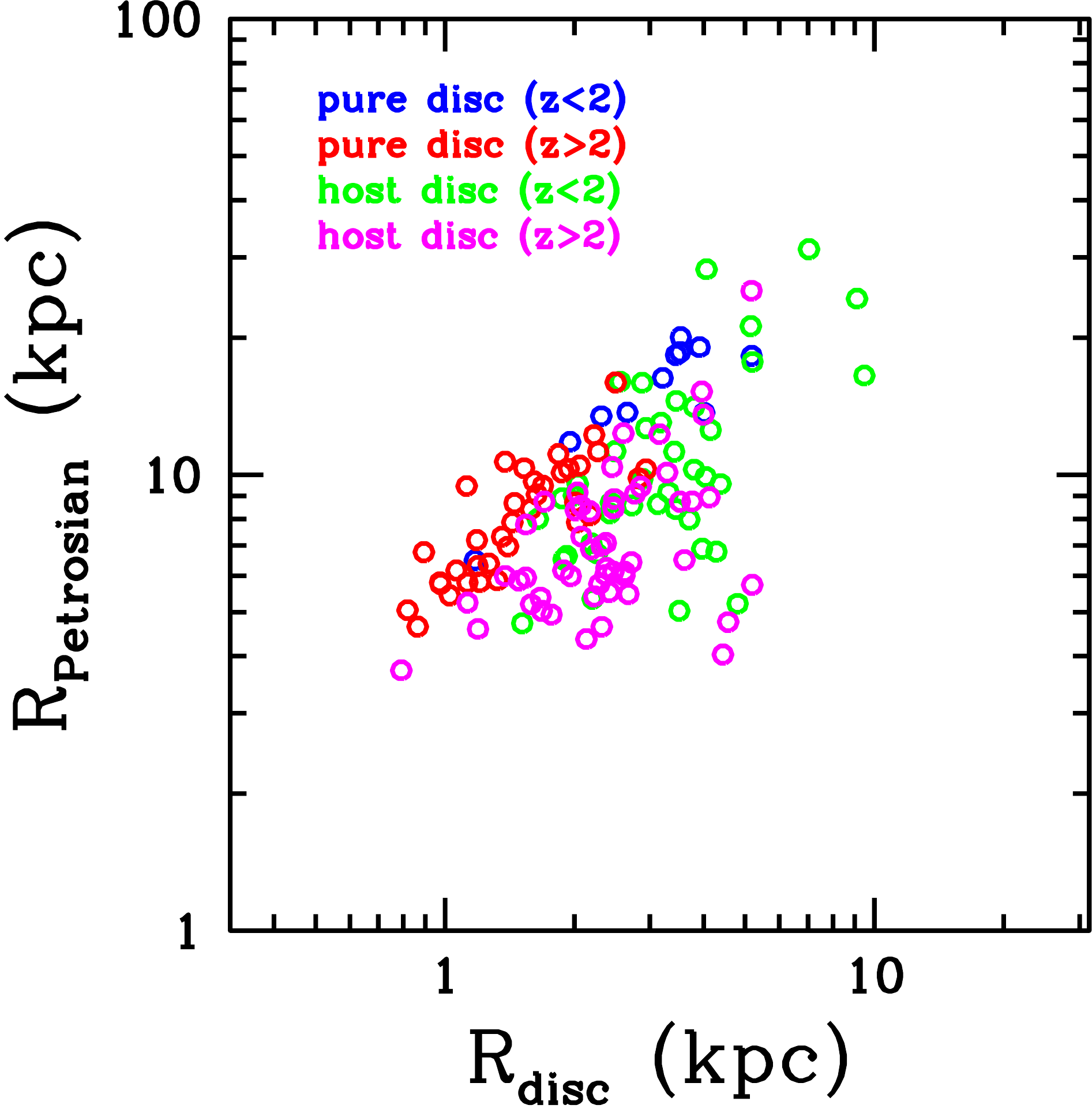}}
\mbox{\includegraphics[width=55mm]{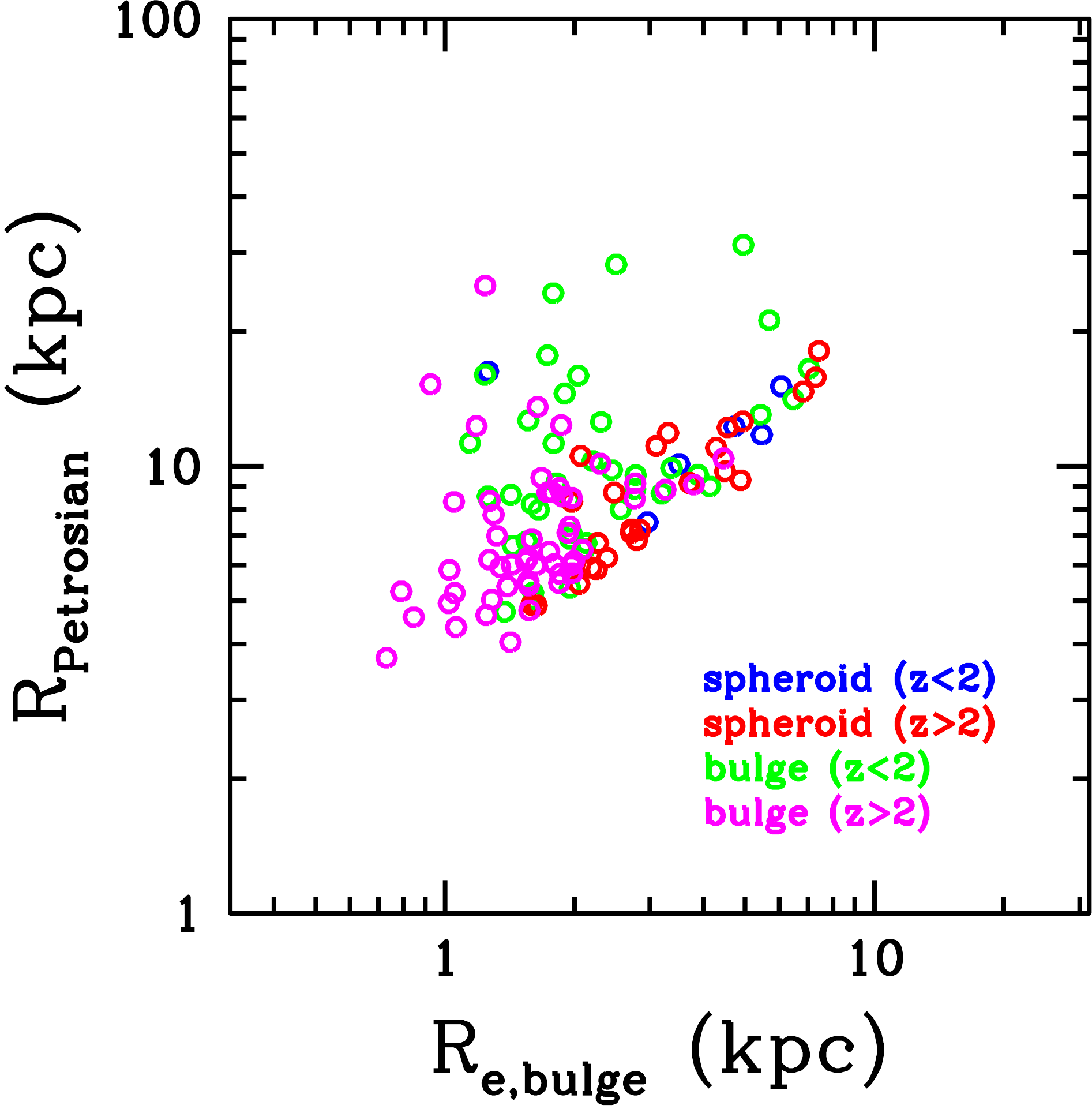}}
\caption{{\bf Non-parametric measures:} {\it Left panel:} The Asymmetry index of the galaxies is plotted against their bulge to total light ratio, for 2 component systems. {\it Middle panel:} The Petrosian radius (non parametric measurement of the full radius) of the galaxies is plotted against their disc scale length, where, the disc can either be pure disc or part of the 2-component system (i.e., host disc). {\it Right panel:} The Petrosian radius of the galaxies is plotted against their bulge effective radius, where, the bulge can either be pure spheroid or part of a 2-component system.}
\label{cas-param}
\end{figure*}

Quite similar trends are visible in the case of ``surviving" (because their population diminishes) pure spheroids and pure discs as well. Surviving pure disc galaxies register an expansion ($\sim$2.5 times) that is significantly more pronounced than that seen for surviving pure spheroids. This growth in disc size, of both pure discs and host-discs, in comparison to that of bulges and pure spheroids, is also the driving factor behind a $\sim$50\% increase in the total size of the full galaxy population, as seen through Petrosian radius (see Fig.~\ref{cas-param}). In addition to this discriminating increase in sizes, evolution in stellar properties (explored in following sections) also suggest that $z$$\sim$2 appears to be more of a disc formation period than that of spheroid formation. This is also evident from the lack of evolution in the distribution of bulge to total ratio (B/T) of the 2-component systems (Fig.~\ref{cas-param}), over the two redshift ranges. While it can be seen that the Asymmetry index is lower for galaxies with higher B/T, it is also apparent that there is no change in the distribution of either Asymmetry or B/T, over the two redshift ranges.

\subsection{Kormendy relation}

Projecting our sample of pure spheroids onto the Kormendy plane, we examine evolution in the Kormendy relation from local and intermediate redshifts to higher redshift ranges (1.5$<$$z$$<$4.0 or $z$$\sim$2). To obtain the local Kormendy relation, we select all bright ($M_B$$<$-20) galaxies from \citet{Simardetal2011} which were well fit by a single S\'ersic component with S\'ersic index higher than an empirically chosen value (we use 3.5) that minimises disc contamination. The relation for intermediate redshift (0.4$<$$z$$<$1.0) ellipticals is from our earlier works \citep{Sachdevaetal2015,Sachdevaetal2017} where we applied identical criteria for their selection.

\begin{figure*}
\mbox{\includegraphics[width=65mm]{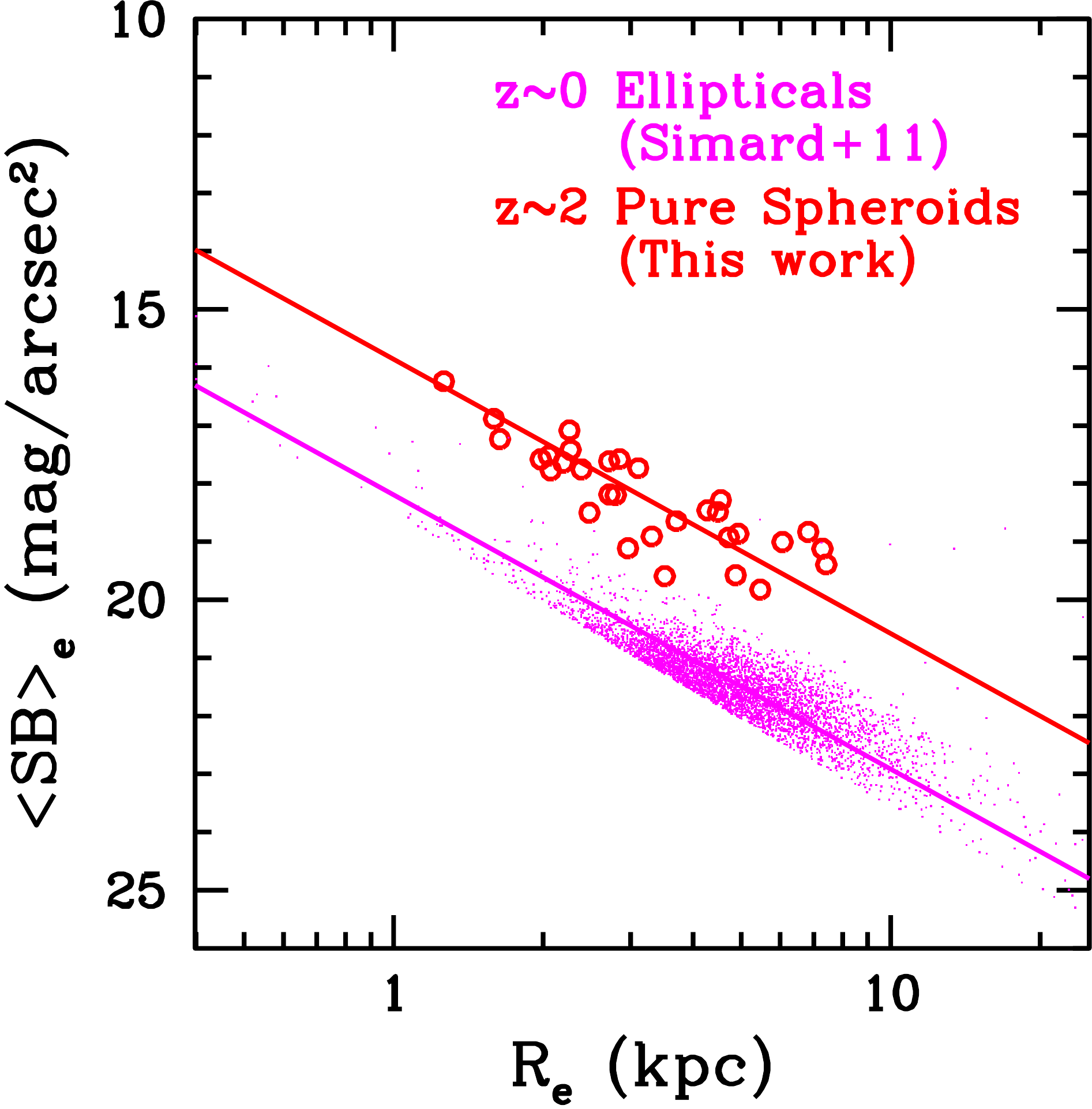}}
\mbox{\includegraphics[width=65mm]{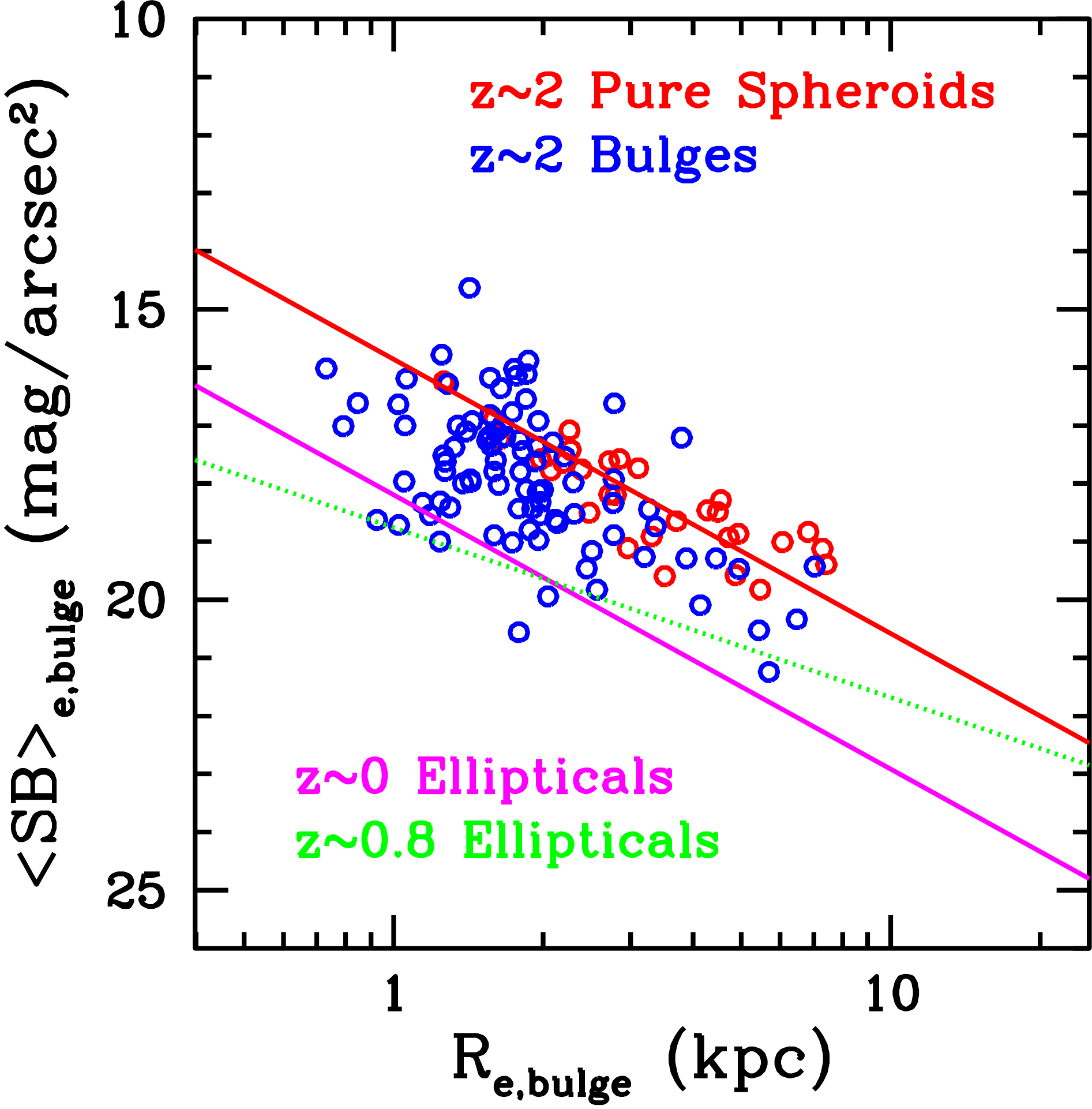}}
\caption{{\bf Kormendy relation:} Distribution of our sample ($z$$\sim$2) of pure spheroids (red circles) and bulges (blue circles) on the Kormendy plane is presented. Kormendy relation for bright local elliptical galaxies (solid magenta line) selected from the sample of \citet{Simardetal2011} is also marked. Distribution of this sample is presented (magenta dots) in the {\it left panel}. The relation followed by pure spheroids in our sample (solid red line) is marked in both the plots. The relation has the same slope as that observed for the local, while the intercept has shifted. In the {\it right panel}, we also mark the Kormendy relation for elliptical galaxies at intermediate redshifts (dashed green line) as found in our earlier works \citep{Sachdevaetal2015,Sachdevaetal2017}.}
\label{kormendy}
\end{figure*}

For local ellipticals, the Kormendy relation is given by the equation

\begin{equation}
<SB>_{e,B} = (4.72 \pm 0.02) \times R_{e,B} + (18.20 \pm 0.02),
\end{equation}

\noindent where, $<$$SB$$>$$_{e,B}$ is the intrinsic average surface brightness inside the effective radius $R_{e,B}$ in rest-frame {\it B}-band. In comparison, we find that high redshift ($z$$\sim$2) spheroids are well fit by this relation,

\begin{equation}
<SB>_{e,B} = (4.72 \pm 0.09) \times R_{e,B} + (15.86 \pm 0.09),
\end{equation}

\noindent where, the slope has the same value as for the local ellipticals but the intercept moves to a $\sim$2.3 mag brighter value. The two relations, marked in Fig.~\ref{kormendy}, suggests that same size galaxies were 5-6 times brighter at $z$$\sim$2 than they are at $z$$\sim$0. However these two might not be representative of the same population at different redshift ranges, i.e., high redshift galaxies might have witnessed significant size increase or surface brightness decrease or both with time. Also, high redshift spheroids may not have necessarily evolved into local elliptical galaxies. For intermediate redshifts, a different slope value (2.92$\pm$0.08) has been observed \citep{LaBarberaetal2003,Longhettietal2007,Sachdevaetal2015}. This relation is also marked in the right panel of Fig.~\ref{kormendy}. It is interesting to note that the high redshift ($z$$>$1.5) ellipticals, unlike the intermediate redshift ones, follow the same slope as their local counterparts.       
 
\subsection{Stellar properties and the main-sequence}

\begin{figure*}
\mbox{\includegraphics[width=65mm]{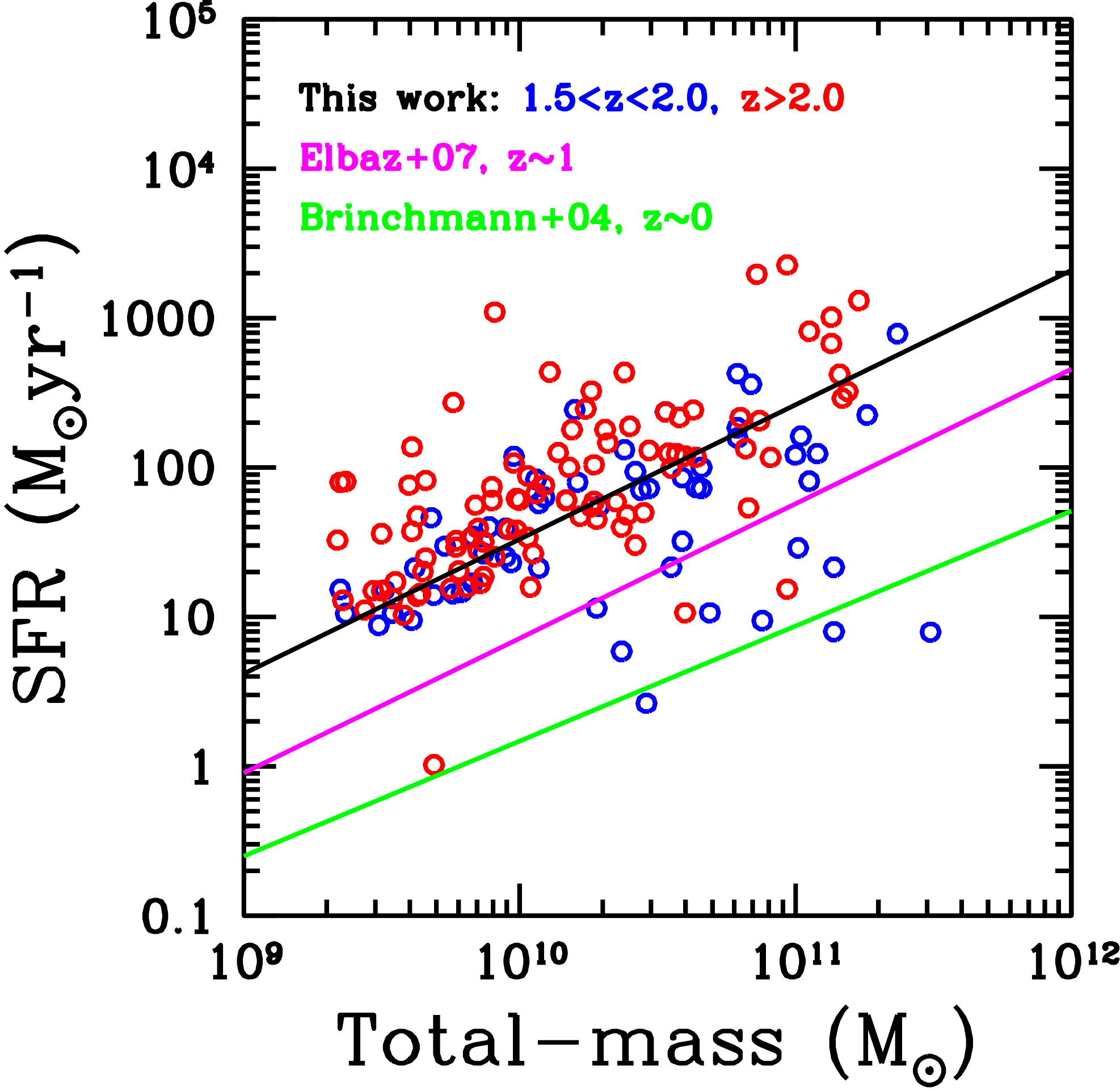}}
\mbox{\includegraphics[width=65mm]{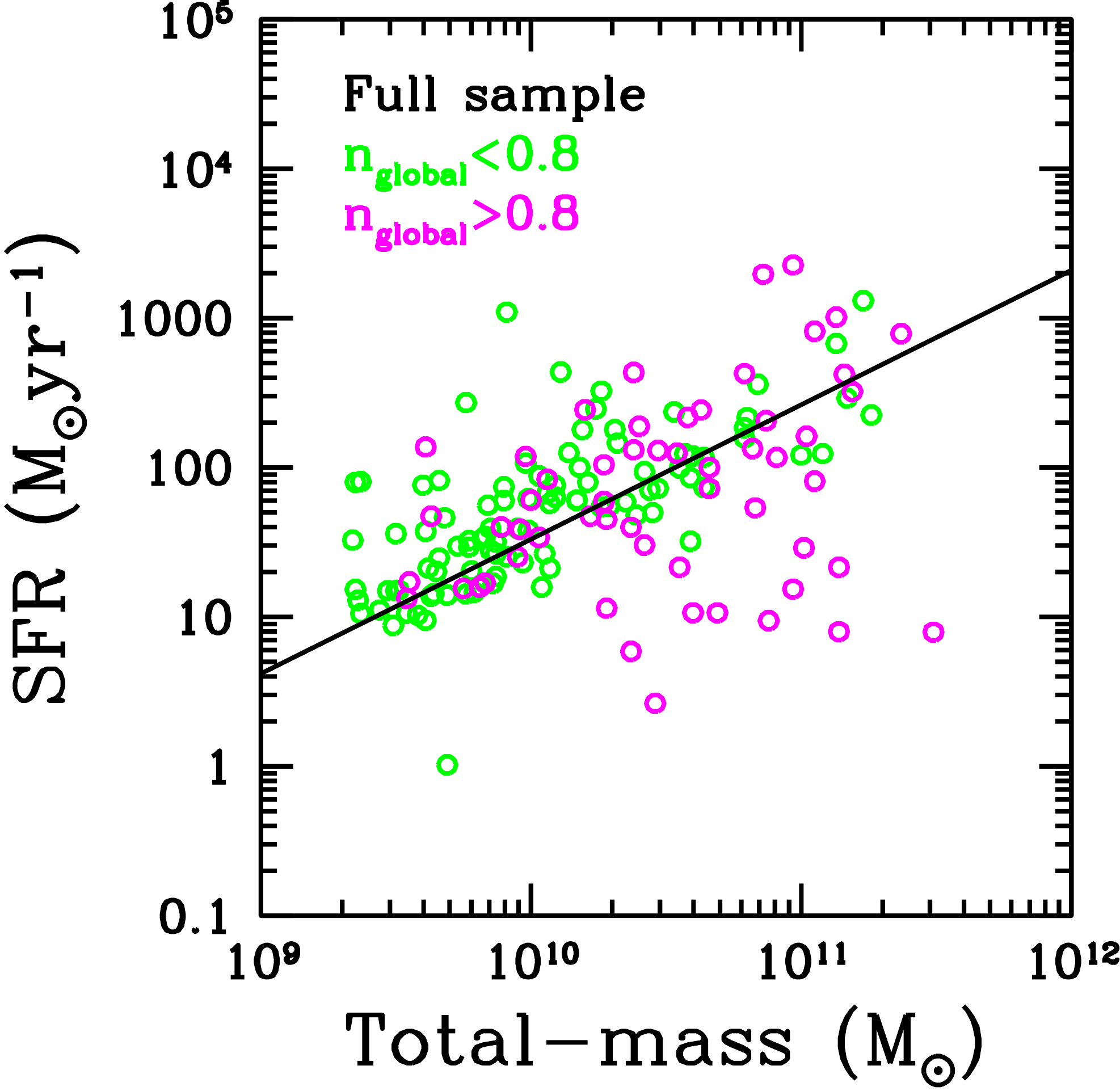}}\\
\mbox{\includegraphics[width=55mm]{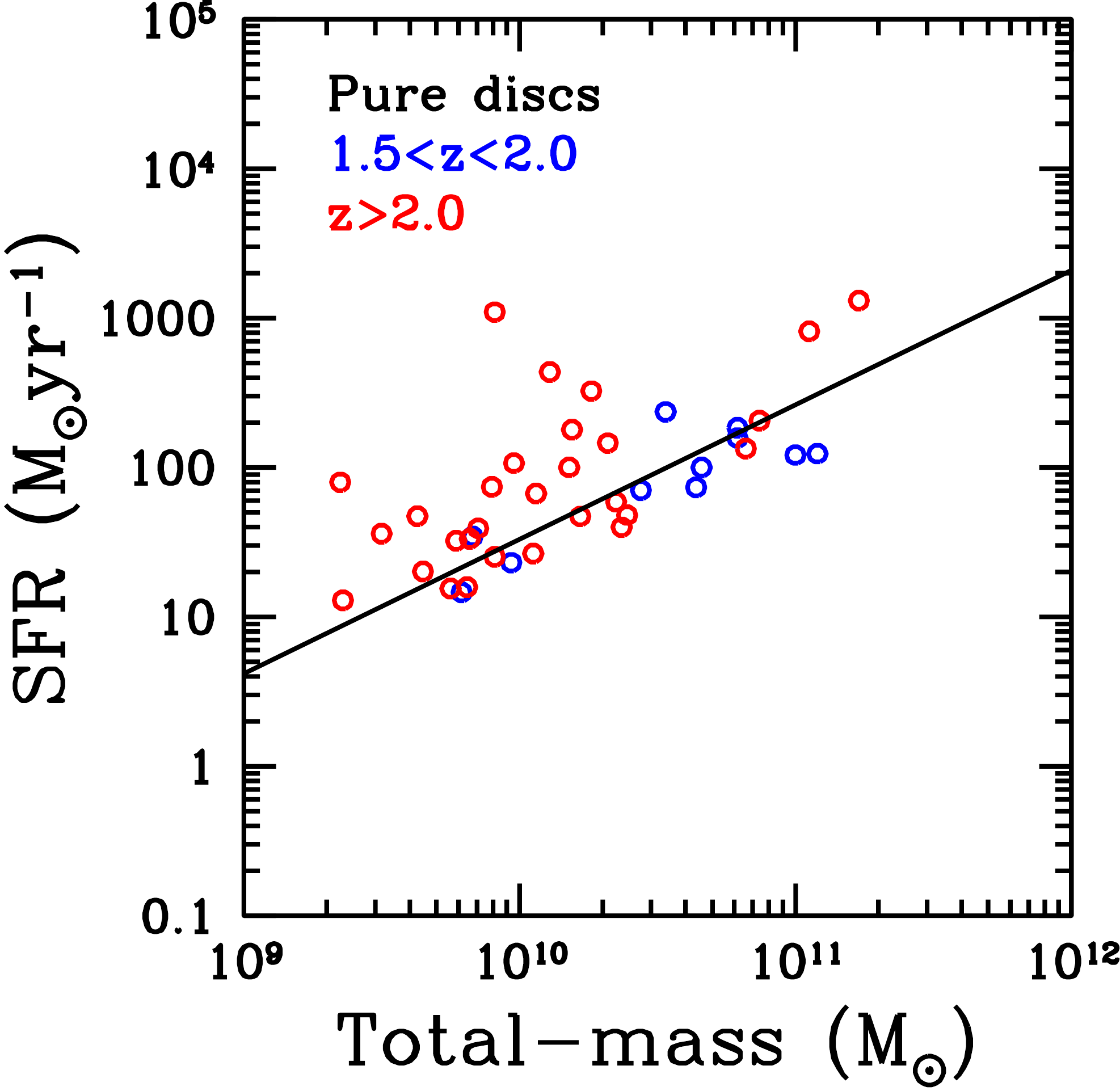}}
\mbox{\includegraphics[width=55mm]{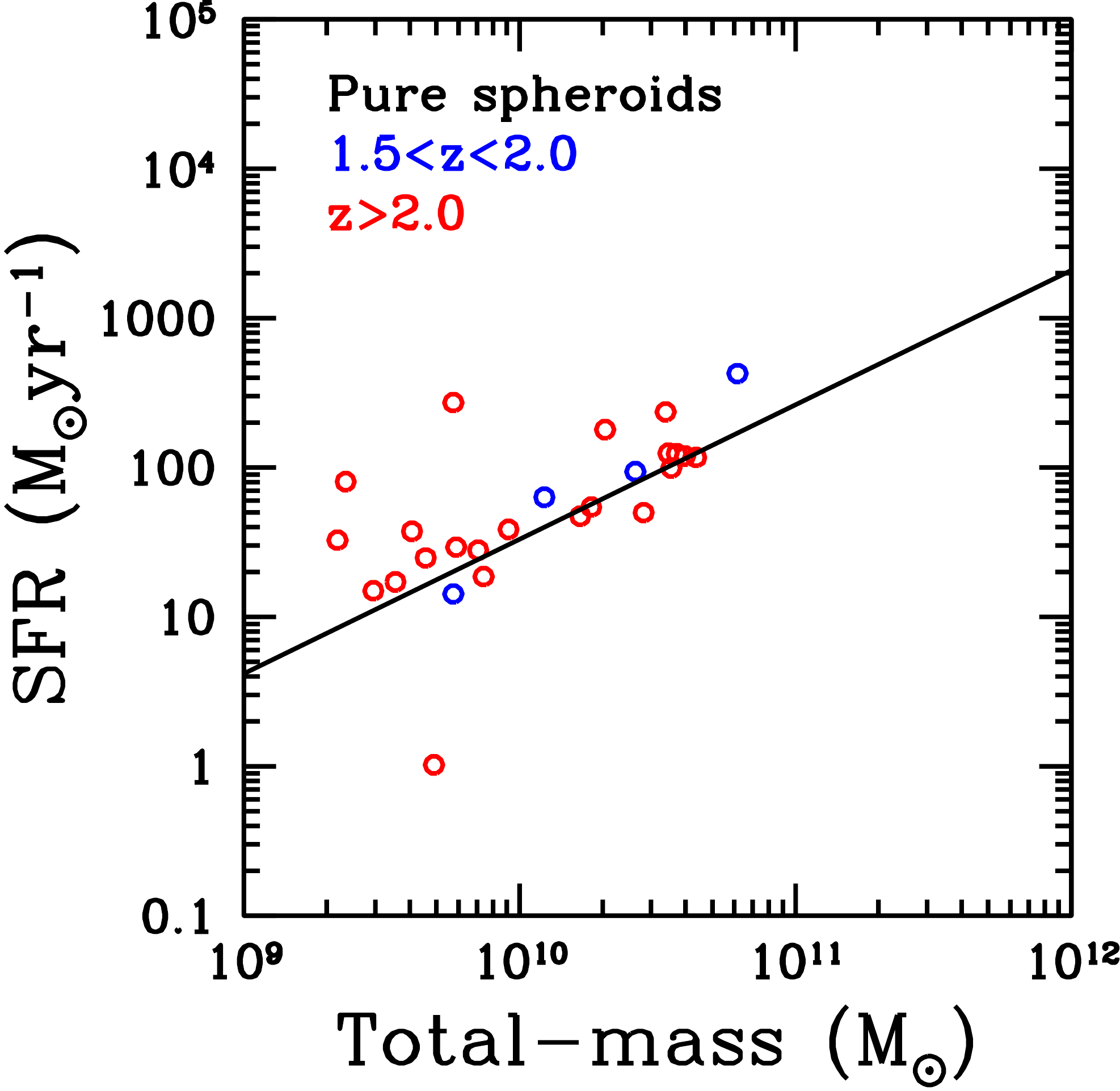}}
\mbox{\includegraphics[width=55mm]{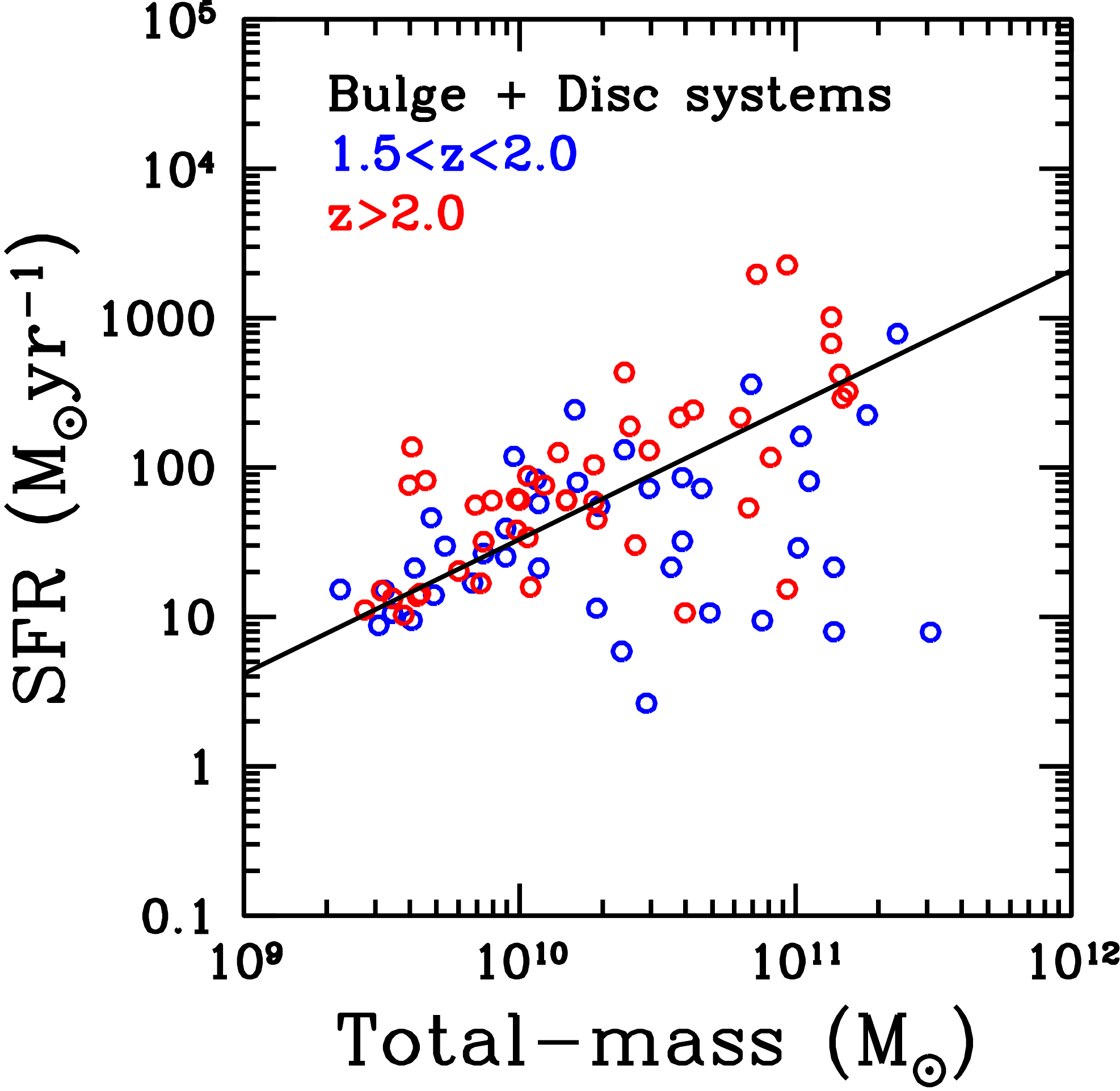}}
\caption{{\bf Main-Sequence:} Distribution of our sample of galaxies on the main-sequence, i.e., SFR-mass plane, is presented. While in the {\it upper left panel} galaxies are divided according to their redshifts, in the {\it upper right panel} they are divided according to their global S\'ersic index as obtained from {\it Galfit} single S\'ersic fitting. The solid black line in {\bf all plots} is the fit to our full sample, indicating the main-sequence relation around $z$$\sim$2. Main-sequence relations observed for star-forming galaxies at other redshifts, i.e., $z$$\sim$1 (solid magenta line, \citet{Elbazetal2007}) and $z$$\sim$0 (solid green line, \citet{Brinchmannetal2004}) are also marked in {\it upper left panel} plot. The three plots in the {\it lower panel} describe the placement of pure discs, pure spheroids and 2-component systems on the main-sequence for the two redshift ranges.}
\label{mainsequence}
\end{figure*}

In Fig.~\ref{mainsequence}, we present the distribution of our full sample on the main-sequence, i.e., SFR-stellar mass plane. Obtaining a fit for our full sample we find that the relation, for $z$$\sim$2, has a similar slope as that observed for star-forming galaxies in the GOODS field for $z$$\sim$1 by \citet{Elbazetal2007}. We have marked the relations observed for star-forming galaxies at $z$$\sim$0 and $z$$\sim$1 \citep{Brinchmannetal2004,Elbazetal2007} along with the relation observed for our sample. These relations are given by

\begin{equation}
\log(SFR/M_{\odot}yr^{-1})^{z\sim0} = 0.77 \times \log(M_*/M_{\odot}) - 7.53,
\end{equation}
\begin{equation}
\log(SFR/M_{\odot}yr^{-1})^{z\sim1} = 0.9 \times \log(M_*/M_{\odot}) - 8.14,
\end{equation}
\begin{equation}
\begin{split}
\log(SFR/M_{\odot}yr^{-1})^{z\sim2} & = (0.9 \pm 0.06) \times \log(M_*/M_{\odot})\\
                                    & \quad - (7.48 \pm 0.05),
\end{split}
\end{equation}

\noindent where, although slope for $z$$\sim$2 is similar to that observed for $z$$\sim$1, intercept shifts towards higher SFR values. It can be seen from Fig.~\ref{mainsequence} that the shift is in similar proportion to that observed from $z$$\sim$0 to $z$$\sim$1. The outliers, towards the quiescent side of the relation, are all in the lower redshift range ($z$$<$2.0) and with higher global S\'ersic indices ($n_g$$>$0.8) obtained from single S\'ersic fitting using {\it Galfit}. The value $n_g$$=$0.8 equally divides the full galaxy sample into two parts, i.e., those with lower ($n_g$$<$0.8) and higher ($n_g$$>$0.8) global S\'ersic indices.

Note that stellar mass and SFR are global measurements, i.e., refer to that of the full galaxy. Measurement of stellar mass and SFR separately for the bulge and disc component would have provided significant insight regarding the build up of stellar mass and quenching (or starburst) associated with the same. Although it is not measurable for the present sample due to resolution and wavelength constraints, future telescopes and dedicated spectroscopic surveys are awaited for the same.

In Fig.~\ref{mainsequence}, we present the placement of pure discs, pure spheroids and 2-component systems on the main-sequence. Both pure discs as well as pure spheroids favour the higher star formation side of the relation. We add a note of caution here that it is possible that their higher stellar activity might have prevented the detection of the other component, making it fainter in comparison to the bright nuclei. All quiescent outliers are 2-component systems and mostly belong to the lower redshift range. Earlier we found that the population of 2-component systems increases from $\sim$45\% (for $z$$>$2.0) to $\sim$70\% (for $z$$<$2.0), at the expense of pure spheroids and pure discs. This indicates that processes involved in the transformation of single component systems to 2-component systems are also responsible for the quenching of their star formation activity.

\begin{figure*}
\mbox{\includegraphics[width=55mm]{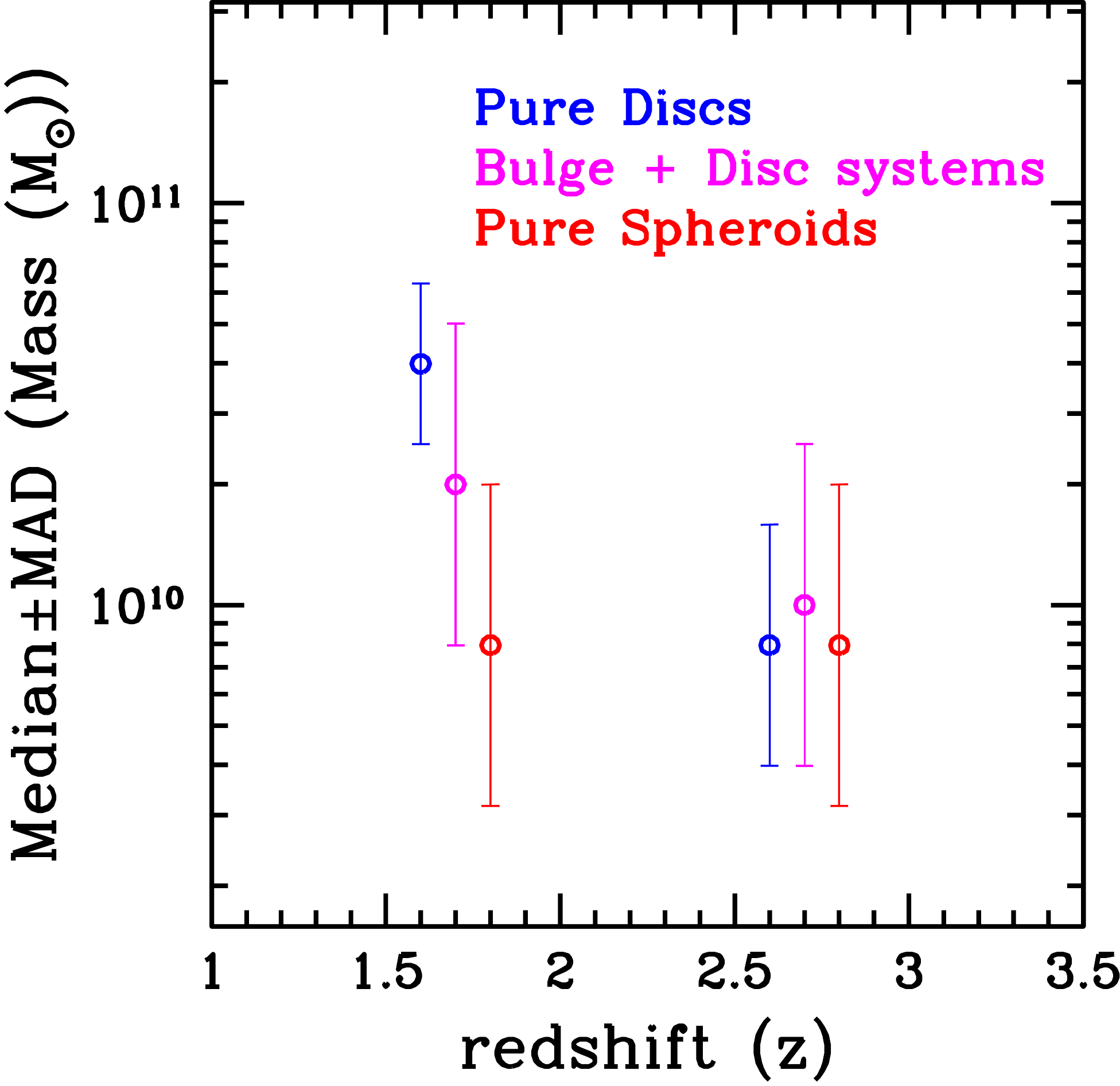}}
\mbox{\includegraphics[width=55mm]{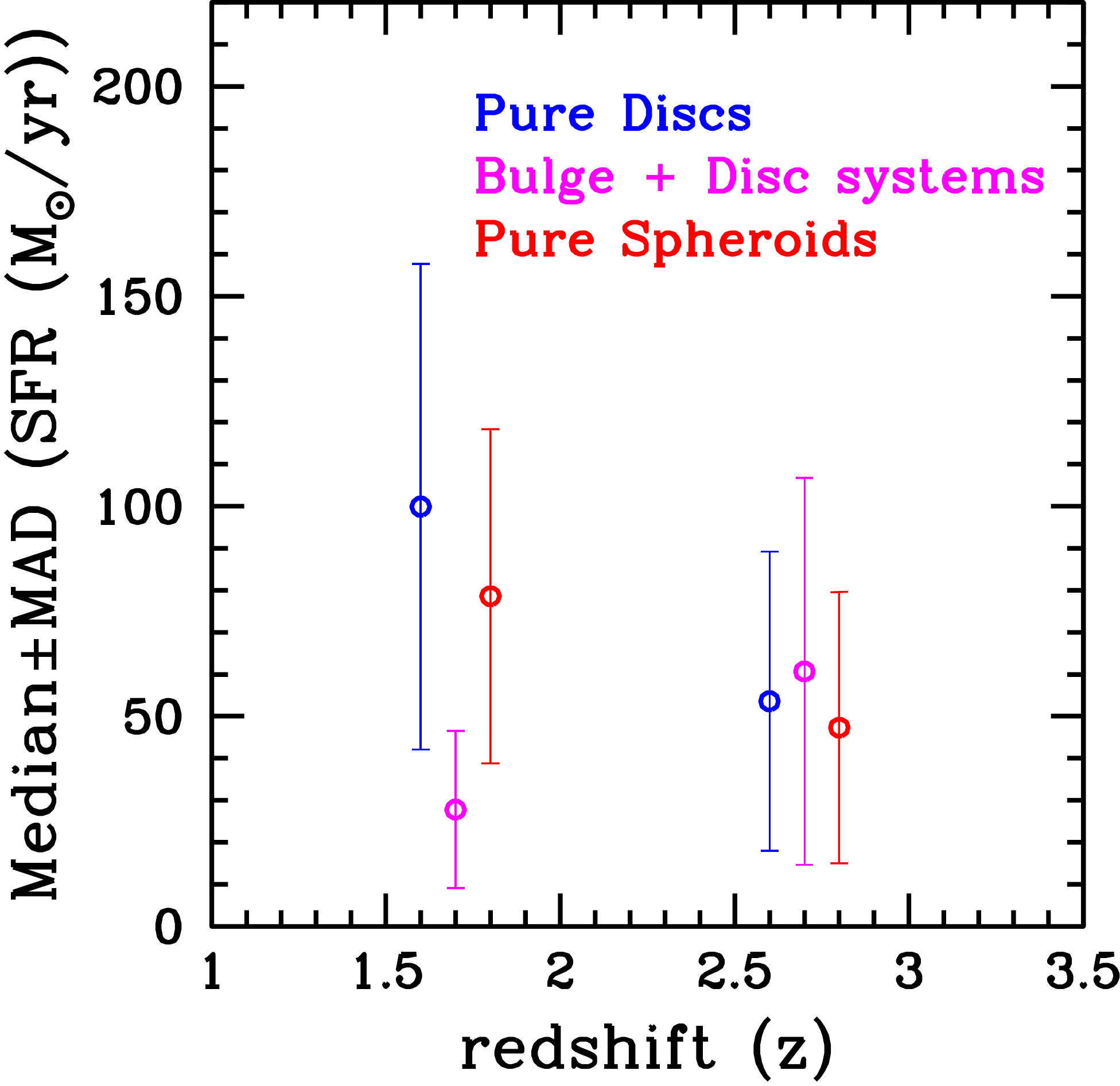}}
\mbox{\includegraphics[width=55mm]{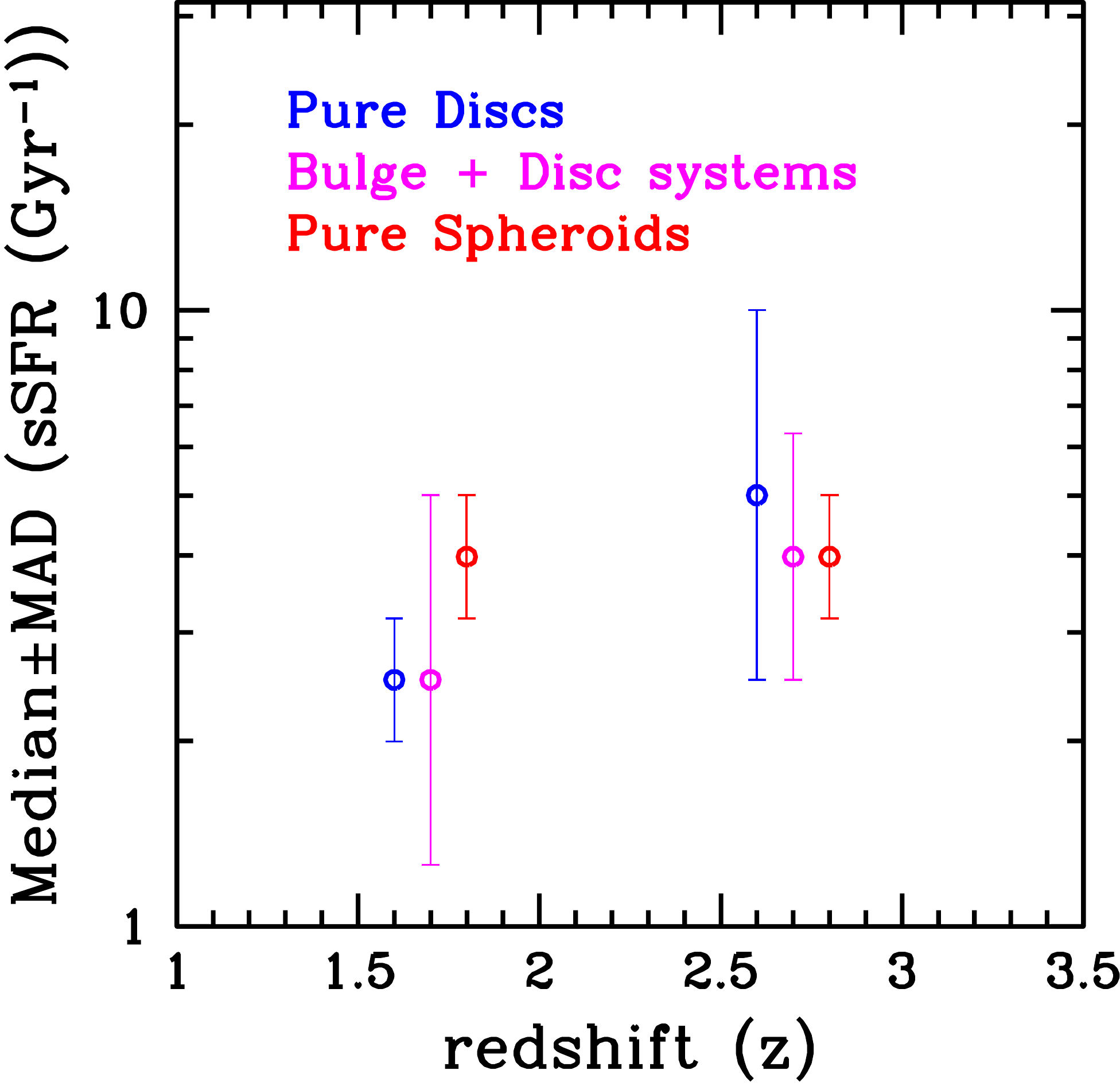}}
\caption{{\bf Median values:} Median values of total stellar mass, star formation rate (SFR) and specific star formation rate (sSFR) for different morphologies (i.e., pure discs, pure spheroids and 2-component systems), for the two redshift ranges, are marked. The error bars in each range refer to the median absolute deviation (MAD) values of that sample.}
\label{stellar-prop}
\end{figure*}

Evidence for this phenomenon is derived from the statistics of their stellar properties as well. In Fig.~\ref{stellar-prop}, median and median absolute deviation (MAD) values of stellar mass, star formation rate (SFR) and specific star formation rate (sSFR) are marked for the three morphological types in the two redshift ranges. While the stellar mass of 2-component systems doubles over the two redshift ranges, their SFR falls by the same factor. Contrary to that, surviving pure spheroids and pure discs register an increase in their star formation rate.  

Note that since pure spheroids and pure discs form a small fraction of the population, especially at the lower redshift range ($\sim$10\% and $\sim$20\% respectively), the overall trend of the full sample resembles that of 2-component systems. Interestingly, evolutionary properties of surviving spheroids are quite dissimilar from that of surviving discs. Crucially, surviving spheroids retain same stellar mass and effective size as their predecessors, whereas, surviving pure discs, in addition to expanding their size by a factor of $\sim$2.5, increase their stellar mass by a factor of $\sim$5 as we reach $z$$\sim$1.5 (Fig.~\ref{stellar-prop}). Their markedly dissimilar behaviour gives support to the argument that spheroids/ellipticals formed prior to $z$$\sim$2 and discs have formed around this epoch.

\section{Discussion}

In this work, we have studied the combined evolution of morphologies and stellar properties of galaxies, around z$\sim$2, by comparing the properties and their correlations at $z$$<$2 (1.5$<$$z$$<$2.0) and $z$$>$2 (2.0$<$$z$$<$4.0). We summarize our major findings below:

\begin{itemize}

\item The fraction of 2-component (bulge$+$disc) systems increases from $\sim$46\% (at $z$$>$2) to $\sim$70\% (at $z$$<$2), at the expense of one component (pure spheroids and pure discs) systems. While for $z$$>$2, 2-component systems dominate more towards the massive end (i.e., $>$ $10^{10.5}$$M_{\odot}$), for $z$$<$2 they dominate at all masses, but more so towards the lower mass end (i.e., $<$ $10^{10}$$M_{\odot}$). 

\item Even while comparing with the most luminous bulge$+$disc systems at $z$$\sim$0, we find that both bulge and host-disc objects in our full sample (1.5$<$$z$$<$4.0) have undergone significant decrease in luminosity ($\sim$3-4 mag) and brightness ($\sim$3-4 mag/arcsec$^2$), along with near doubling of their sizes, from $z$$\sim$2 to the present epoch.

\item Pure spheroids at $z$$\sim$2 follow a Kormendy relation with the same slope as that observed for ellipticals at $z$$\sim$0. But there is a shift in the intercept which signifies that same size galaxies were $\sim$5-6 mag brighter at $z$$\sim$2 than they are at the present epoch.

\item As we move from $z$$>$2 to $z$$<$2, pure-discs register a large increase their size ($\sim$2.5 times) and stellar mass ($\sim$5 times). This is in stark contrast to pure spheroids which maintain similar sizes and stellar mass as their predecessors. Additionally, while host-disc galaxies witness an expansion in scale length ($\sim$1.3 times), their bulge sizes and bulge-to-total light ratios see no evolution. Overall it appears that $z$$\sim$2 is more of a disc formation period, while bulges/spheroids mostly formed prior to this epoch.

\item The main-sequence relation for our sample ($z$$\sim$2) has the same slope as that observed for star-forming galaxies at $z$$\sim$1, with a marked shift (factor of $\sim$5) towards higher SFRs. In our sample, quiescent outliers are all 2-component systems, lying towards the lower redshift range (i.e., $z$$<$2), suggesting that the mechanisms involved in transformation of galaxies are also responsible for the quenching of their star formation activity.   

\end{itemize}

Quantitative morphological decomposition provides evidence for the transformation of one component systems (pure spheroids and pure discs) to 2-component (bulge $+$ disc) systems, around $z$$\sim$2. This is in agreement with \citet{Bruceetal2012} and \citet{Bruceetal2014} who also marked $z$$\sim$2 as the key transition epoch for the formation of 2-component systems. They argued that pure discs do not directly transform into spheroids, rather spheroids are rare all the way till $z$$\sim$1. This is in agreement with our findings at the lower redshift range ($z$$<$2), where the fraction of pure spheroids is significantly less. 

However, for the higher redshift range ($z$$>$2), we do report for the presence ($\sim$20\%) of pure spheroids, which is more in agreement with \citet{Huertas-Companyetal2015} who report that pure spheroids form $\sim$30\% of their sample for $z$$>$2. According to our analysis, pure spheroids are not dense (contrary to the rare red nuggets), with quite low S\'ersic indices ($\sim$0.6) compared to local ellipticals and hence are prone to be classified as pure discs. Their nature gets revealed through their light profile which is not disc like and on the Kormendy plane where they exhibit a tight relation with the same slope as that observed for local ellipticals.

We find that the transformation to 2-component systems is prevalent more for the population of pure spheroids (whose fraction drops to half) than for pure discs (whose fraction falls by 1/3rd). In addition to this, other evidence also indicates $z$$\sim$2 to be more of a disc formation epoch. In agreement with \citet{Margalef-Bentaboletal2016}, we report that while host discs witness a significant expansion in their size, their bulges maintain roughly the same size. We also find that the B/T distribution of 2-component systems overlaps for the two redshift ranges ($z$$<$2, $z$$>$2). Evidence for the discriminative growth of discs, over that of spheroids, emerges directly from the evolution of pure discs and pure spheroids. While pure discs register a marked increase in their size ($\sim$2.5 times) and mass ($\sim$5 times), pure spheroids register no evolution in mass or size.

In agreement with earlier works \citep{Langetal2014,Bruceetal2014,Huertas-Companyetal2015}, we report that transformation of morphologies appears to be linked to the quenching of star formation activity. We find that while SFR increases for the surviving population of pure spheroids and pure discs, it drops down for 2-component systems. All the quiescent outliers to the main-sequence are 2-component systems, lying in the lower redshift range. If formation of the disc is expected to increase stellar activity, quiescent populations might solely be comprising of those 2-component systems which either existed prior to $z$$\sim$2 or formed out of pure discs witnessing bulge formation.

Thus, according to our findings, there are two channels for morphological transformations, i.e., the formation of 2-component systems, around $z$$\sim$2. The first channel is through the formation of bulges inside pure discs, which also leads to the quenching of stellar activity in these systems \citep{Bruceetal2014,Langetal2014,Huertas-Companyetal2015,Margalef-Bentaboletal2018}. This is mostly understood to be the result of the migration of massive clumps, formed in gas rich high redshift discs, to the centre of the discs \citep{Bournaudetal2011,Hopkinsetal2012}. 

The other channel, and according to our results a crucial one, is through the formation of discs around pre-existing pure spheroids. This newly accreted matter (disc) from the halo might have been the residual from a merger or ejected material from stellar/AGN feedback. Addressing the problem that discs are ubiquitous even after large number of mergers, modellers and simulators constantly try to improvise gas physics and feedback mechanisms to prevent all baryonic matter from losing angular momentum and forming a spheroid \citep{Hopkinsetal2009,Governatoetal2010,Scannapiecoetal2012}. Additionally, simulators now agree that regulation of supply to the ISM, mainly through feedback, is critically involved in galaxy formation processes \citep{Schayeetal2010,Duttonetal2010,Vogelsbergeretal2013,Crainetal2015}. With improved algorithms for feedback prescriptions in simulations and observations of gradients of star formation in galaxies, this picture should emerge with greater clarity.
 
\section*{Acknowledgments}

We are thankful to Sandra M. Faber, Luis C. Ho, Christopher J. Conselice, Ran Wang, Francesco Shankar, Yingjie Peng and Jing Wang for useful discussions. RG acknowledges support from the Associateship Program of IUCAA, Pune. AK acknowledges support from the Raja Ramanna Felowship of Department of Atomic Energy, India. We express our gratitude to the referee for insightful suggestions which have improved the quality of this work. 


\bsp

\label{lastpage}

\begin{thebibliography}{69}
\expandafter\ifx\csname natexlab\endcsname\relax\def\natexlab#1{#1}\fi

\bibitem[{{Akhlaghi} \& {Ichikawa}(2015)}]{AkhlaghiandIchikawa2015}
{Akhlaghi} M., {Ichikawa} T., 2015, ApJS, 220, 1

\bibitem[{{Barnes} \& {Hernquist}(1996)}]{BarnesandHernquist1996}
{Barnes} J.~E., {Hernquist} L., 1996, ApJ, 471, 115

\bibitem[{{Bauer} {et~al}\mbox{.}(2011){Bauer}, {Conselice},
  {P{\'e}rez-Gonz{\'a}lez}, {Gr{\"u}tzbauch}, {Bluck}, {Buitrago}, \&
  {Mortlock}}]{Baueretal2011}
{Bauer} A.~E., {Conselice} C.~J., {P{\'e}rez-Gonz{\'a}lez} P.~G.,
  {Gr{\"u}tzbauch} R., {Bluck} A.~F.~L., {Buitrago} F., {Mortlock} A., 2011,
  MNRAS, 417, 289

\bibitem[{{Bertin} \& {Arnouts}(1996)}]{BertinandArnouts1996}
{Bertin} E., {Arnouts} S., 1996, A\&AS, 117, 393

\bibitem[{{Bluck} {et~al}\mbox{.}(2012){Bluck}, {Conselice}, {Buitrago},
  {Gr{\"u}tzbauch}, {Hoyos}, {Mortlock}, \& {Bauer}}]{Blucketal2012}
{Bluck} A.~F.~L., {Conselice} C.~J., {Buitrago} F., {Gr{\"u}tzbauch} R.,
  {Hoyos} C., {Mortlock} A., {Bauer} A.~E., 2012, ApJ, 747, 34

\bibitem[{{Bournaud} {et~al}\mbox{.}(2011){Bournaud}, {Chapon}, {Teyssier},
  {Powell}, {Elmegreen}, {Elmegreen}, {Duc}, {Contini}, {Epinat}, \&
  {Shapiro}}]{Bournaudetal2011}
{Bournaud} F. {et~al.}, 2011, ApJ, 730, 4

\bibitem[{{Bournaud}, {Jog} \& {Combes}(2005){Bournaud}, {Jog}, \&
  {Combes}}]{Bournaudetal2005}
{Bournaud} F., {Jog} C.~J., {Combes} F., 2005, A\&A, 437, 69

\bibitem[{{Brammer} {et~al}\mbox{.}(2012){Brammer}, {van Dokkum}, {Franx},
  {Fumagalli}, {Patel}, {Rix}, {Skelton}, {Kriek}, {Nelson}, {Schmidt},
  {Bezanson}, {da Cunha}, {Erb}, {Fan}, {F{\"o}rster Schreiber}, {Illingworth},
  {Labb{\'e}}, {Leja}, {Lundgren}, {Magee}, {Marchesini}, {McCarthy},
  {Momcheva}, {Muzzin}, {Quadri}, {Steidel}, {Tal}, {Wake}, {Whitaker}, \&
  {Williams}}]{Brammeretal2012}
{Brammer} G.~B. {et~al.}, 2012, ApJS, 200, 13

\bibitem[{{Brinchmann} {et~al}\mbox{.}(2004){Brinchmann}, {Charlot}, {White},
  {Tremonti}, {Kauffmann}, {Heckman}, \& {Brinkmann}}]{Brinchmannetal2004}
{Brinchmann} J., {Charlot} S., {White} S.~D.~M., {Tremonti} C., {Kauffmann} G.,
  {Heckman} T., {Brinkmann} J., 2004, MNRAS, 351, 1151

\bibitem[{{Bruce} {et~al}\mbox{.}(2012){Bruce}, {Dunlop}, {Cirasuolo},
  {McLure}, {Targett}, {Bell}, {Croton}, {Dekel}, {Faber}, {Ferguson},
  {Grogin}, {Kocevski}, {Koekemoer}, {Koo}, {Lai}, {Lotz}, {McGrath}, {Newman},
  \& {van der Wel}}]{Bruceetal2012}
{Bruce} V.~A. {et~al.}, 2012, MNRAS, 427, 1666

\bibitem[{{Bruce} {et~al}\mbox{.}(2014){Bruce}, {Dunlop}, {McLure},
  {Cirasuolo}, {Buitrago}, {Bowler}, {Targett}, {Bell}, {McIntosh}, {Dekel},
  {Faber}, {Ferguson}, {Grogin}, {Hartley}, {Kocevski}, {Koekemoer}, {Koo}, \&
  {McGrath}}]{Bruceetal2014}
{Bruce} V.~A. {et~al.}, 2014, MNRAS, 444, 1001

\bibitem[{{Byun} \& {Freeman}(1995)}]{ByunandFreeman1995}
{Byun} Y.~I., {Freeman} K.~C., 1995, ApJ, 448, 563

\bibitem[{{Calzetti} {et~al}\mbox{.}(2000){Calzetti}, {Armus}, {Bohlin},
  {Kinney}, {Koornneef}, \& {Storchi-Bergmann}}]{Calzettietal2000}
{Calzetti} D., {Armus} L., {Bohlin} R.~C., {Kinney} A.~L., {Koornneef} J.,
  {Storchi-Bergmann} T., 2000, ApJ, 533, 682

\bibitem[{{Chabrier}(2003)}]{Chabrier2003}
{Chabrier} G., 2003, PASP, 115, 763

\bibitem[{{Ciambur}(2016)}]{Ciambur2016}
{Ciambur} B.~C., 2016, PASA, 33, e062

\bibitem[{{Conselice}(2003)}]{Conselice2003}
{Conselice} C.~J., 2003, ApJS, 147, 1

\bibitem[{{Conselice}(2006)}]{Conselice2006}
{Conselice} C.~J., 2006, ApJ, 638, 686

\bibitem[{{Conselice}(2014)}]{Conselice2014}
{Conselice} C.~J., 2014, ARA\&A, 52, 291

\bibitem[{{Conselice} {et~al}\mbox{.}(2013){Conselice}, {Mortlock}, {Bluck},
  {Gr{\"u}tzbauch}, \& {Duncan}}]{Conseliceetal2013}
{Conselice} C.~J., {Mortlock} A., {Bluck} A.~F.~L., {Gr{\"u}tzbauch} R.,
  {Duncan} K., 2013, MNRAS, 430, 1051

\bibitem[{{Crain} {et~al}\mbox{.}(2015){Crain}, {Schaye}, {Bower}, {Furlong},
  {Schaller}, {Theuns}, {Dalla Vecchia}, {Frenk}, {McCarthy}, {Helly},
  {Jenkins}, {Rosas-Guevara}, {White}, \& {Trayford}}]{Crainetal2015}
{Crain} R.~A. {et~al.}, 2015, MNRAS, 450, 1937

\bibitem[{{de Jong}(1996)}]{deJong1996}
{de Jong} R.~S., 1996, A\&AS, 118, 557

\bibitem[{{Dutton}, {van den Bosch} \& {Dekel}(2010){Dutton}, {van den Bosch},
  \& {Dekel}}]{Duttonetal2010}
{Dutton} A.~A., {van den Bosch} F.~C., {Dekel} A., 2010, MNRAS, 405, 1690

\bibitem[{{Elbaz} {et~al}\mbox{.}(2007){Elbaz}, {Daddi}, {Le Borgne},
  {Dickinson}, {Alexander}, {Chary}, {Starck}, {Brandt}, {Kitzbichler},
  {MacDonald}, {Nonino}, {Popesso}, {Stern}, \& {Vanzella}}]{Elbazetal2007}
{Elbaz} D. {et~al.}, 2007, A\&A, 468, 33

\bibitem[{{Freeman}(1977)}]{Freeman1977}
{Freeman} K.~C., 1977, in Evolution of Galaxies and Stellar Populations,
  {Tinsley} B.~M., {Larson} D.~Campbell R.~B.~G., eds., p. 133

\bibitem[{{Giovanelli} {et~al}\mbox{.}(1994){Giovanelli}, {Haynes}, {Salzer},
  {Wegner}, {da Costa}, \& {Freudling}}]{Giovanellietal1994}
{Giovanelli} R., {Haynes} M.~P., {Salzer} J.~J., {Wegner} G., {da Costa} L.~N.,
  {Freudling} W., 1994, AJ, 107, 2036

\bibitem[{{Governato} {et~al}\mbox{.}(2010){Governato}, {Brook}, {Mayer},
  {Brooks}, {Rhee}, {Wadsley}, {Jonsson}, {Willman}, {Stinson}, {Quinn}, \&
  {Madau}}]{Governatoetal2010}
{Governato} F. {et~al.}, 2010, Nature, 463, 203

\bibitem[{{Graham}(2001)}]{Graham2001}
{Graham} A.~W., 2001, MNRAS, 326, 543

\bibitem[{{Graham} \& {Driver}(2005)}]{GrahamandDriver2005}
{Graham} A.~W., {Driver} S.~P., 2005, PASA, 22, 118

\bibitem[{{Grogin} {et~al}\mbox{.}(2011){Grogin}, {Kocevski}, {Faber},
  {Ferguson}, {Koekemoer}, {Riess}, {Acquaviva}, {Alexander}, {Almaini},
  {Ashby}, {Barden}, {Bell}, {Bournaud}, {Brown}, {Caputi}, {Casertano},
  {Cassata}, {Castellano}, {Challis}, {Chary}, {Cheung}, {Cirasuolo},
  {Conselice}, \& {Roshan Cooray}}]{Groginetal2011}
{Grogin} N.~A. {et~al.}, 2011, ApJS, 197, 35

\bibitem[{{Hopkins} {et~al}\mbox{.}(2009){Hopkins}, {Cox}, {Younger}, \&
  {Hernquist}}]{Hopkinsetal2009}
{Hopkins} P.~F., {Cox} T.~J., {Younger} J.~D., {Hernquist} L., 2009, ApJ, 691,
  1168

\bibitem[{{Hopkins} {et~al}\mbox{.}(2012){Hopkins}, {Kere{\v s}}, {Murray},
  {Quataert}, \& {Hernquist}}]{Hopkinsetal2012}
{Hopkins} P.~F., {Kere{\v s}} D., {Murray} N., {Quataert} E., {Hernquist} L.,
  2012, MNRAS, 427, 968

\bibitem[{{Huertas-Company} {et~al}\mbox{.}(2016){Huertas-Company}, {Bernardi},
  {P{\'e}rez-Gonz{\'a}lez}, {Ashby}, {Barro}, {Conselice}, {Daddi}, {Dekel},
  {Dimauro}, {Faber}, {Grogin}, {Kartaltepe}, {Kocevski}, {Koekemoer}, {Koo},
  {Mei}, \& {Shankar}}]{Huertas-Companyetal2016}
{Huertas-Company} M. {et~al.}, 2016, MNRAS, 462, 4495

\bibitem[{{Huertas-Company} {et~al}\mbox{.}(2015){Huertas-Company},
  {P{\'e}rez-Gonz{\'a}lez}, {Mei}, {Shankar}, {Bernardi}, {Daddi}, {Barro},
  {Cabrera-Vives}, {Cattaneo}, {Dimauro}, \&
  {Gravet}}]{Huertas-Companyetal2015}
{Huertas-Company} M. {et~al.}, 2015, ApJ, 809, 95

\bibitem[{{Jedrzejewski}(1987)}]{Jedrzejewski1987}
{Jedrzejewski} R.~I., 1987, MNRAS, 226, 747

\bibitem[{{Jogee} {et~al}\mbox{.}(2009){Jogee}, {Miller}, {Penner}, {Skelton},
  {Conselice}, {Somerville}, {Bell}, {Zheng}, {Rix}, {Robaina}, {Barazza},
  {Barden}, {Borch}, {Beckwith}, {Caldwell}, {Peng}, {Heymans}, {McIntosh},
  {H{\"a}u{\ss}ler}, {Jahnke}, {Meisenheimer}, {Sanchez}, {Wisotzki}, {Wolf},
  \& {Papovich}}]{Jogeeetal2009}
{Jogee} S. {et~al.}, 2009, ApJ, 697, 1971

\bibitem[{{Karim} {et~al}\mbox{.}(2011){Karim}, {Schinnerer},
  {Mart{\'{\i}}nez-Sansigre}, {Sargent}, {van der Wel}, {Rix}, {Ilbert},
  {Smol{\v c}i{\'c}}, {Carilli}, {Pannella}, {Koekemoer}, {Bell}, \&
  {Salvato}}]{Karimetal2011}
{Karim} A. {et~al.}, 2011, ApJ, 730, 61

\bibitem[{{Kent}(1986)}]{Kent1986}
{Kent} S.~M., 1986, AJ, 91, 1301

\bibitem[{{Koekemoer} {et~al}\mbox{.}(2011){Koekemoer}, {Faber}, {Ferguson},
  {Grogin}, {Kocevski}, {Koo}, {Lai}, {Lotz}, {Lucas}, {McGrath}, {Ogaz},
  {Rajan}, {Riess}, {Rodney}, {Strolger}, {Casertano}, {Castellano}, {Dahlen},
  {Dickinson}, {Dolch}, {Fontana}, {Giavalisco}, {Grazian}, {Guo}, {Hathi},
  {Huang}, {van der Wel}, {Yan}, {Acquaviva}, \&
  {Alexander}}]{Koekemoeretal2011}
{Koekemoer} A.~M. {et~al.}, 2011, ApJS, 197, 36

\bibitem[{{Kormendy}(1977)}]{Kormendy1977}
{Kormendy} J., 1977, ApJ, 217, 406

\bibitem[{{Kormendy} {et~al}\mbox{.}(2010){Kormendy}, {Drory}, {Bender}, \&
  {Cornell}}]{Kormendyetal2010}
{Kormendy} J., {Drory} N., {Bender} R., {Cornell} M.~E., 2010, ApJ, 723, 54

\bibitem[{{Kriek} {et~al}\mbox{.}(2009){Kriek}, {van Dokkum}, {Labb{\'e}},
  {Franx}, {Illingworth}, {Marchesini}, \& {Quadri}}]{Krieketal2009}
{Kriek} M., {van Dokkum} P.~G., {Labb{\'e}} I., {Franx} M., {Illingworth}
  G.~D., {Marchesini} D., {Quadri} R.~F., 2009, ApJ, 700, 221

\bibitem[{{La Barbera} {et~al}\mbox{.}(2003){La Barbera}, {Busarello},
  {Merluzzi}, {Massarotti}, \& {Capaccioli}}]{LaBarberaetal2003}
{La Barbera} F., {Busarello} G., {Merluzzi} P., {Massarotti} M., {Capaccioli}
  M., 2003, ApJ, 595, 127

\bibitem[{{Lang} {et~al}\mbox{.}(2014){Lang}, {Wuyts}, {Somerville},
  {F{\"o}rster Schreiber}, {Genzel}, {Bell}, {Brammer}, {Dekel}, {Faber},
  {Ferguson}, {Grogin}, {Kocevski}, {Koekemoer}, {Lutz}, {McGrath}, {Momcheva},
  {Nelson}, {Primack}, {Rosario}, {Skelton}, {Tacconi}, {van Dokkum}, \&
  {Whitaker}}]{Langetal2014}
{Lang} P. {et~al.}, 2014, ApJL, 788, 11

\bibitem[{{Lauberts} \& {Valentijn}(1989)}]{LaubertsandValentijn1989}
{Lauberts} A., {Valentijn} E.~A., 1989, {The surface photometry catalogue of
  the ESO-Uppsala galaxies}

\bibitem[{{Longhetti} {et~al}\mbox{.}(2007){Longhetti}, {Saracco},
  {Severgnini}, {Della Ceca}, {Mannucci}, {Bender}, {Drory}, {Feulner}, \&
  {Hopp}}]{Longhettietal2007}
{Longhetti} M. {et~al.}, 2007, MNRAS, 374, 614

\bibitem[{{Madau} \& {Dickinson}(2014)}]{MadauandDickinson2014}
{Madau} P., {Dickinson} M., 2014, ARA\&A, 52, 415

\bibitem[{{Margalef-Bentabol} {et~al}\mbox{.}(2016){Margalef-Bentabol},
  {Conselice}, {Mortlock}, {Hartley}, {Duncan}, {Ferguson}, {Dekel}, \&
  {Primack}}]{Margalef-Bentaboletal2016}
{Margalef-Bentabol} B., {Conselice} C.~J., {Mortlock} A., {Hartley} W.,
  {Duncan} K., {Ferguson} H.~C., {Dekel} A., {Primack} J.~R., 2016, MNRAS, 461,
  2728

\bibitem[{{Margalef-Bentabol} {et~al}\mbox{.}(2018){Margalef-Bentabol},
  {Conselice}, {Mortlock}, {Hartley}, {Duncan}, {Kennedy}, {Kocevski}, \&
  {Hasinger}}]{Margalef-Bentaboletal2018}
{Margalef-Bentabol} B., {Conselice} C.~J., {Mortlock} A., {Hartley} W.,
  {Duncan} K., {Kennedy} R., {Kocevski} D.~D., {Hasinger} G., 2018, MNRAS, 473,
  5370

\bibitem[{{Momcheva} {et~al}\mbox{.}(2016){Momcheva}, {Brammer}, {van Dokkum},
  {Skelton}, {Whitaker}, {Nelson}, {Fumagalli}, {Maseda}, {Leja}, {Franx},
  {Rix}, {Bezanson}, {Da Cunha}, {Dickey}, {F{\"o}rster Schreiber},
  {Illingworth}, {Kriek}, {Labb{\'e}}, {Ulf Lange}, {Lundgren}, {Magee},
  {Marchesini}, {Oesch}, {Pacifici}, {Patel}, {Price}, {Tal}, {Wake}, {van der
  Wel}, \& {Wuyts}}]{Momchevaetal2016}
{Momcheva} I.~G. {et~al.}, 2016, ApJS, 225, 27

\bibitem[{{Mortlock} {et~al}\mbox{.}(2013){Mortlock}, {Conselice}, {Hartley},
  {Ownsworth}, {Lani}, {Bluck}, {Almaini}, {Duncan}, {van der Wel},
  {Koekemoer}, {Dekel}, {Dav{\'e}}, {Ferguson}, {de Mello}, {Newman}, {Faber},
  {Grogin}, {Kocevski}, \& {Lai}}]{Mortlocketal2013}
{Mortlock} A. {et~al.}, 2013, MNRAS, 433, 1185

\bibitem[{{Peebles} \& {Nusser}(2010)}]{PeeblesandNusser2010}
{Peebles} P.~J.~E., {Nusser} A., 2010, Nature, 465, 565

\bibitem[{{Peng} {et~al}\mbox{.}(2002){Peng}, {Ho}, {Impey}, \&
  {Rix}}]{Pengetal2002}
{Peng} C.~Y., {Ho} L.~C., {Impey} C.~D., {Rix} H.-W., 2002, AJ, 124, 266

\bibitem[{{Putman}(2017)}]{Putman2017}
{Putman} M.~E., 2017, in Astrophysics and Space Science Library, Vol. 430, Gas
  Accretion onto Galaxies, {Fox} A., {Dav{\'e}} R., eds., p.~1

\bibitem[{{Sachdeva}(2013)}]{Sachdeva2013}
{Sachdeva} S., 2013, MNRAS, 435, 1186

\bibitem[{{Sachdeva} {et~al}\mbox{.}(2015){Sachdeva}, {Gadotti}, {Saha}, \&
  {Singh}}]{Sachdevaetal2015}
{Sachdeva} S., {Gadotti} D.~A., {Saha} K., {Singh} H.~P., 2015, MNRAS, 451, 2

\bibitem[{{Sachdeva} \& {Saha}(2016)}]{SachdevaandSaha2016}
{Sachdeva} S., {Saha} K., 2016, ApJL, 820, L4

\bibitem[{{Sachdeva} \& {Saha}(2018)}]{SachdevaandSaha2018}
{Sachdeva} S., {Saha} K., 2018, MNRAS, 478, 41

\bibitem[{{Sachdeva}, {Saha} \& {Singh}(2017){Sachdeva}, {Saha}, \&
  {Singh}}]{Sachdevaetal2017}
{Sachdeva} S., {Saha} K., {Singh} H.~P., 2017, ApJ, 840, 79

\bibitem[{{Santini} {et~al}\mbox{.}(2014){Santini}, {Maiolino}, {Magnelli},
  {Lutz}, {Lamastra}, {Li Causi}, {Eales}, {Andreani}, {Berta}, {Buat},
  {Cooray}, {Cresci}, {Daddi}, {Farrah}, {Fontana}, {Franceschini}, {Genzel},
  {Granato}, {Grazian}, {Le Floc'h}, {Magdis}, {Magliocchetti}, {Mannucci},
  {Menci}, {Nordon}, {Oliver}, {Popesso}, {Pozzi}, {Riguccini}, {Rodighiero},
  {Rosario}, {Salvato}, {Scott}, {Silva}, {Tacconi}, {Viero}, {Wang}, {Wuyts},
  \& {Xu}}]{Santinietal2014}
{Santini} P. {et~al.}, 2014, A\&A, 562, A30

\bibitem[{{Scannapieco} {et~al}\mbox{.}(2012){Scannapieco}, {Wadepuhl},
  {Parry}, {Navarro}, {Jenkins}, {Springel}, {Teyssier}, {Carlson}, {Couchman},
  {Crain}, {Dalla Vecchia}, {Frenk}, {Kobayashi}, {Monaco}, {Murante},
  {Okamoto}, {Quinn}, {Schaye}, {Stinson}, {Theuns}, {Wadsley}, {White}, \&
  {Woods}}]{Scannapiecoetal2012}
{Scannapieco} C. {et~al.}, 2012, MNRAS, 423, 1726

\bibitem[{{Schaye} {et~al}\mbox{.}(2010){Schaye}, {Dalla Vecchia}, {Booth},
  {Wiersma}, {Theuns}, {Haas}, {Bertone}, {Duffy}, {McCarthy}, \& {van de
  Voort}}]{Schayeetal2010}
{Schaye} J. {et~al.}, 2010, MNRAS, 402, 1536

\bibitem[{{Scoville} {et~al}\mbox{.}(2016){Scoville}, {Sheth}, {Aussel},
  {Vanden Bout}, {Capak}, {Bongiorno}, {Casey}, {Murchikova}, {Koda},
  {{\'A}lvarez-M{\'a}rquez}, {Lee}, {Laigle}, {McCracken}, {Ilbert}, {Pope},
  {Sanders}, {Chu}, {Toft}, {Ivison}, \& {Manohar}}]{Scovilleetal2016}
{Scoville} N. {et~al.}, 2016, ApJ, 820, 83

\bibitem[{{Simard} {et~al}\mbox{.}(2011){Simard}, {Mendel}, {Patton},
  {Ellison}, \& {McConnachie}}]{Simardetal2011}
{Simard} L., {Mendel} J.~T., {Patton} D.~R., {Ellison} S.~L., {McConnachie}
  A.~W., 2011, ApJS, 196, 11

\bibitem[{{Skelton} {et~al}\mbox{.}(2014){Skelton}, {Whitaker}, {Momcheva},
  {Brammer}, {van Dokkum}, {Labb{\'e}}, {Franx}, {van der Wel}, {Bezanson}, {Da
  Cunha}, {Fumagalli}, {F{\"o}rster Schreiber}, {Kriek}, {Leja}, {Lundgren},
  {Magee}, {Marchesini}, {Maseda}, {Nelson}, {Oesch}, {Pacifici}, {Patel},
  {Price}, {Rix}, {Tal}, {Wake}, \& {Wuyts}}]{Skeltonetal2014}
{Skelton} R.~E. {et~al.}, 2014, ApJS, 214, 24

\bibitem[{{Springel} \& {Hernquist}(2005)}]{SpringelandHernquist2005}
{Springel} V., {Hernquist} L., 2005, ApJL, 622, L9

\bibitem[{{Trujillo} {et~al}\mbox{.}(2001){Trujillo}, {Aguerri}, {Cepa}, \&
  {Guti{\'e}rrez}}]{Trujilloetal2001}
{Trujillo} I., {Aguerri} J.~A.~L., {Cepa} J., {Guti{\'e}rrez} C.~M., 2001,
  MNRAS, 328, 977

\bibitem[{{Vogelsberger} {et~al}\mbox{.}(2013){Vogelsberger}, {Genel},
  {Sijacki}, {Torrey}, {Springel}, \& {Hernquist}}]{Vogelsbergeretal2013}
{Vogelsberger} M., {Genel} S., {Sijacki} D., {Torrey} P., {Springel} V.,
  {Hernquist} L., 2013, MNRAS, 436, 3031

\bibitem[{{Whitaker} {et~al}\mbox{.}(2014){Whitaker}, {Franx}, {Leja}, {van
  Dokkum}, {Henry}, {Skelton}, {Fumagalli}, {Momcheva}, {Brammer}, {Labb{\'e}},
  {Nelson}, \& {Rigby}}]{Whitakeretal2014}
{Whitaker} K.~E. {et~al.}, 2014, ApJ, 795, 104

\bibitem[{{Wuyts} {et~al}\mbox{.}(2008){Wuyts}, {Labb{\'e}}, {F{\"o}rster
  Schreiber}, {Franx}, {Rudnick}, {Brammer}, \& {van Dokkum}}]{Wuytsetal2008}
{Wuyts} S., {Labb{\'e}} I., {F{\"o}rster Schreiber} N.~M., {Franx} M.,
  {Rudnick} G., {Brammer} G.~B., {van Dokkum} P.~G., 2008, ApJ, 682, 985

\end{thebibliography}
\end{document}